\documentclass[aps,prd,twocolumn,superscriptaddress,nofootinbib]{revtex4-1}
\usepackage{amsmath}
\usepackage[scientific-notation=true]{siunitx}
\usepackage{graphicx}
\usepackage{hyperref}
\usepackage{placeins}
\usepackage{bm}
\usepackage{enumitem}
\global\hyphenpenalty=450
\global\interfootnotelinepenalty=10000
\let\vec\bm
\DeclareMathOperator{\sinc}{sinc}
\DeclareMathOperator{\erf}{erf}
\newcommand{\appropto}{\mathrel{\vcenter{
			\offinterlineskip\halign{\hfil$##$\cr
				\propto\cr\noalign{\kern2pt}\sim\cr\noalign{\kern-2pt}}}}}

\begin{document}
\title{Predicting the density profiles of the first halos}
\author{M. Sten Delos}
\email{Electronic address: delos@unc.edu}
\author{Margie Bruff}
\affiliation{Department of Physics and Astronomy, University of North Carolina at Chapel Hill, Phillips Hall CB3255, Chapel Hill, North Carolina 27599, USA}
\author{Adrienne L. Erickcek}
\email{Electronic address: erickcek@physics.unc.edu}
\affiliation{Department of Physics and Astronomy, University of North Carolina at Chapel Hill, Phillips Hall CB3255, Chapel Hill, North Carolina 27599, USA}

\begin{abstract}
The first dark matter halos form by direct collapse from peaks in the matter density field, and evidence from numerical simulations and other analyses suggests that the dense inner regions of these objects largely persist today.  These halos would be the densest dark matter structures in the Universe, and their abundance can probe processes that leave imprints on the primordial density field, such as inflation or an early matter-dominated era.  They can also probe dark matter through its free-streaming scale.  The first halos are qualitatively different from halos that form by hierarchical clustering, as evidenced by their $\rho\propto r^{-3/2}$ inner density profiles.  In this work, we present and tune models that predict the density profiles of these halos from properties of the density peaks from which they collapsed.  These models predict the coefficient $A$ of the $\rho=Ar^{-3/2}$ small-radius asymptote of the density profile along with the maximum circular velocity $v_\mathrm{max}$ and associated radius $r_\mathrm{max}$.  These models are universal; they can be applied to any cosmology, and we confirm this by validating them using six $N$-body simulations carried out in wildly disparate cosmological scenarios.  We find that these models can even predict the full population of halos with reasonable accuracy in scenarios with narrowly supported power spectra, although for broader power spectra, an understanding of the impact of halo mergers is needed.  With their connection to the primordial density field established, the first dark matter halos will serve as probes of the early Universe and the nature of dark matter.
\end{abstract}

                              
\maketitle

\section{Introduction}

Decades of work have been devoted to understanding the halos that form by the gravitational collapse of collisionless dark matter.  In the cold dark matter (CDM) model, $N$-body simulations have demonstrated that the radial density profiles $\rho(r)$ of these halos are well described by a remarkably universal form, known as the Navarro-Frenk-White (NFW) profile \cite{navarro1996structure,navarro1997universal}, which is a double power law that transitions from $\rho\propto r^{-1}$ at small radii to $\rho\propto r^{-3}$ at large radii.  This profile has received only minor corrections since its introduction (e.g., Ref.~\cite{navarro2010diversity}).  In the CDM picture, all halos form by hierarchical clustering of smaller halos, and the NFW profile appears to be the generic consequence.

CDM represents an idealized scenario, however.  In reality, dark matter particles are expected to have a nonzero temperature, and the corresponding random particle motions wash out density fluctuations smaller than a characteristic free-streaming scale.  The first halos form by direct collapse of overdense regions at this scale, and $N$-body simulations with sufficient resolution show that these halos possess a markedly different density profile that asymptotes to $\rho\propto r^{-3/2}$ at small radii \cite{ishiyama2010gamma,anderhalden2013density,*anderhalden2013erratum,ishiyama2014hierarchical,polisensky2015fingerprints,ogiya2017sets,angulo2017earth}.  This profile is stable; it does not relax to the NFW profile, at least in the absence of halo mergers.

The first halos subsequently merge to produce successively larger halos, and Refs.~\cite{ogiya2016dynamical,angulo2017earth} find that these mergers gradually drive the halos' density profiles toward the NFW form.  As these works argue, the shallowing of the inner density profile is likely a consequence of violent relaxation \cite{lynden1967statistical} during merger events.  However, there are multiple lines of evidence suggesting that such relaxation does not erase the memory of prior states.  For nearly equal-mass mergers, the density profile of the merger remnant depends sensitively on the profiles of its progenitors \cite{polisensky2015fingerprints,angulo2017earth}.  Successive mergers either raise or leave unaltered a halo's characteristic density \cite{ogiya2016dynamical,angulo2017earth,drakos2018major}; since the first halos are the densest, this trend preserves them.  Finally, for highly unequal-mass mergers, the smaller halo is generally expected to survive as a subhalo of the larger \cite{berezinsky2008remnants}.  The smaller halo's central density profile may even be mostly unaffected by the merger \cite{hayashi2003structural,penarrubia2010impact,ogiya2019dash}.

In this way, it is broadly plausible that the dense central regions of the first halos survive the hierarchical clustering process.  These halos, forming during a denser epoch, should be the densest dark matter objects in the Universe, and such density leads to observational prospects.  Signals from dark matter annihilation are dramatically enhanced by the high density within these halos (e.g., Refs.~\cite{diemand2007dark,strigari2007precise,pieri2008dark,jeltema2008searching,bartels2015boosting,ando2019halo}).  Gravitational signatures, whether through microlensing (e.g., Refs.~ \cite{boden1998astrometric,dominik2000astrometric,erickcek2011astrometric,van2018halometry}), timing delays (e.g., Ref.~\cite{siegel2007probing}), or stellar dynamics (e.g., Refs.~\cite{erkal2015forensics,erkal2016number,buschmann2018stellar}), are also enhanced.  Meanwhile, since these halos form by direct collapse from overdense patches, they carry sensitive information about the primordial density field on scales that are inaccessible to other probes.  Our goal is to develop a model that extracts this information; we aim to connect the properties of the first halos to the statistics of the primordial matter density field.

The matter density field is intimately connected to some of the most fundamental questions of cosmology.  The power spectrum $\mathcal{P}(k)$, which quantifies the power in density fluctuations at scale wave number $k$, has been precisely measured at scales above approximately $1$ Mpc using the cosmic microwave background \cite{hlozek2012atacama} and the Lyman-$\alpha$ forest \cite{bird2011minimally}.  However, with a few exceptions \cite{josan2009generalized,chluba2012probing}, it is largely unconstrained at submegaparsec scales.  Since density fluctuations are thought to have been seeded during inflation, $\mathcal{P}(k)$ serves as a valuable probe of inflationary models \cite{lidsey1997reconstructing}.  The power spectrum is also sensitive to the thermal history of the Universe after inflation.  An early matter-dominated era (EMDE), driven by an unstable heavy relic, would amplify small-scale density fluctuations \cite{erickcek2011reheating,barenboim2014structure,fan2014nonthermal,erickcek2015dark}, as would a period of domination by a fast-rolling scalar field \cite{redmond2018growth}.  The power spectrum $\mathcal{P}(k)$ thereby supplies one of the few windows into the Universe prior to big bang nucleosynthesis.  Separately, its sensitivity to the free-streaming scale makes it a probe of the nature of dark matter.

Numerous previous works have explored the prospects of using the first dark matter halos to probe the small-scale primordial power spectrum \cite{josan2010gamma,bringmann2012improved,li2012new,yang2013neutrino,yang2013dark,yang2014constraints,clark2015investigatingII,*clark2017erratumII,yang2017tau,nakama2018constraints}.  However, most treatments assumed that these halos, if they form sufficiently early, possess a particularly compact $\rho\propto r^{-9/4}$ density profile \cite{ricotti2009new} derived from self-similar theory \cite{bertschinger1985self}.  References \cite{gosenca20173d} and~\cite{delos2018ultracompact} (Paper I) used $N$-body simulations to show that this profile does not arise from a realistic formation scenario.  In Ref.~\cite{delos2018density}, hereafter Paper II, we showed that, despite possessing shallower $\rho\propto r^{-3/2}$ density profiles, the first halos can still supply competitive constraints on the power spectrum through nondetection of their observable signals.  In that work, we used scaling arguments to model the population of the first halos given a particular family of power spectra.  Each peak in the density field was mapped to a collapsed halo at later time.

Our present work represents a natural extension of that model to arbitrary power spectra.  It is based on the notion that the density profile of a halo forming by direct collapse is uniquely related to the properties of its precursor density peak.  Numerous prior works have explored the problem of explaining a halo's density profile in terms of the structure of the density peak whence it collapsed.  In work that pioneered the so-called spherical infall model, Refs.~\cite{gunn1972infall,gott1975formation} approximated the density profile of a collapsed halo by employing the simplifying assumption that particle orbits are unaltered after accretion.  Subsequent works extended this model by including the contraction of orbiting material due to new accretion \cite{gunn1977massive,fillmore1984self,bertschinger1985self,hoffman1985local,ryden1987galaxy,zaroubi1993gravitational,lokas2000formation,dalal2010origin}, relaxing the assumption of spherical symmetry \cite{ryden1993self,lithwick2011self}, and modeling nonradial motions \cite{white1992models,nusser2001self,ascasibar2004physical,lu2006origin,ascasibar2007secondary,zukin2010self,zukin2010velocity}.  Because of the difficult, nonlinear nature of the matter infall problem, every treatment employs simplifying assumptions.  Exact solutions only exist for the self-similar case \cite{fillmore1984self,bertschinger1985self,ryden1993self,lithwick2011self}, in which the primordial mass excess is a power law in radius.  More general treatments employ Ans\"atze related to angular momenta and orbital contraction.  As an alternative to tracking each orbit, other works have employed an Ansatz related to virialization \cite{salvador2012theoretical,juan2014fixing}.

Whereas spherical infall models relate halo density profiles to the precursor mass distribution, a complementary paradigm empirically studies the distribution of halo density profiles as a function of cosmology and redshift.  In this paradigm, cosmological $N$-body simulations are used to tune a parametric model that describes the density profile of a halo as a function of its mass, and models built in this paradigm are known as concentration-mass relations.\footnote{The concentration is a parameter in the NFW profile (and extended to other profiles) describing how centrally distributed the halo's mass is.}  The distribution of halo masses can be subsequently obtained using Press-Schechter theory \cite{press1974formation,sheth2001ellipsoidal,sheth2002excursion}, so these models can predict the full population of halos and their density profiles.
Concentration-mass relations have been studied extensively.  The simplest models describe the halo distribution in a particular cosmological scenario and at a particular time of interest \cite{avila1999density,jing2000density,colin2004dwarf,dolag2004numerical,avila2005dependence,neto2007statistics,duffy2008dark,*duffy2011erratum,gao2008redshift,maccio2008concentration,klypin2011dark,munoz2011redshift,bhattacharya2013dark,dutton2014cold,heitmann2015q,klypin2016multidark,hellwing2016copernicus,angel2016dark,child2018halo}.  Other works have framed a halo's density profile in terms of its age or assembly history \cite{navarro1996structure,navarro1997universal,bullock2001profiles,eke2001power,wechsler2002concentrations,zhao2003mass,zhao2009accurate,giocoli2012formation,ludlow2013mass,ludlow2014mass,bosch2014coming,correa2015accretion,ludlow2016mass}, and progress has been made in isolating the physical variables most directly relevant for predicting density profiles \cite{prada2012halo,okoli2015concentration,diemer2015universal,diemer2019accurate}.  Nevertheless, due to their empirical nature, concentration-mass relations do not readily extend beyond the cosmological scenarios, times, and halo-mass ranges over which they are tuned.

Broadly, spherical infall models attempt to explain the structures of halos from first principles, while concentration-mass relations endeavor to predict these structures pragmatically.  Our analysis constitutes a hybrid between these two procedures that is specialized to the first halos.  Forming by direct collapse, these halos are well  suited to the spherical infall description.  At the smallest radii, we use ellipsoidal collapse arguments \cite{bond1996peak,sheth2001ellipsoidal,sheth2002excursion} to predict the coefficient of the $\rho\propto r^{-3/2}$ inner profile.  Beyond the inner asymptote, we employ the simplest spherical infall models to predict the larger profile, parametrized by the maximum circular velocity $v_\mathrm{max}$ and the radius $r_\mathrm{max}$ at which it is attained \cite{ascasibar2008dynamical}.  By building from such first-principles descriptions, our predictive models are valid in any cosmological scenario; we demonstrate this by validating and tuning the models using six high-resolution cosmological $N$-body simulations carried out in wildly disparate cosmological scenarios.  Our models nominally predict a halo's density profile from the density peak whence it collapsed.  However, modulo the influence of halo mergers (which we discuss), the statistics of peaks \cite{bardeen1986statistics} may be applied to thereby predict the full halo population at a given time.

This paper is structured as follows.  In Sec.~\ref{sec:sims}, we detail our simulations and the procedure we use to connect halos with density peaks.  Section~\ref{sec:asymptote} develops a model that predicts the small-radius asymptote $\rho\propto r^{-3/2}$ of a halo's density profile, while Sec.~\ref{sec:rmax} compares models that predict the profile at larger radii.  In Sec.~\ref{sec:summary}, we discuss prospects for predicting whole populations of halos, including discussion of halo mergers in Sec.~\ref{sec:merge}.  Section~\ref{sec:conclusion} concludes and discusses avenues for future work.  Appendixes \ref{sec:halos} and~\ref{sec:peaks} further detail how we extract simulation data.  Finally, in Appendix~\ref{sec:stats}, we present a procedure to directly sample halos from a power spectrum using our model.

\section{Simulations}\label{sec:sims}

\begin{figure}[t]
	\centering
	\includegraphics[width=\columnwidth]{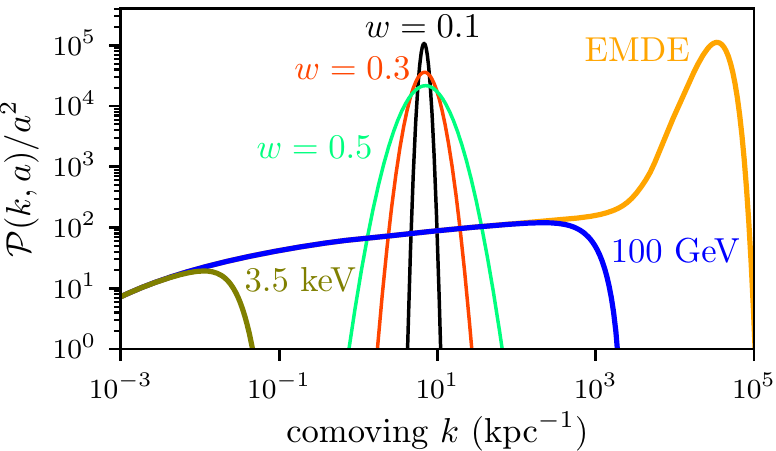}\\
	\includegraphics[width=\columnwidth]{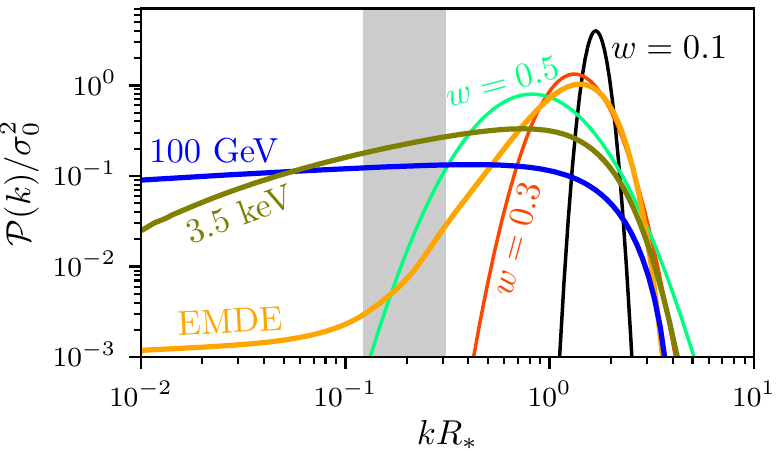}
	\caption{\label{fig:power} Top: The (linear-theory) dimensionless matter power spectra of our six simulations during matter domination.  In linear theory, $\mathcal{P}(k,a)\propto a^2$ when matter dominates, which motivates the $y$-axis scaling (with $a=1$ today).  Bottom: The same power spectra, but each is scaled to its density variance and characteristic correlation length (see the text).  The shaded region marks the range of box sizes for the six simulations: larger modes (smaller $k$) have wavelengths longer than the simulation box, so they are not sampled.}
\end{figure}

We first build a halo catalogue on which to test our model.  For this purpose, we simulated six different simulation boxes drawn from different initial power spectra.  These power spectra are shown in Fig.~\ref{fig:power}.  Three of these power spectra are constructed as ``spikes'' centered at the scale $k_s=6.8$ kpc$^{-1}$ with the form
\begin{equation}\label{spike}
\mathcal{P}(k) = \frac{\mathcal{A}}{\sqrt{2\pi}w}\exp\!\left[-\frac{1}{2}\left(\frac{\ln(k/k_s)}{w}\right)^2\right]
\end{equation}
for different values of $w$.  These spectra are primarily intended as artificial test beds for halo formation, but they do have qualitative motivations in inflationary phenomenology \cite{salopek1989designing,starobinsky1992,ivanov1994inflation,randall1996supernatural,starobinsky1998beyond,martin2000nonvacuum,chung2000probing,barnaby2009particle,barnaby2010features,bugaev2011curvature}.  A fourth power spectrum represents the impact of an EMDE with reheat temperature $T_\mathrm{RH}=100$ MeV and ratio $k_\mathrm{cut}/k_\mathrm{RH}=20$ between the free-streaming cutoff and the largest scale affected by the EMDE (see Ref.~\cite{erickcek2015dark}).  Finally, the last two power spectra include only the free-streaming cutoffs associated with cold dark matter with mass $m_\chi=100$ GeV and kinetic decoupling temperature $T_\mathrm{kd}=33$ MeV (corresponding to a typical weakly interacting massive particle \cite{jungman1996supersymmetric}) and warm dark matter (WDM)\footnote{For WDM, we also take the dark matter particle to be fermionic with 2 degrees of freedom (so, $g_X=1.5$ in Ref.~\cite{bode2001halo}).} with mass $m_\chi=3.5$ keV (close to lower bounds from the Lyman-$\alpha$ forest \cite{viel2013warm}), respectively.  The particular parameter choices for these power spectra are intended to represent very different cosmological scenarios so that we can test the broad applicability of our models.

We generate the three spiked power spectra from Eq.~(\ref{spike}) at a starting redshift of $z=10^4$, normalizing $\mathcal{A}=2\times 10^{-7}$ at this time in order to effect abundant halo formation by $z\sim 100$.  To generate the latter three power spectra, we begin with a primordial curvature power spectrum of amplitude $\mathcal{A}_s=2.142\times 10^{-9}$ and spectral index $n_s = 0.9667$ \cite{2016planck}.  We then use the Boltzmann solver \textsc{Camb Sources} \cite{challinor2011linear,lewis200721} to produce a matter power spectrum at $z=500$ in a scenario with no thermal dark matter motion.  Next, we apply transfer functions from Refs. \cite{erickcek2015dark}, \cite{green2004power}, and~\cite{bode2001halo} to create, respectively, the EMDE, CDM, and WDM spectra.  Finally, we use linear theory \cite{hu1996small} to evolve these power spectra to their initial redshifts.  Note that we include radiation in our simulations, and the linear evolution accounts for this; see Paper II for details.\footnote{Modes that were subhorizon during an EMDE grow in an altered way during radiation domination \cite{erickcek2015dark}, and we modify the linear growth function for this scenario accordingly.}  We also include a cosmological constant, although it is of minimal relevance.

The starting redshift $z_\mathrm{init}$, ending redshift $z_\mathrm{final}$, and periodic box size of each simulation are listed in Table~\ref{tab:sims}.  Also listed are the rms density variance $\sigma_0$ and comoving characteristic correlation length $R_*$ associated with each power spectrum in linear theory.  These spectral parameters are defined as \cite{bardeen1986statistics}
\begin{align}\label{sigmaj}
\sigma_j &= \left(\int_0^\infty \frac{dk}{k}\mathcal{P}(k) k^{2j}\right)^{1/2}
\\\label{Rstar}
R_* &= \sqrt 3 \frac{\sigma_1}{\sigma_2}.
\end{align}
To illustrate the significance of these quantities, Fig.~\ref{fig:power} also shows the correspondingly rescaled power spectra.  This rescaling is useful because it factors out the scale differences between the spectra, leaving only their shapes.  For this reason, we will use these spectral parameters as our units.  The mass unit is $m_\mathrm{unit}\equiv \bar\rho_0 R_*^3$, where $\bar\rho_0$ is the background matter density today.  Meanwhile, the physical distance unit is $r_\mathrm{unit}\equiv [a/\sigma_0(a)]R_*$, where $\sigma_0(a)$ is evaluated using linear theory during matter domination (so $\sigma_0\propto a$) and $a=1$ today.  For each simulation, the ending redshift is tuned so that $\sigma_0\sim 3$ (in linear theory) at simulation termination, and the comoving box size is tuned to be roughly 30 to 80 times $R_*$.  The initial redshift is chosen so that the largest fractional density excesses ${\delta\equiv(\rho-\bar\rho)/\bar\rho}$ are of order $0.2$ or smaller.

\begin{table}[t]
	\caption{\label{tab:sims} The simulation list with basic parameters.  Also listed are $\sigma_0$, the rms density variance, and $R_*$, the comoving characteristic correlation length, both calculated in linear theory.  When matter dominates, $\sigma(a)\propto a$, which motivates the scaling in this column (with $a=1$ today).}
	\begin{ruledtabular}
		\begin{tabular}{cccccc}
			Simulation & $z_\mathrm{init}$ & $z_\mathrm{final}$ & Box (kpc) & $\sigma_0(a)/a$ & $R_*$ (kpc) \\
			\hline\vspace{-3mm}\\
			$w=0.1$ & $10^4$ & $50$ & $7.4$ & $164$ & $0.25$ \\
			$w=0.3$ & $10^4$ & $50$ & $7.4$ & $164$ & $0.19$ \\
			$w=0.5$ & $10^4$ & $50$ & $7.4$ & $164$ & $0.12$ \\
			EMDE & $3\times 10^4$ & $100$ & $1.5\times 10^{-3}$ & $330$ & $4.1\times 10^{-5}$ \\
			$100$ GeV & $500$ & $9$ & $0.15$ & $30$ & $1.9\times 10^{-3}$ \\
			$3.5$ keV & $150$ & $2$ & $4.4\times 10^3$ & $7.6$ & $66$ \\
		\end{tabular}
	\end{ruledtabular}
\end{table}

For each simulation, we draw a random density field from the matter power spectrum at $z_\mathrm{init}$ and generate initial conditions using the Zel'dovich approximation modified to account for radiation as described in Paper II.\footnote{For the EMDE, the methods in Paper II were adapted to the post-EMDE growth function.}  Finally, we carry out the simulations using a version of the cosmological simulation code $\textsc{Gadget-2}$ \cite{springel2005cosmological,springel2001gadget} that we modified to include the effects of radiation (see Paper II).  All simulations employed $1024^3$ particles and a comoving force softening length set at $3\%$ of the initial interparticle spacing (with forces becoming non-Newtonian at 2.8 times this length).

\begin{figure*}[t]
	\centering
	\includegraphics[width=.333\linewidth]{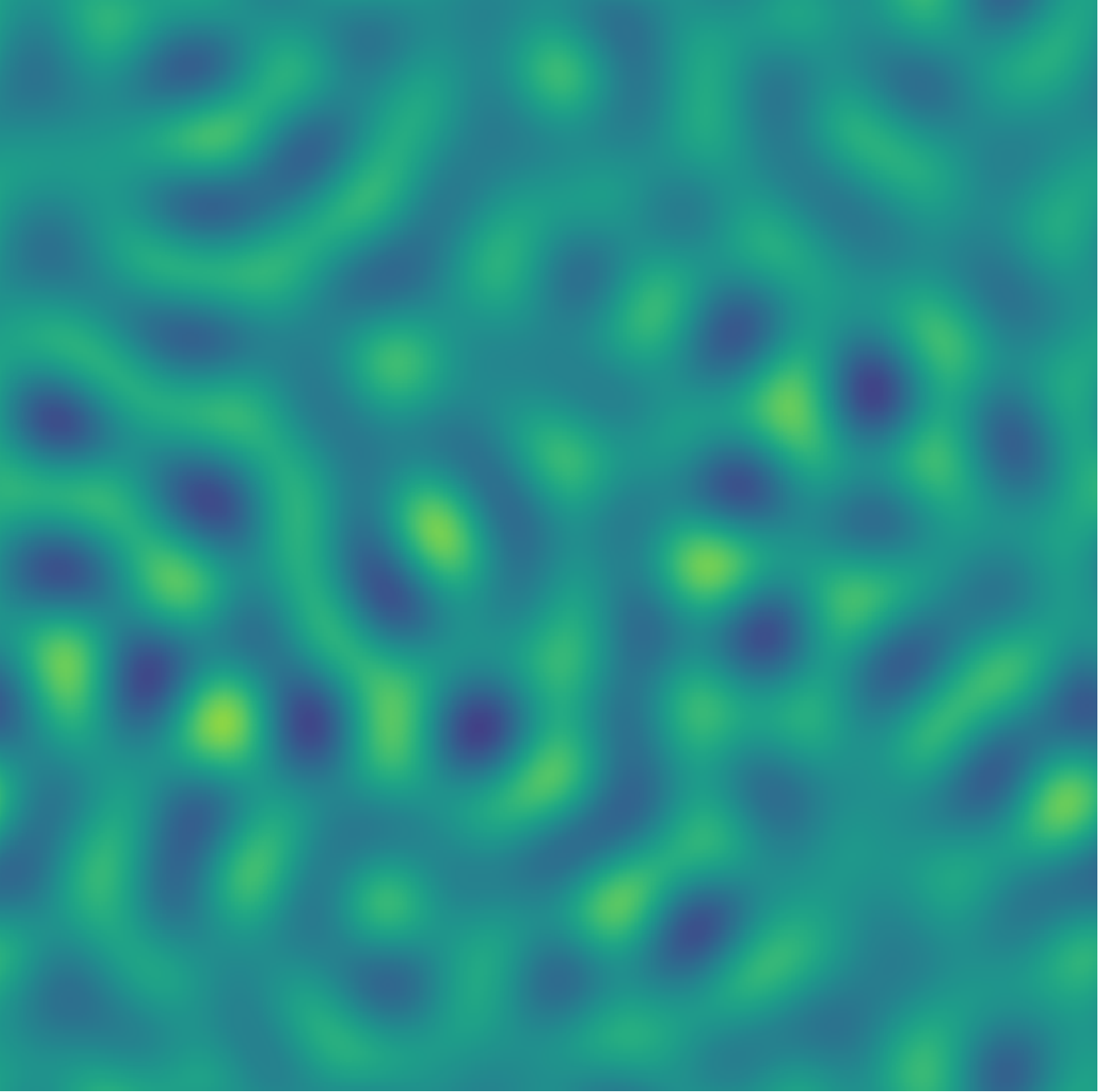}\hspace{-1mm}
	\includegraphics[width=.333\linewidth]{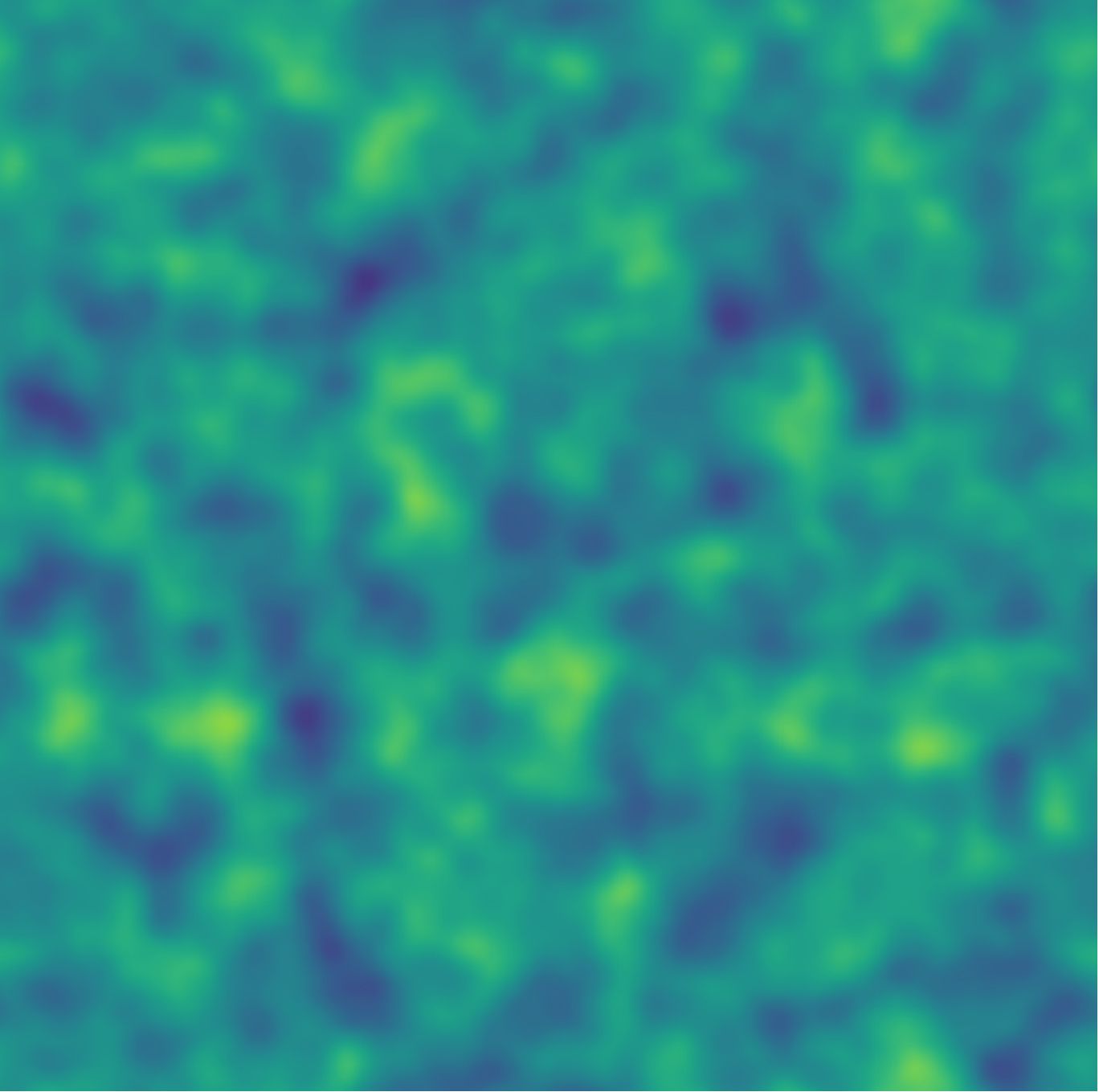}\hspace{-1mm}
	\includegraphics[width=.333\linewidth]{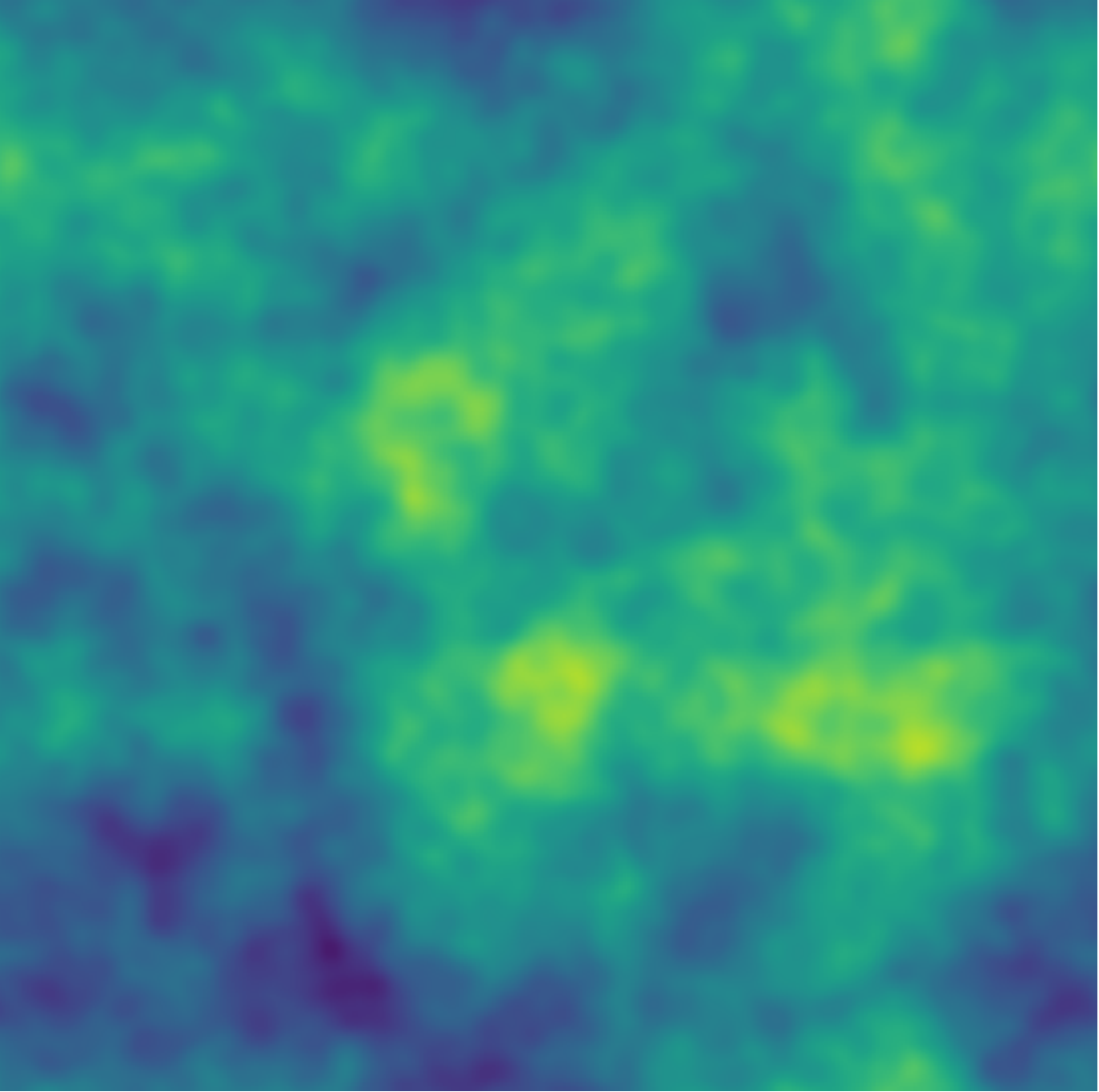}\\
	\includegraphics[width=.333\linewidth]{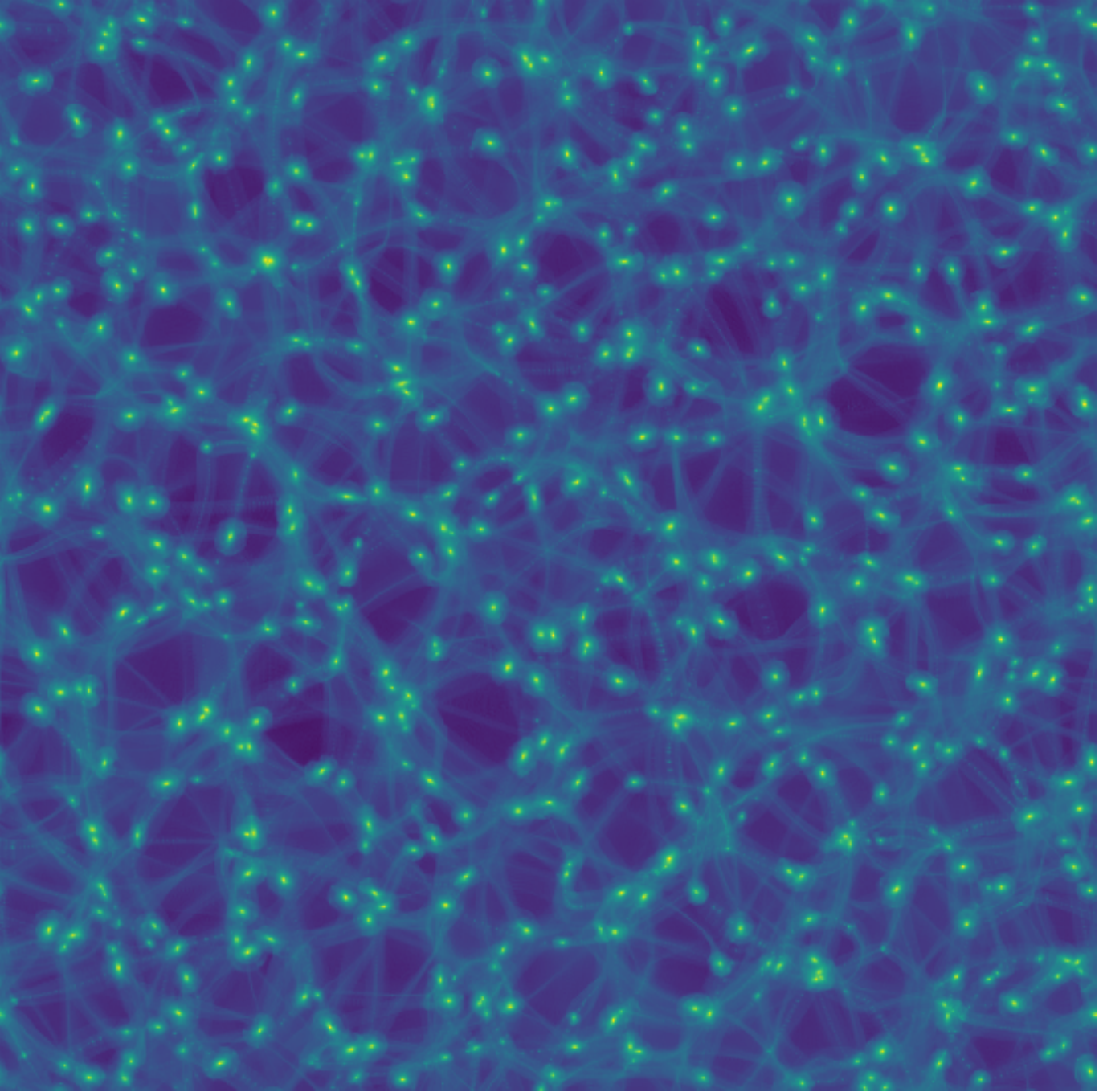}\hspace{-1mm}
	\includegraphics[width=.333\linewidth]{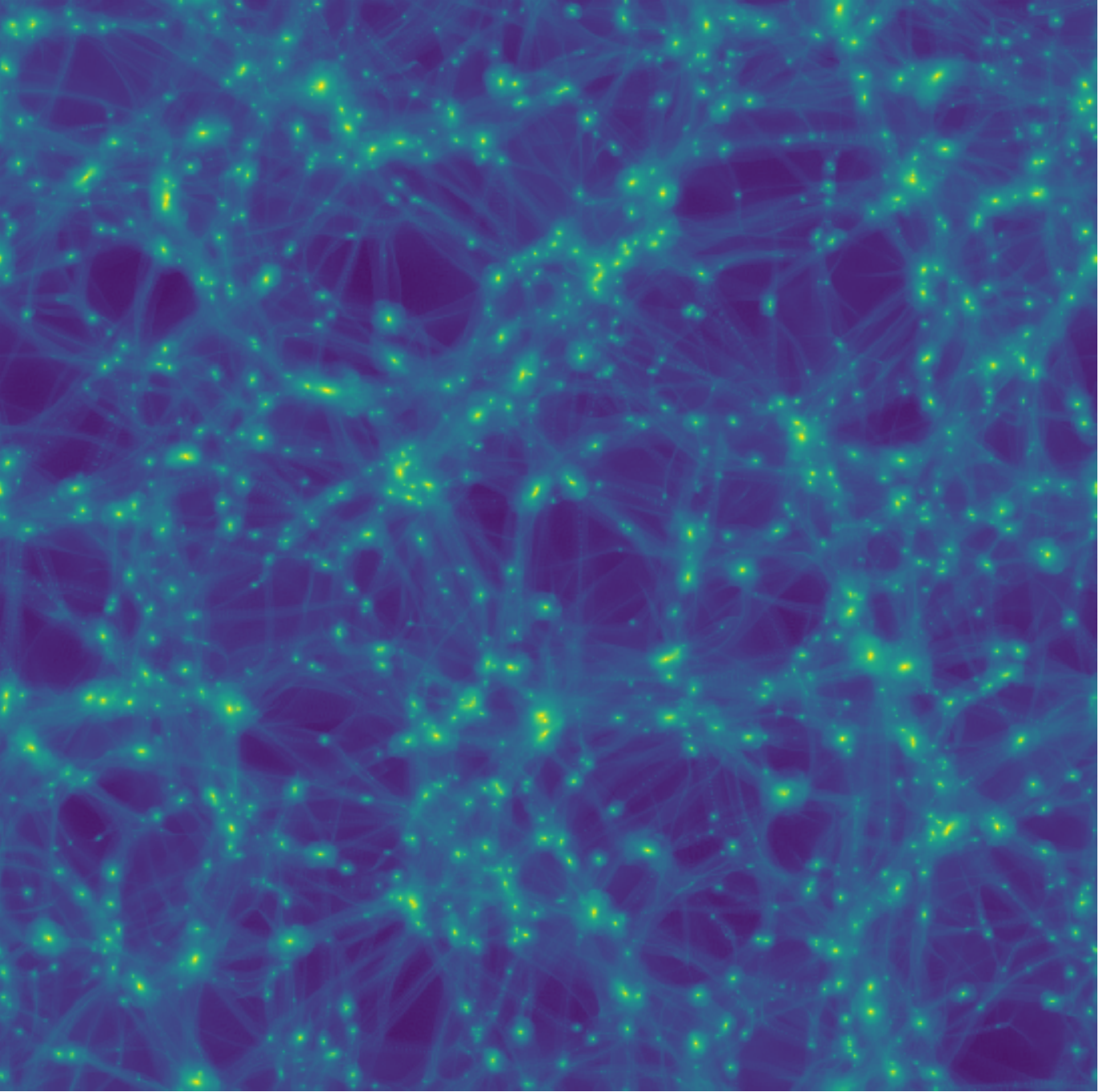}\hspace{-1mm}
	\includegraphics[width=.333\linewidth]{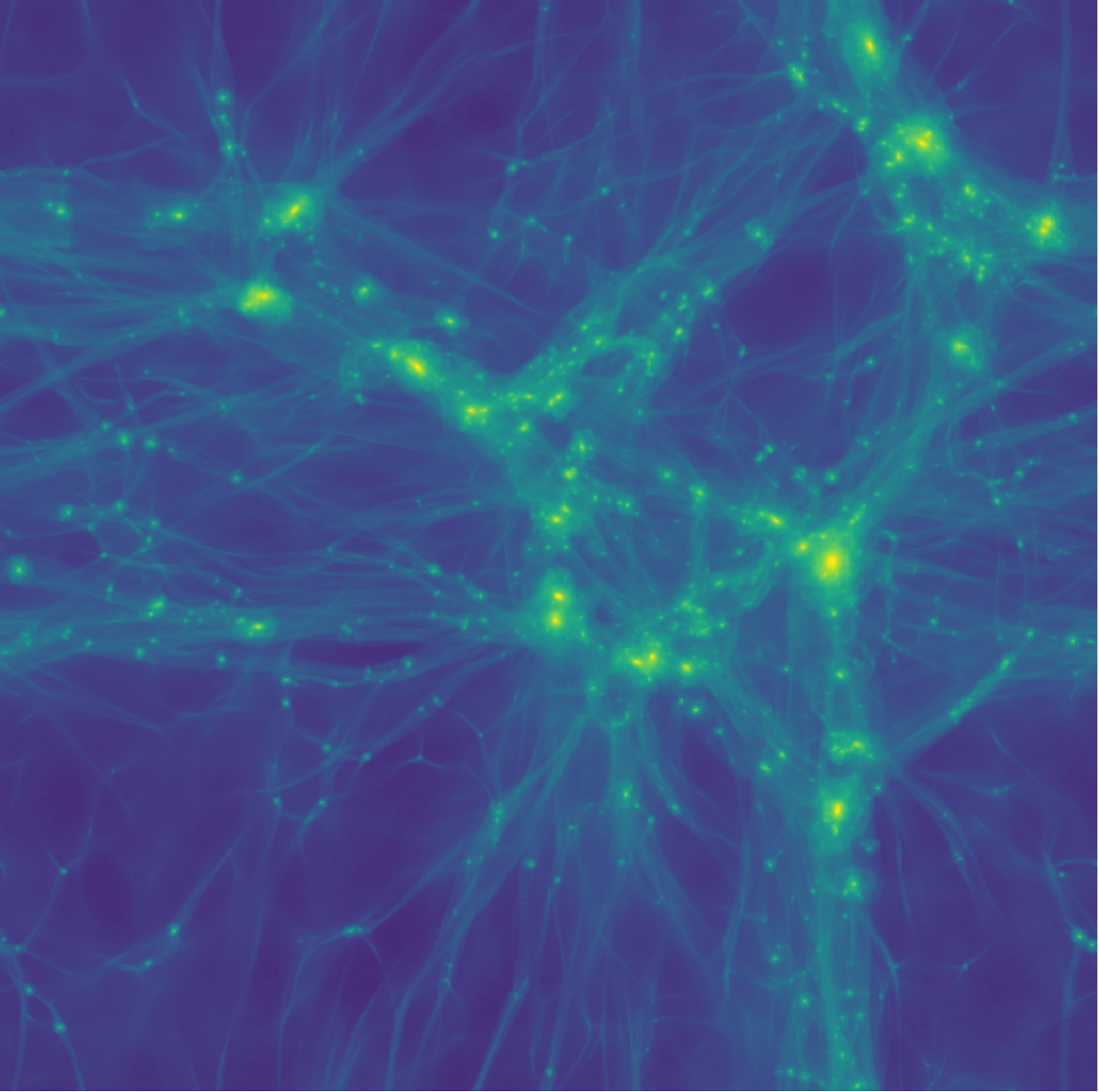}
	\caption{\label{fig:fields} The projected initial (top) and final (bottom) density fields for the $w=0.1$ (left), $w=0.5$ (middle), and ${m_\chi=100}$~GeV (right) simulations.  The color scale is logarithmic; lighter is denser.}
\end{figure*}

Figure~\ref{fig:fields} shows the initial and final density fields for three of these simulations.  This sample illustrates the characteristic differences between the density fields and halo populations that result from the different power spectra.  The $w=0.1$ spectrum produces fluctuations only on a characteristic scale, and the resulting halo population is relatively uniform in size and separation.  Meanwhile, the $w=0.5$ spectrum produces fluctuations on a wider range of scales, yielding very small halos as well as larger halo clusters and voids.  Finally, the 100 GeV power spectrum includes fluctuations at scales up to and exceeding the box size.\footnote{In fact, the entire boxes of the 100 GeV and 3.5 keV simulations should be collapsing by the final redshift, a fact that cannot be reflected in these simulations.  Thus, we do not expect these two simulations to accurately capture the large-scale dynamics; for example, halo mergers are probably under-represented. We focus on understanding the halos that form by direct collapse from peaks in the density field, so we do not expect this shortcoming to impact our main results.}  Accordingly, the bulk of the halos in the box reside within a few clusters, and some halos are growing exceedingly large.

At the final redshift of each simulation, we use the \textsc{Rockstar} halo finder \cite{behroozi2012rockstar} to identify every halo with mass larger than a cutoff $M_\mathrm{cut} = (4\pi/3)k_\mathrm{max}^{-3}\bar\rho_0$, where $k_\mathrm{max}$ is the wave number $k$ at which $\mathcal{P}(k)$ is maximized.  This cutoff is intended to exclude most of the artificial fragments \cite{angulo2013warm,lovell2014properties} that are normally expected to appear in simulations with a small-scale cutoff in density fluctuations.  Roughly speaking, fluctuations at the scale $k_\mathrm{max}$ are the first to collapse, so we do not expect to find many real halos at smaller scales.  Table~\ref{tab:halo_count} lists the number $N_\mathrm{halo}$ of such halos found in each simulation.

\begin{table}[t]
	\caption{\label{tab:halo_count} For each simulation, we list the halo count $N_\mathrm{halo}$ above the mass cutoff at the final redshift, the number of peaks $N_\mathrm{peak}$ in the linear density field, and the number $N_\mathrm{match}$ of successful halo-peak matches.  Of these matches, also listed are the number $N_\mathrm{A}$ with density profiles resolved at small radii, the number $N_\mathrm{r}$ with $r_\mathrm{max}<r_\mathrm{vir}$, and the subsets $N_\mathrm{A}^{<3\mathrm{MM}}$ and $N_\mathrm{r}^{<3\mathrm{MM}}$ that also underwent fewer than three major mergers.}
	\begin{ruledtabular}
		\begin{tabular}{cccccccc}
			Simulation & $N_\mathrm{halo}$ & $N_\mathrm{peak}$ & $N_\mathrm{match}$ & $N_\mathrm{A}$ & $N_\mathrm{A}^{<3\mathrm{MM}}$ & $N_\mathrm{r}$ & $N_\mathrm{r}^{<3\mathrm{MM}}$\\
			\hline\vspace{-3mm}\\
			$w=0.1$ & 696 & 439 & 411 & 408 & 350 & 404 & 347 \\
			$w=0.3$ & 801 & 933 & 565 & 561 & 481 & 537 & 457 \\
			$w=0.5$ & 1011& 3896& 795 & 790 & 683 & 738 & 634 \\
			EMDE    & 610 & 730 & 424 & 416 & 366 & 388 & 339 \\
			100 GeV & 216 & 6108& 157 & 153 & 141 & 82  &  73 \\
			3.5 keV & 581 & 4401& 463 & 455 & 430 & 322 & 303 \\
		\end{tabular}
	\end{ruledtabular}
\end{table}

Next, we wish to associate each halo with the density peak whence it collapsed.  To do this, we use \textsc{Rockstar} to generate halo catalogues at intervals of roughly $5\%$ in the scale factor, and we use the \textsc{Consistent Trees} merger tree code \cite{behroozi2012gravitationally} to improve the consistency of halo tracking through time.  Starting at the final redshift, we trace each halo back to its formation time, defined as the time when its mass first exceeded $M_\mathrm{cut}$.  In case of a merger, we follow the larger progenitor.  We then identify all particles in the halo at its formation time, find their positions in the initial particle grid at $z_\mathrm{init}$, and compute the center of mass $\vec x_\mathrm{CM}$ and rms spread $r_\mathrm{CM}$ of these positions.  Finally, we consider the initial grid $\delta(\vec x)$ of fractional density contrasts and find the largest value of $\delta(\vec x)$ within a sphere of radius $r_\mathrm{CM}$ about $\vec x_{CM}$.  If this maximum is also a local maximum in $\delta(\vec x)$ (a condition that could fail if the maximum lies at the sphere's boundary), we take it to be the initial density peak for this halo.  Otherwise, we discard the halo from further consideration, suspecting it to be an artificial fragment.\footnote{We explored the alternative procedure of following the density gradient to a local maximum, but the only consequence was the introduction of unphysically distant halo-peak associations.}  Moreover, when this procedure associates multiple halos in the final box with the same initial peak, we discard all but the most massive halo.  Table~\ref{tab:halo_count} lists both the number $N_\mathrm{peak}$ of peaks in the initial density field and the number $N_\mathrm{match}$ of peak-halo identifications made through this procedure.

To test our models, we must still make further cuts to the halo-peak population.  For the small-radius asymptote of the density profile (Sec.~\ref{sec:asymptote}), we consider the subset of our peak-matched halos that have well-resolved inner density profiles\footnote{In particular, we require that at least 100 particles reside within the simulation softening length.  Otherwise, the central density is small, usually implying that the density profile was not centered on the halo's cusp, which is only true of a handful of halos not found by \textsc{Rockstar} but filled in by \textsc{Consistent Trees}.  Due to the minuscule fraction of halos affected, we did not make further efforts to find the true center.}.  The number $N_A$ of such halos in each simulation is listed in Table~\ref{tab:halo_count}.  Moreover, Ref.~\cite{ogiya2016dynamical} found that successive major mergers alter the inner asymptote of a halo's density profile, and we confirm this effect in Sec.~\ref{sec:merge}.  We aim to model the initial density profile of a halo before it undergoes any disruptive dynamics, leaving a treatment of mergers for future work.\footnote{We prefer to exclude the impact of mergers in this work because mergers are a continuing process, so their impact is sensitive to the arbitrarily chosen simulation termination redshift.}  Thus, we also restrict our sample to halos that underwent fewer than three major mergers, which we define to be mergers between two halos with a mass ratio smaller than 3. The number $N_A^{<3\mathrm{MM}}$ of these halos is also listed in Table~\ref{tab:halo_count}.  The particular threshold of three major mergers is chosen as a compromise between minimizing the impact of mergers on our results and maximizing the sample of halos from which our results are drawn.  As we will see in Sec.~\ref{sec:merge}, the first two major mergers only marginally alter halo density profiles.

Meanwhile, for the larger-radius density profile (Sec.~\ref{sec:rmax}), we consider the subset of our peak-matched halos that have $r_\mathrm{max}$, the radius at which the circular velocity is maximized, inside their virial radius, defined as the radius $r_\mathrm{vir}$ enclosing average density 200 times the background.\footnote{The requirement $r_\mathrm{max}<r_\mathrm{vir}$ culls a significant fraction of the halo population in the 100 GeV simulation, but this is not a serious concern.  It turns out that the models we discuss predict very large $r_\mathrm{max}$ for these halos as well, and $r_\mathrm{max}>r_\mathrm{vir}$ only implies that the halo's outermost profile has not yet steepened to the point that $\mathrm{d}\ln M/\mathrm{d}\ln r<1$.}  The number $N_\mathrm{r}$ of these halos is listed for each simulation in Table~\ref{tab:halo_count}.  Mergers only minimally alter $r_\mathrm{max}$ (see Sec.~\ref{sec:merge}), so when testing models for $r_\mathrm{max}$, it is not necessary to restrict to halos that underwent few mergers.  However, mergers can significantly alter the mass $M(r_\mathrm{max})$ enclosed within $r_\mathrm{max}$ (or equivalently $v_\mathrm{max}$), so to test our mass predictions, we restrict the sample to the subset $N_\mathrm{r}^{<3\mathrm{MM}}$ of halos that underwent fewer than three major mergers.  This number is also listed in Table~\ref{tab:halo_count}.  Altogether, it is evident from Table~\ref{tab:halo_count} that only a small fraction of the full peak and halo populations is used to test our models.  Section~\ref{sec:summary} explores the impact of these restrictions by returning to the full peak and halo populations.

We now have a catalogue that matches peaks in an initial density field to collapsed halos at a much later time.  All that remains is to collect the halo density profiles and the parameters of each peak, and we detail these processes in Appendixes \ref{sec:halos} and~\ref{sec:peaks}, respectively.  With these data, we are now prepared to test any model relating the structure of a dark matter halo to its precursor density peak.  In the following sections, we develop such a model.

\section{Density profile at small radii}\label{sec:asymptote}

\begin{figure}[t]
	\centering
	\includegraphics[width=\columnwidth]{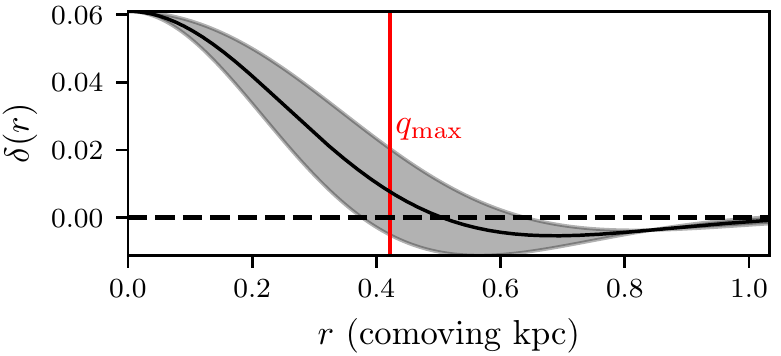}
	\caption{\label{fig:collapse_init} The radial profile of a density peak drawn from the $w=0.3$ power spectrum at redshift $z=10^6$.  The peak is ellipsoidal, and the shading indicates the variation in the profile along different axes.  The vertical line marks $q_\mathrm{max}$ computed using the turnaround model (Sec.~\ref{sec:frozen}).}
\end{figure}

We first study the coefficients $A$ of the $\rho=A r^{-3/2}$ asymptotes of the density profiles of the first halos.  It is of prime importance to accurately predict the density profiles in this regime because these radii source the bulk of the prospective signal from dark matter annihilation, and this remains true even if these halos relax toward ${\rho\propto r^{-1}}$ profiles due to mergers.  Moreover, there is another reason to study small radii separately: the ${\rho\propto r^{-3/2}}$ inner density profile is established almost immediately after collapse.  To illustrate this fact, we simulate\footnote{This simulation employed about 9 million particles in a comoving vacuum-bounded sphere of radius 1.5 kpc.  The starting redshift was $z=10^6$.} the collapse of the isolated density peak shown in Fig.~\ref{fig:collapse_init}.  This peak represents a typical $3\sigma$ peak drawn from the $w=0.3$ power spectrum using the statistics of peaks as described in Ref.~\cite{bardeen1986statistics}.  The shallowing of this profile toward $r=0$ is associated with the absence of small-scale fluctuations, and this is the single feature common to density peaks drawn from all of the power spectra we are studying.  Indeed, it has been suggested that this feature is responsible for the development of the ${\rho\propto r^{-3/2}}$ profile \cite{ogiya2017sets}.

\begin{figure}[t]
	\centering
	\includegraphics[width=\columnwidth]{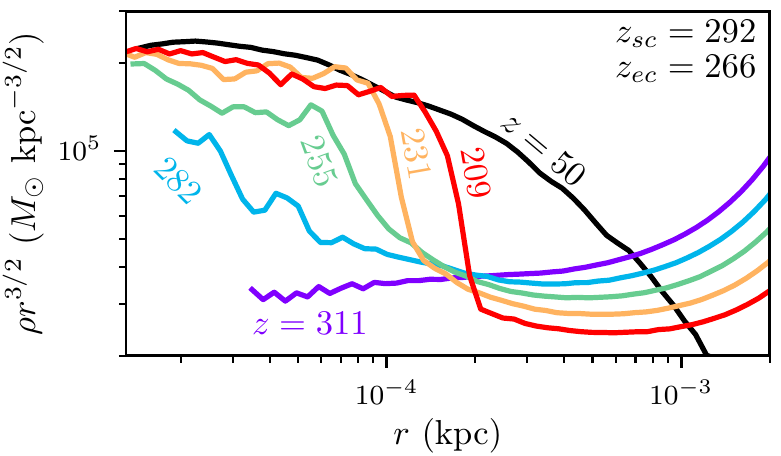}
	\caption{\label{fig:collapse} The density profile of the resulting halo during and after collapse of the peak in Fig.~\ref{fig:collapse_init}.  The numbers denote the redshifts $z$ at which the density profile is plotted, while $z_\mathrm{sc}$ and $z_\mathrm{ec}$ are the redshifts of spherical and ellipsoidal collapse, respectively, computed from the initial peak.  The black curve shows the density profile long after collapse.}
\end{figure}

Figure~\ref{fig:collapse} shows the halo resulting from this initial peak in the moments after collapse.  We also compute the expected collapse time from the initial peak using both the spherical collapse model, $\delta(a_\mathrm{sc})=1.686$, and the ellipsoidal collapse model described in Ref.~\cite{sheth2001ellipsoidal}.  The redshifts $z_\mathrm{sc}$ and $z_\mathrm{ec}$ of spherical and ellipsoidal collapse, respectively, are indicated in Fig.~\ref{fig:collapse}.  The critical observation is that the $\rho\propto r^{-3/2}$ asymptote of the density profile develops almost immediately after collapse, matching closely the late-time density profile as early as $a \simeq 1.15 a_\mathrm{ec}$.  Subsequently, the profile grows outward in radius alone, gradually steepening as accretion of new material slows.  The interval $\Delta a/a \simeq 0.15$ over which the profile develops is not arbitrary.  The characteristic dynamical time $\Delta t=\sqrt{3\pi/(16G\rho)}$ \cite{binney1987galactic} of a virialized region with density $200$ times the background is $\Delta t = 0.16/H$, where $H$ is the Hubble rate.  This interval corresponds to $\Delta a/a= 0.16$, implying that the inner density profile develops over a single dynamical time interval.

\subsection{Spherical collapse}\label{sec:sc}

\begin{figure*}[t]
	\includegraphics[width=\linewidth]{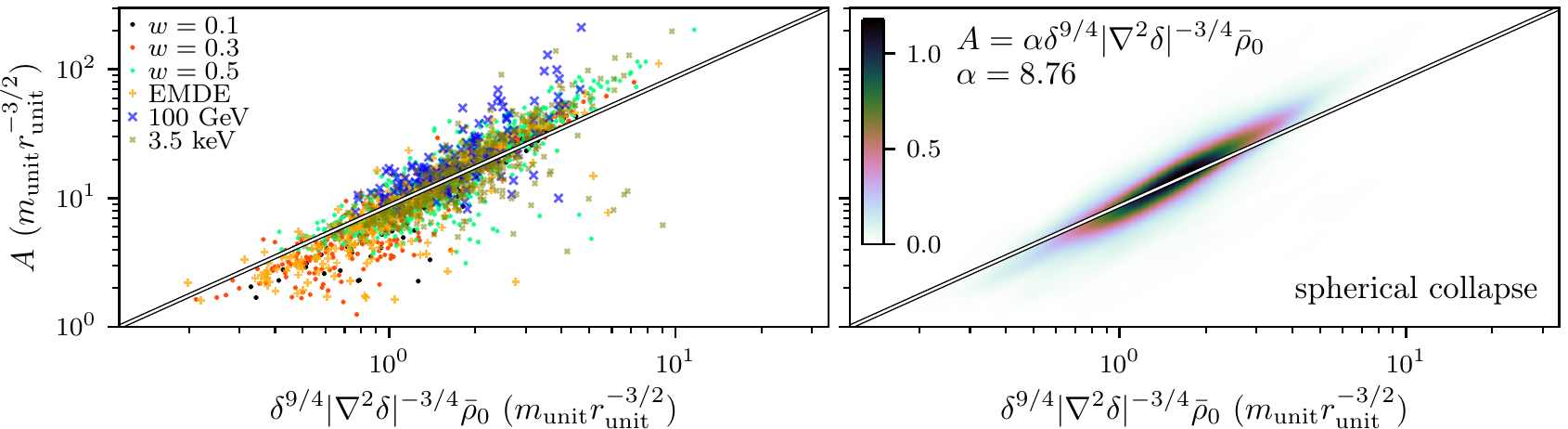}\\
	\includegraphics[width=\linewidth]{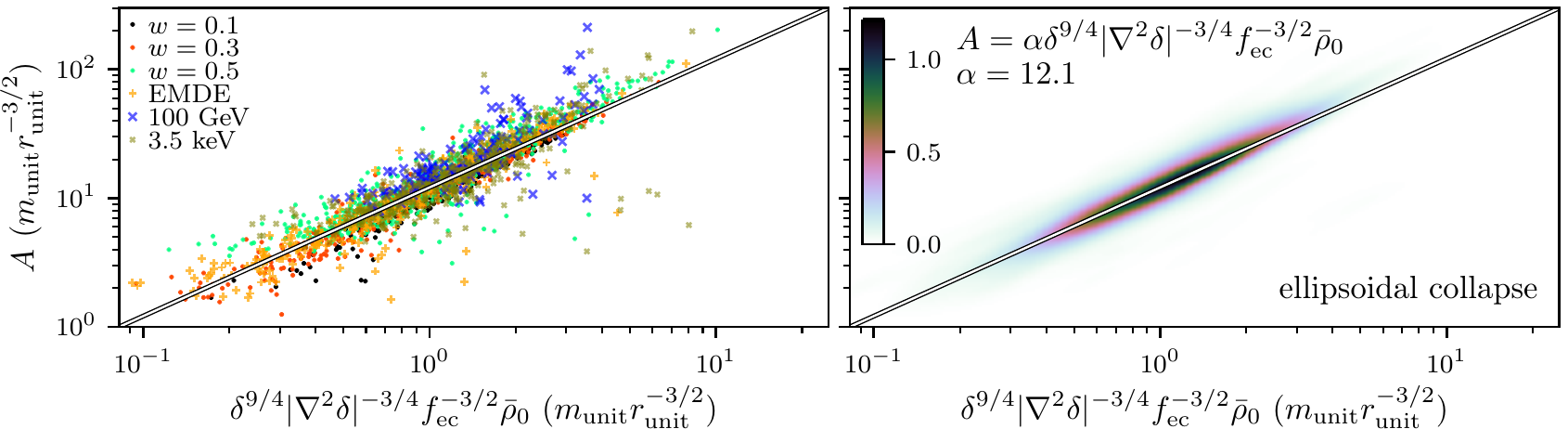}
	\caption{\label{fig:model_asy} The coefficient $A$ of the $\rho=A r^{-3/2}$ density profile asymptote, plotted against the prediction from our models.  Top: The spherical model, Eq.~(\ref{asymp}).  Bottom: The ellipsoidal model, Eq.~(\ref{asymp_e}).  The left panels plot all of the halos, while the right panels plot a density estimate: the color scale indicates the density $f(\ln A,\ln A_\mathrm{pred})$ of the distribution, where $A_\mathrm{pred}$ is the x-axis quantity.  We also plot the line corresponding to the median proportionality constant for each model.}
\end{figure*}

The rapid development of a halo's inner $\rho=A r^{-3/2}$ profile implies that the coefficient $A$ can only be influenced by the immediate neighborhood of the precursor density peak.  To make this argument precise, we define\footnote{After this section, we will omit the tilde, and any quantity related to the linear density field should be understood to be evaluated on the scaled linear density field $\tilde\delta(\vec x)$ unless otherwise specified.}
\begin{equation}
\tilde\delta(\vec x)\equiv \delta(\vec x,a)/a,
\end{equation}
where $\delta(\vec x,a)$ is evaluated using linear theory during matter domination (when $\delta\propto a$).  Here, $\vec x$ is a comoving coordinate, and we define $a=1$ today.  We now claim that the inner $\rho\propto r^{-3/2}$ asymptote is only sensitive to the local properties $\tilde\delta$ and $\partial_i\partial_j\tilde\delta$ of the precursor density peak.  If we neglect deviations from spherical symmetry, the peak reduces to two parameters: its amplitude $\tilde\delta$ and curvature $|\nabla^2\tilde\delta|$.  During matter domination, spherical collapse theory implies that $a_\mathrm{sc}\propto\tilde\delta^{-1}$.  Meanwhile, $|\nabla^2\tilde\delta|$ defines a comoving length scale $q_\mathrm{pk}\equiv(\tilde\delta/|\nabla^2\tilde\delta|)^{1/2}$ associated with the peak.  At the time of collapse, this comoving length corresponds to the physical length scale $a_\mathrm{sc} q_\mathrm{pk}\propto (\tilde\delta|\nabla^2\tilde\delta|)^{-1/2}$.  There is also a physical density scale $\bar\rho_0 a_\mathrm{sc}^{-3}\propto \bar\rho_0\tilde\delta^3$, i.e., the cosmological background density at collapse.  If these are the only scales, then up to a constant coefficient, there is a unique prediction for the coefficient $A$ of the $\rho=A r^{-3/2}$ asymptote of the density profile,
\begin{equation}\label{asymp}
A = \alpha \bar\rho_0 \delta^{9/4} |\nabla^2\delta|^{-3/4},
\end{equation}
where $\alpha$ is a proportionality constant.  We omit the tildes in Eq.~(\ref{asymp}), but all quantities related to the linear density field are understood to be evaluated on $\tilde\delta(\vec x)$.  This convention applies to the remainder of this work.  The parameter $\alpha$ must be calibrated by simulations, but this calibration is only necessary once; it should be the same for any power spectrum.

We test this model on the $N_A^{<3\mathrm{MM}}$ peak-matched halos in each simulation that have well-resolved inner density profiles and underwent fewer than three major mergers (see Sec.~\ref{sec:sims}).  For these halos, Fig.~\ref{fig:model_asy} plots the left-hand side against the right-hand side of Eq.~(\ref{asymp}) in order to test the model.  Evidently, our model works well for how simple it is, predicting the asymptotes with reasonable success in all six simulations.  There does, however, appear to be a correlated effect wherein halos with the densest predicted asymptotes tend to exceed that prediction, and vice versa at the less dense end.  In fact, this effect is caused by the assumption of spherical collapse.  Peaks of smaller amplitude tend to be less spherical, thereby collapsing later and forming less dense halos than their amplitude would suggest.  We next account for this effect.

\subsection{Ellipsoidal collapse}\label{sec:ellipsoid}

Equation~(\ref{asymp}) neglects deviations from spherical symmetry.  However, it can be immediately generalized in the following way.  Since $A\propto a_\mathrm{sc}^{-3/2}$, we can use the theory of ellipsoidal collapse \cite{bond1996peak,sheth2001ellipsoidal,sheth2002excursion} to predict how the three-dimensional shape of a peak alters its collapse time.  In particular, ellipsoidal collapse occurs later than spherical collapse by a factor $f_\mathrm{ec}(e,p)\equiv a_\mathrm{ec}/a_\mathrm{sc}$, which is a function of the ellipticity $e$ and prolateness $p$ of the gravitational potential $\phi$ in the vicinity of the peak (see Appendix~\ref{sec:peaks} for definitions).  To compute $f_\mathrm{ec}$, we use the approximation \cite{sheth2001ellipsoidal}
\begin{equation}\label{ec}
f_\mathrm{ec}(e,p) = 1+0.47\left[5(e^2-p|p|)f_\mathrm{ec}^2(e,p)\right]^{0.615}.
\end{equation}
Accounting for ellipsoidal collapse,
\begin{equation}\label{asymp_e}
A = \alpha \bar\rho_0 \delta^{9/4} |\nabla^2\delta|^{-3/4}f_\mathrm{ec}^{-3/2}(e,p),
\end{equation}
where $\alpha$ is a proportionality constant.

\begin{figure}[t]
	\includegraphics[width=\columnwidth]{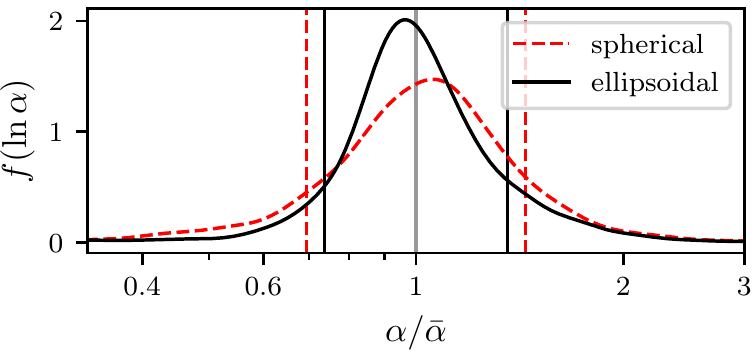}
	\caption{\label{fig:hist_A} A comparison of the scatter from the two asymptote models.  For each model, we plot a histogram of the coefficient $\alpha$ between the asymptote $A$ and its model prediction, scaled to the mean coefficient for all halos.  Vertical lines mark $1\sigma$ deviations from the mean, computed in log space.}
\end{figure}

Figure~\ref{fig:model_asy} compares the predicted asymptotes in the ellipsoidal collapse model to their measured values.  The correlated scatter associated with the spherical model is no longer apparent, confirming that the spherical collapse assumption was the source.  Figure~\ref{fig:hist_A} compares the overall scatter between these two models; the ellipsoidal collapse model is clearly superior.  We remark that not all of the scatter depicted is physical; there is numerical noise in our simulated density profiles and finite density grids.  Nevertheless, there are still clear sources of physical scatter.  Absent an understanding of why the $\rho\propto r^{-3/2}$ profile develops, we constructed our model using purely scaling arguments, and these arguments can break down in two ways.  First, the inner profile does not develop instantly, so it is sensitive to a small region about the density peak instead of only its immediate neighborhood.  Also, as depicted in Fig.~\ref{fig:collapse}, the inner asymptote is not completely static but instead grows marginally after it develops, implying that it exhibits some sensitivity to the larger density field.  Second, we only accounted for nonspherical peak shapes by altering the time of collapse.  The sensitivity of the asymptotic coefficient $A$ to the shape of the peak may be more complicated.

Nevertheless, it is a significant success to obtain a mere approximately $30\%$ scatter across such a broad range of cosmologies from a model as simple as the one presented here.  The proportionality coefficients and their statistics for our models are summarized in Table~\ref{tab:params}.  Thus calibrated, the model developed in this section may be employed to predict the inner asymptotes of the first halos' density profiles in any cosmological scenario.

\section{Density profile at large radii}\label{sec:rmax}

We next study the density profile beyond the inner $\rho\propto r^{-3/2}$ asymptote.  These larger radii, containing the bulk of the halo mass, are relevant to gravitational lensing signatures as well as to the dynamical evolution of the halo through mergers and tidal stripping.  In this section, we discuss and validate models that can predict the density profile at large radii.  We focus on three physical models:
\begin{enumerate}[label={(\arabic*)}]
	\item a ``turnaround'' model, tracing back to Ref.~\cite{gunn1972infall}, in which each mass shell freezes at a fraction of its turnaround radius, or the radius of first apocenter;
	\item a ``contraction'' model, expressed most concisely by Ref.~\cite{dalal2010origin}, that accounts for contraction of halo particle orbits due to subsequent infall;
	\item a ``virialization'' model, put forward by Ref.~\cite{salvador2012theoretical}, in which the final radius of a mass shell is determined by enforcing that the shell's enclosed energy be distributed according to the virial theorem.
\end{enumerate}

All of these models rely on the assumption that material is accreted gradually.  Hence, they are not applicable to the density profiles at small radii, and none of them predicts the $\rho\propto r^{-3/2}$ asymptote.  We will explore in Sec.~\ref{sec:rdis} precisely where these models are accurate.  In the meantime, we validate and tune the models by using them to predict the radius $r_\mathrm{max}$ at which the halo's circular velocity is maximized along with the mass $M(r_\mathrm{max})$ within this radius.  The maximum circular velocity itself follows as $v_\mathrm{max}=\sqrt{G M(r_\mathrm{max})/r_\mathrm{max}}$.

The radius $r_\mathrm{max}$ characterizes where the profile bends away from its inner asymptote.\footnote{For example, if $\rho(r) = \rho_s(r/r_s)^{-3/2}(1+r/r_s)^{-3/2}$ \cite{delos2018ultracompact,moore1999cold} (where $\rho_s$ and $r_s$ are scale parameters), $r_\mathrm{max}=1.055r_s$.}  It is more common in the literature to study the scale radius $r_s$, often defined to be where $\mathrm{d}\ln\rho/\mathrm{d}\ln r=-2$, instead of $r_\mathrm{max}$, which is where $\mathrm{d}\ln M/\mathrm{d}\ln r=1$.  However, we favor $r_\mathrm{max}$ over $r_s$ for several reasons.  First,  $r_\mathrm{max}$ can be read from a simulation more robustly than $r_s$, since the mass profile is less noisy than the density profile.  Second, $r_\mathrm{max}$ turns out to be cleaner to predict from the linear density field.  Finally, since $r_\mathrm{max}$ characterizes the total mass rather than the local density, it is the more relevant quantity for understanding halo dynamics \cite{ascasibar2008dynamical} and for predicting gravitational lensing signatures.  The combination of $A$ and $r_\mathrm{max}$ serves to mostly characterize the density profile of a halo, and the mass $M(r_\mathrm{max})$ (or velocity $v_\mathrm{max}$) supplies an additional constraint.

\subsection{Turnaround}\label{sec:frozen}

To develop a model that can predict the full density profile of a collapsed halo, we must relate this profile to the spherically averaged fractional density excess profile $\delta(q)$, where $q$ is the comoving radius, about the corresponding peak in the linear density field.  Following the convention established in Sec.~\ref{sec:asymptote}, we define $\delta(q)\equiv\delta(q,a)/a$ evaluated in linear theory during matter domination (with $a=1$ today).  We will also use 
\begin{equation}\label{Delta_int}
\Delta(q)=\frac{3}{q^3}\int_0^q \delta(q^\prime)q^{\prime 2}\mathrm{d}q^\prime,
\end{equation}
the fractional enclosed mass excess defined under the same convention.  As in Ref.~\cite{gunn1972infall}, we consider the simplified spherical infall model in which each spherical shell freezes at a fixed fraction of its turnaround radius.  A mass shell initially at comoving radius $q$ turns around at physical radius $r_\mathrm{ta}=(3/5)q/\Delta(q)$.  Hence, the final physical radius $r$ of this shell is
\begin{equation}\label{r_f}
r = \beta q/\Delta(q),
\end{equation}
where $\beta$ is a proportionality constant to be measured in simulations.  But in this model, mass shells never cross, so the mass enclosed within this shell, now at $r$, is still
\begin{equation}\label{M_f}
M(q)=\beta_M \frac{4\pi}{3}q^3\bar\rho_0
\end{equation}
(at zeroth order in $\Delta$) with $\beta_M=1$.  This equation gives the mass profile $M(r)$ of the collapsed halo if $q$ is obtained from $r$ by inverting Eq.~(\ref{r_f}).

To find $r_\mathrm{max}$ (up to the proportionality constant $\beta$), we may maximize $M(r)/r$.  Alternatively, we can write
\begin{equation}\label{Mslope}
\frac{\mathrm{d}\ln M}{\mathrm{d}\ln r} = \frac{3}{1+3\epsilon(q)},
\end{equation}
where we define\footnote{We define $\epsilon(q)$ as a generalization of the index $\epsilon$ of $\delta M/M\propto M^{-\epsilon}$ in the self-similar theory \cite{fillmore1984self}.  Consequently, Eq.~(\ref{Mslope}) has exactly the same form as its analogue in the self-similar theory.}
\begin{equation}\label{epsilon}
\epsilon(q) \equiv -\frac{1}{3}\frac{\mathrm{d}\ln \Delta}{\mathrm{d}\ln q} = 1-\frac{\delta(q)}{\Delta(q)}.
\end{equation}
In this case, $r_\mathrm{max}$ is obtained as the solution to $\mathrm{d}\ln M/\mathrm{d}\ln r = 1$ or, equivalently, ${\epsilon(q_\mathrm{max})=2/3}$ with $r_\mathrm{max}$ computed from $q_\mathrm{max}$ using Eq.~(\ref{r_f}).

This model, at first glance, seems to be far divorced from a realistic description.  However, it turns out to be a reasonable approximation of late accretion.  At late times, halo density profiles are stable in time; see, e.g., Papers I and II.  This observation is explained by noting that once the halo is established and accretion begins to slow, newly accreted mass only contributes significantly to the outskirts of the halo; there is too little new matter, spending too small a fraction of its orbital period at small radii, to substantially raise the interior density.  Since the density profile is static, a newly accreted particle settles into a stable orbit with time-averaged radius proportional to its orbital apocenter, the turnaround radius.\footnote{The orbital apocenter actually decays over time, an effect for which the contraction model accounts.}  The final radius in this model is to be understood as that orbital average.

\begin{figure}[t]
	\includegraphics[width=\columnwidth]{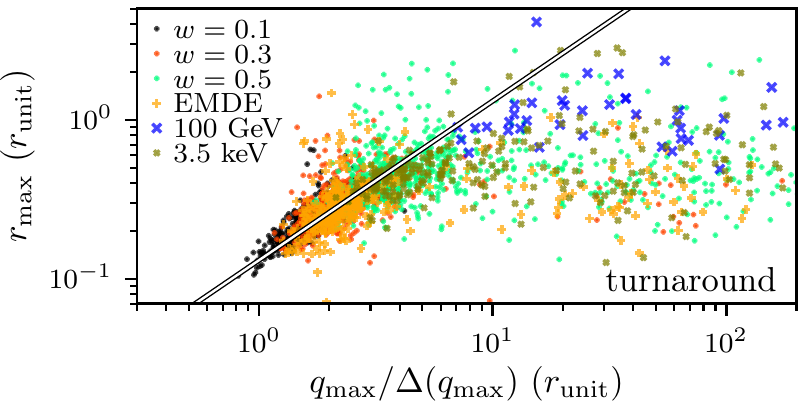}
	\caption{\label{fig:model_rmax0} For the turnaround model, this figure plots the values of $r_\mathrm{max}$ measured in our simulations against the predicted values.  For visual reference, an example proportionality curve is plotted (not a fit).  The model dramatically overpredicts $r_\mathrm{max}$ for a subpopulation of the halos, a problem that we attribute to the finite time at which the simulated density profiles are measured (see the text).}
\end{figure}

We test this model on the $N_\mathrm{r}$ peak-matched halos for which $r_\mathrm{max}<r_\mathrm{vir}$ (see Sec.~\ref{sec:sims}).  In the top panel of Fig.~\ref{fig:model_rmax0}, we plot the measured $r_\mathrm{max}$ against the prediction from this model.  The model appears to work well for the bulk of the halos.  However, there is a significant population of halos, especially coming from the broader 100 GeV, 3.5 keV, EMDE, and $w=0.5$ power spectra, for which the predicted $r_\mathrm{max}$ is much larger than the measured value.  In fact, in these simulations, there are many peaks for which the mass shell with $\epsilon=2/3$ has not yet accreted onto the halo by the final redshift.  In these cases, it makes no sense to use the properties of this shell to predict $r_\mathrm{max}$.

\begin{figure*}[hbtp]
	\includegraphics[width=\linewidth]{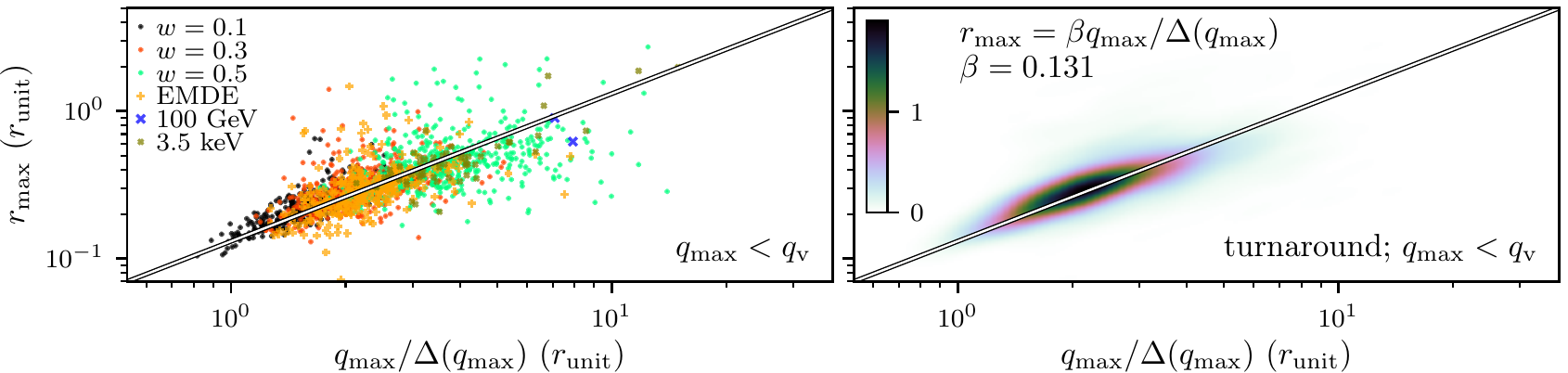}\\
	\includegraphics[width=\linewidth]{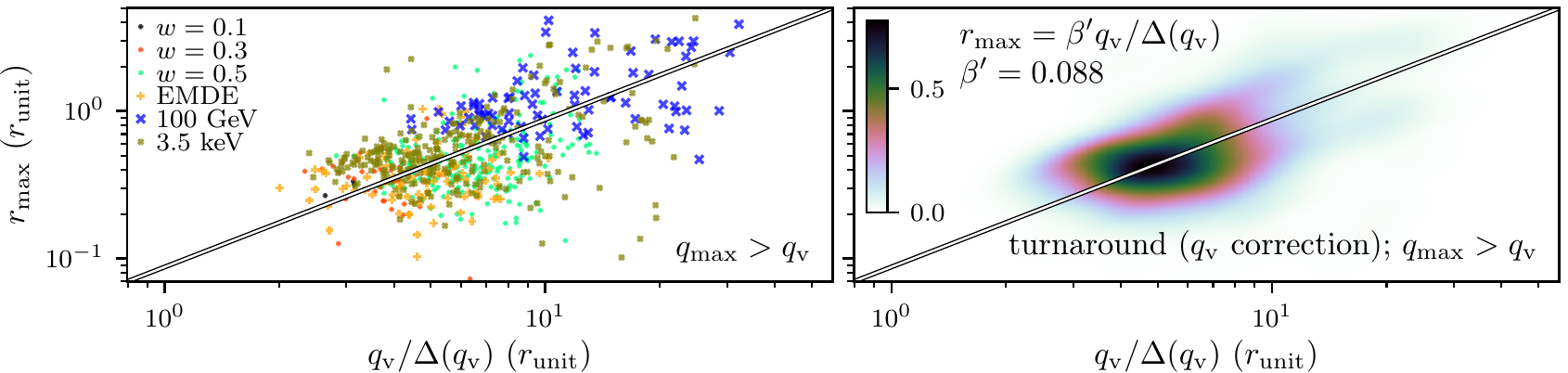}
	\caption{\label{fig:model_rmaxta} For the corrected turnaround model, we plot the values of $r_\mathrm{max}$ measured in our simulations against the predicted values.  At the top, we show the turnaround model with halos with $q_\mathrm{max}>q_\mathrm{v}$ excluded, while at bottom, we use $q_\mathrm{v}$ to predict $r_\mathrm{max}$ for the previously excluded halos.  This separation corrects the discrepancy seen in Fig.~\ref{fig:model_rmax0}.  The left panels plot all of the halos, while the right panels show density estimates; the color scale indicates the density $f(\ln r_\mathrm{max},\ln r_\mathrm{max,pred})$ of the distribution, where $r_\mathrm{max,pred}$ is the x-axis quantity.  The solid line corresponds to the median proportionality constant.}
\end{figure*}

\begin{figure}[t]
	\includegraphics[width=\columnwidth]{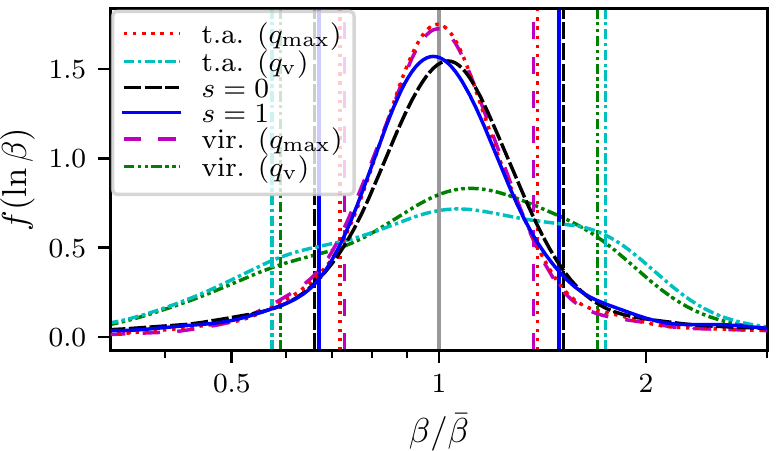} \\
	\includegraphics[width=\columnwidth]{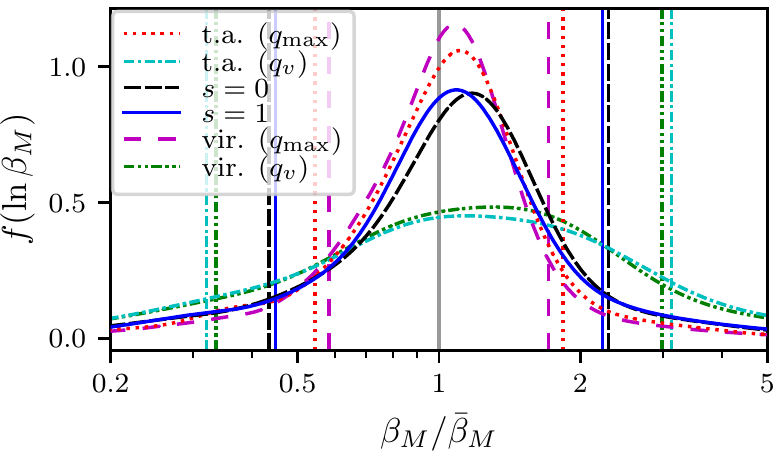}
	\caption{\label{fig:hist_rmax} Comparisons of the scatter in the $r_\mathrm{max}$ (top) and $M(r_\mathrm{max})$ (bottom) predictions.  For each model, we plot the distribution of the coefficient $\beta$ or $\beta_M$ between the halo quantity and its model prediction, scaled to the mean coefficient for all halos.  ``t.a. ($q_\mathrm{max}$)'' and ``t.a. ($q_\mathrm{v}$)'' denote the turnaround models using $q_\mathrm{max}$ and $q_\mathrm{v}$ and applied only to halos with $q_\mathrm{max}<q_\mathrm{v}$ and $q_\mathrm{max}>q_\mathrm{v}$, respectively.  ``$s=0$'' and ``$s=1$'' denote the respective contraction models.  ``vir. ($q_\mathrm{max}$)'' and ``vir. ($q_\mathrm{v}$)'' denote the virialization models using $q_\mathrm{max}$ and $q_\mathrm{v}$ and applied only to halos with $q_\mathrm{max}<q_\mathrm{ta}$ and $q_\mathrm{max}>q_\mathrm{ta}$, respectively.  Note that the contraction models have more scatter than the turnaround and virialization ($q_\mathrm{max}$) models only because the latter describe a smaller range of halos; see the footnotes in Table~\ref{tab:params}.}
\end{figure}

A minimal correction, for these halos, is to instead relate $r_\mathrm{max}$ to the outer boundary of the halo.  In the spherical collapse model, a mass shell virializes when it falls to half of its turnaround radius; at this point, its energy is distributed according to the virial theorem.  The halo's outer boundary may be taken to be the physical radius of the last shell to virialize in this way, and we define $q_\mathrm{v}$ to be the Lagrangian radius of this shell.  If we are considering the halo population at scale factor $a$, then $q_\mathrm{v}$ is the smallest $q$ that satisfies $\Delta(q,a)=a\Delta(q)\leq\delta_v$, where $\delta_v=1.583$.  If $r_\mathrm{max}$ is proportional to the halo boundary so defined, then
\begin{equation}\label{rmax_c}
r_\mathrm{max} = \beta^\prime q_\mathrm{v}/\Delta(q_\mathrm{v}),
\end{equation}
where $\beta^\prime$ is another proportionality constant, and this model is understood to apply only to the halos of which the $\epsilon=2/3$ mass shells have not yet virialized, i.e., $q_\mathrm{max}>q_\mathrm{v}$.  In Fig.~\ref{fig:model_rmaxta}, we plot the turnaround model for only those halos of which the $\epsilon=2/3$ shells have virialized, separately plotting the $q_\mathrm{v}$ model for the remainder of the halos.  Evidently, the halos of which the $\epsilon=2/3$ shells had not yet virialized were indeed the population of which the $r_\mathrm{max}$ was severely overpredicted, and the $\epsilon=2/3$ model exhibits significantly less scatter with them excluded.  Meanwhile, using $q_\mathrm{v}$ to predict $r_\mathrm{max}$ for these halos works reasonably well, although the scatter here is significantly larger.  The top panel of Fig.~\ref{fig:hist_rmax} depicts these differences in scatter more transparently.

Finally, we test how well this model predicts $M(r_\mathrm{max})$ in similar fashion by using Eq.~(\ref{M_f}) and allowing $\beta_M$ to float.  In this case, we employ the $N_\mathrm{r}^{<3\mathrm{MM}}$ halos with $r_\mathrm{max}<r_\mathrm{vir}$ that also underwent fewer than three major mergers (see Sec.~\ref{sec:sims}).  As above, we set $q=q_\mathrm{max}$ when $q_\mathrm{max}<q_\mathrm{v}$ and $q=q_\mathrm{v}$ otherwise.  The bottom panel of Fig.~\ref{fig:hist_rmax} shows the scatter in these predictions.  For both $r_\mathrm{max}$ and $M(r_\mathrm{max})$, Table~\ref{tab:params} lists the tuned coefficients $\beta$ and $\beta_M$ and their statistics.

\subsection{Contraction}\label{sec:adiabatic}

\begin{figure}[t]
	\centering
	\includegraphics[width=\columnwidth]{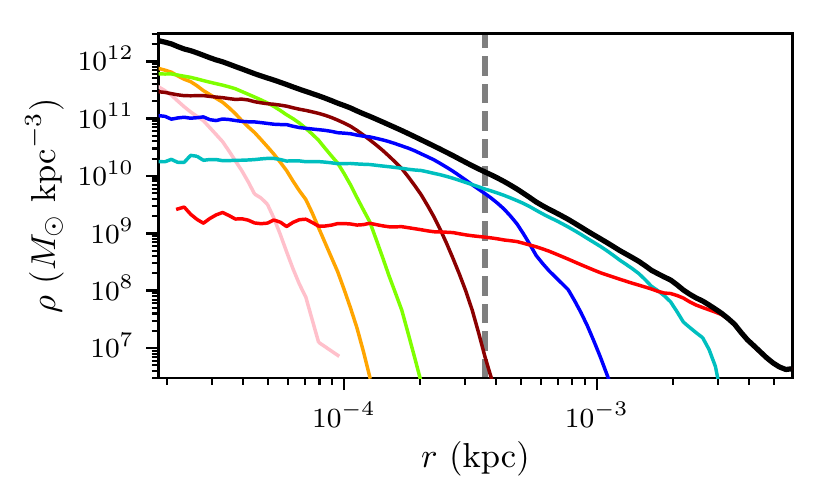}
	\caption{\label{fig:shells} The density profile of a halo (black curve) along with the density profiles contributed by different initial mass shells (colored curves).  The lowest shell contains the mass initially in the comoving radius band 0.044 to 0.066 kpc, the second shell corresponds to the band 0.066 to 0.10 kpc, and so on.  The dashed vertical line marks $r_\mathrm{max}$.}
\end{figure}

The turnaround model assumed that each new shell freezes at a fixed fraction of its turnaround radius, contributing mass to that radius alone.  In reality, a shell contributes mass to a large range of radii.  Figure~\ref{fig:shells} shows the density profiles\footnote{This figure depicts a halo that collapsed from an isolated $3\sigma$ peak drawn from the $w=0.3$ power spectrum.  This peak was simulated with about 70 million particles in a comoving vacuum-bounded sphere of radius 1 kpc.} laid down by a range of initial mass shells.  These shells consist of successive factors of 1.5 in initial radius, so that the lowest shell contains the mass initially in the comoving radius band 0.044 to 0.066 kpc; the second shell contains the mass initially in the band 0.066 to 0.10 kpc; and so on.  Notably, each shell has a characteristic radius within the final halo below which it contributes roughly constant density.  As Ref.~\cite{dalal2010origin} argues, the constant-density contribution follows from the notion that particles from large-radius mass shells cross the lower radii at such high velocity that their motions are effectively unaccelerated.

Consider a halo particle initially orbiting with apocenter radius $r$ that encloses halo mass $M(r)$.  As time goes on, newly accreted shells contribute to the enclosed mass, increasing $M(r)$.  In a spherically symmetric and self-similar system, the quantity $\oint v_r dr\propto [M(r)r]^{1/2}$ is an adiabatic invariant, implying that as the enclosed mass grows $r$ shrinks as $r\propto 1/M(r)$.  While the quantity $Mr$ need not be conserved in a more general picture, Ref.~\cite{lithwick2011self} found that it remains nearly invariant for each particle even with spherical symmetry relaxed.

For a mass shell with apocenter $r^\prime$, let $f(r,r^\prime)$ be the fraction of the shell's mass that is within $r$.  It is readily seen that the mass enclosed within a shell with Lagrangian radius $q$ and final apocenter $r(q)$ increases by the factor
\begin{equation}\label{X}
X(q) \equiv 1+\frac{3}{q^3}
\int_{q}^{q_\mathrm{v}}q^{\prime 2}\mathrm{d}q^\prime f\!\left[r(q),r(q^\prime)\right].
\end{equation}
due to the contribution of shells with $q^\prime>q$.  As before, $q_\mathrm{v}$ is the Lagrangian radius of the latest shell to virialize; it is the smallest $q$ that satisfies $a\Delta(q)\leq \delta_v$, where $a$ is the scale factor at which we wish to characterize the halo population (and $a=1$ today).  If orbital apocenters shrink according to $r\propto 1/M$, then
\begin{equation}\label{ra}
r(q) = \beta\frac{q}{\Delta(q)}\frac{1}{X(q)}
\end{equation}
describes the apocenter of the $q$ shell after contraction, with $\beta=3/5$.

\begin{figure*}[t]
	\includegraphics[width=\linewidth]{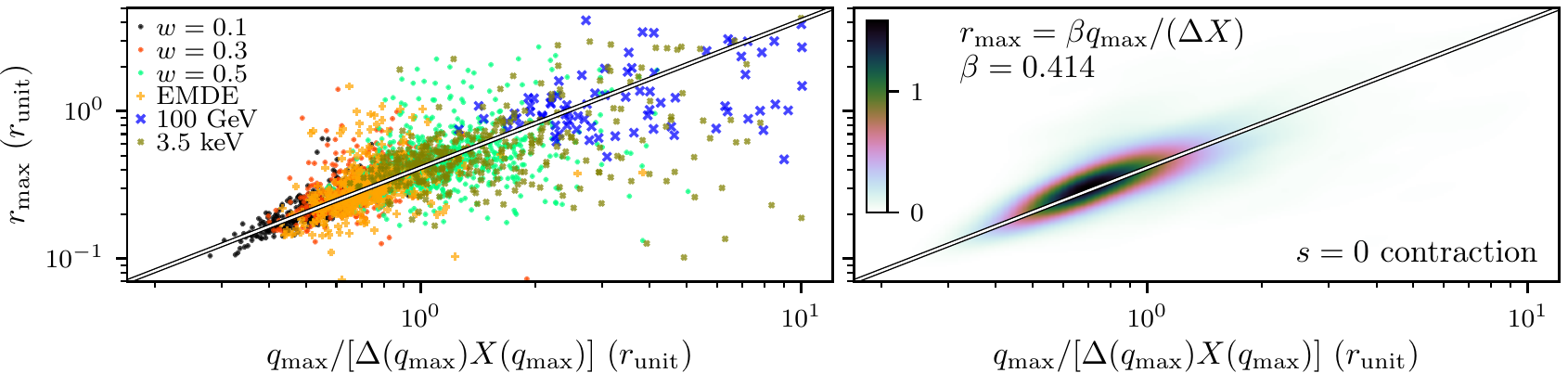}\\
	\includegraphics[width=\linewidth]{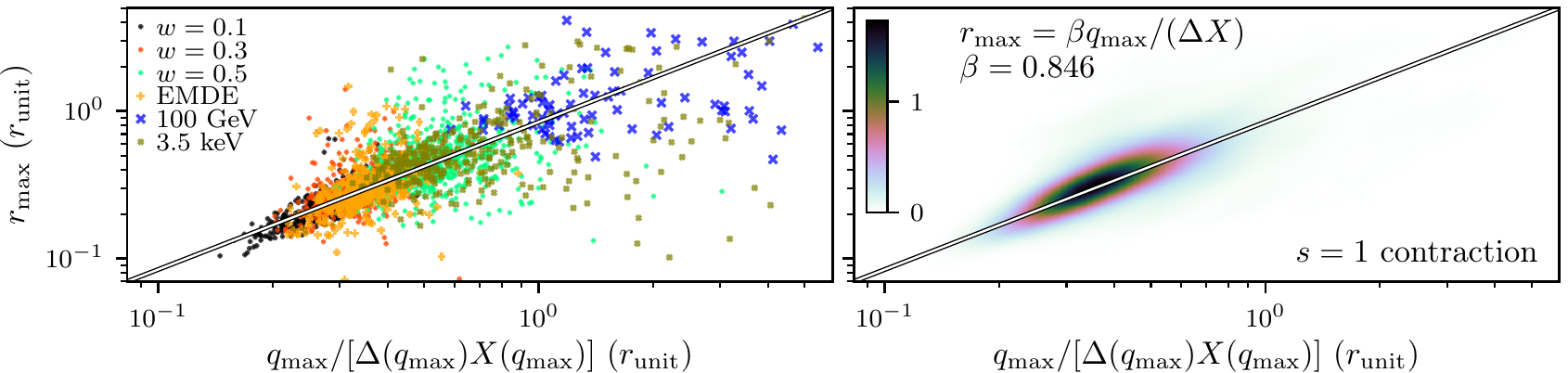}
	\caption{\label{fig:model_rmaxac} For the $s=0$ (top) and $s=1$ (bottom) contraction models, this figure plots the values of $r_\mathrm{max}$ measured in our simulations against the predicted values.  The left panels plot all of the halos, while the right panels show density estimates; the color scale indicates the density $f(\ln r_\mathrm{max},\ln r_\mathrm{max,pred})$ of the distribution, where $r_\mathrm{max,pred}$ is the x-axis quantity.  The solid line corresponds to the median proportionality constant.}
\end{figure*}

As a simple model, let us assume that each shell contributes density $\rho(r)\propto r^{-s}$ below its apocenter radius and 0 above it, so $f(r,r^\prime)=(r/r^\prime)^{3-s}$.  With this shell profile, Eq.~(\ref{X}) yields the ordinary differential equation
\begin{equation}\label{dX}
\frac{\mathrm{d}\ln X}{\mathrm{d}\ln q} = -\frac{3-\left[3(3-s)\epsilon(q)-s\right](X-1)}{1+(4-s)(X-1)}
\end{equation}
with initial condition $X(q_\mathrm{v})=1$.  This equation is equivalent to the simpler expression in Ref.~\cite{dalal2010origin}, but since it is expressed with respect to the variable $q$, it is more straightforward to integrate over a numerically tabulated peak profile $\epsilon(q)$.  The enclosed mass profile after contraction is now
\begin{equation}\label{Ma}
M(q)=\beta_M (4\pi/3)q^3\bar\rho_0 X(q)
\end{equation}
(to zeroth order in $\Delta$) with $\beta_M=1$.  Note that $X$ ranges from $1$ (at $q_\mathrm{v}$) to $\mathcal{O}(10)$.  Using these equations, one may maximize $M(q)/r(q)$ to obtain $q_\mathrm{max}$, after which $r_\mathrm{max}$ and $M(r_\mathrm{max})$ are obtained immediately.  Alternatively,
\begin{equation}\label{key}
\frac{\mathrm{d}\ln M}{\mathrm d\ln r}
=\frac{3+\mathrm{d}\ln X/\mathrm{d}\ln q} {1+3\epsilon(q)-\mathrm{d}\ln X/\mathrm{d}\ln q}
\end{equation}
with $\mathrm{d}\ln X/\mathrm{d}\ln q$ given in Eq.~(\ref{dX}), so one may solve $\mathrm{d}\ln M/\mathrm{d}\ln r = 1$ to obtain $q_\mathrm{max}$.  Finally, since we do not expect to employ the correct form of $f(r,r^\prime)$, we will allow $\beta$ and $\beta_M$ to vary, tuning them to simulations.

We test the contraction model using $s=0$ as well as $s=1$.  While Fig.~\ref{fig:shells} suggests that $s=0$ at small radii, a nonzero choice of $s$ is motivated by noting that for a given mass shell $q$ most of the enclosed mass contributed by higher shells comes from shells just slightly above $q$.  Meanwhile, Fig.~\ref{fig:shells} shows that the density profile contributed by a mass shell does not level off to a constant value until well below its apocenter radius.  Therefore, the bulk of the mass contribution comes from shell density profiles $\rho\propto r^{-s}$ with $s>0$.  The particular choice $s=1$ is partly arbitrary, but it is roughly the slope of the shell profiles in Fig.~\ref{fig:shells} slightly below their apocenters.\footnote{The choice $s=1$ also corresponds roughly to the $\rho^{1/2}$ model in Ref.~\cite{dalal2010origin}, since $\rho$ is approximately proportional to $r^{-2}$ at ${r=r_\mathrm{max}}$.}

Figure~\ref{fig:model_rmaxac} plots $r_\mathrm{max}$ against its model prediction for both contraction models using the $N_\mathrm{r}$ halos with ${r_\mathrm{max}<r_\mathrm{vir}}$.  Both models successfully predict, with some scatter, the values of $r_\mathrm{max}$ for the halos in all six simulations.  In fact, both models work equally well, as Fig.~\ref{fig:hist_rmax} demonstrates.  Moreover, comparing the results of these models to those of the turnaround models, which used the $\epsilon=2/3$ or $q_\mathrm{v}$ shells to predict $r_\mathrm{max}$, we see that the main difference is that the contraction models can handle all of the halos in a single model requiring just one tuned parameter.  Adiabatic contraction can produce a bend in the density profile, and hence an $r_\mathrm{max}$, near the halo outskirts because the outskirts are uncontracted while the rest of the halo is contracted.  Other than this, there is no significant advantage to the contraction models for predicting $r_\mathrm{max}$, as the top panel of Fig.~\ref{fig:hist_rmax} shows.

We also use the $s=0$ and $s=1$ contraction models to predict $M(r_\mathrm{max})$ using Eq.~(\ref{Ma}) with $\beta_M$ allowed to float.  For this test, we employ the $N_\mathrm{r}^{<3\mathrm{MM}}$ halos with $r_\mathrm{max}<r_\mathrm{vir}$ that underwent fewer than three major mergers.  The scatter in these predictions is depicted in the bottom panel of Fig.~\ref{fig:hist_rmax}.  The tuned values of $\beta$ and $\beta_M$ and their statistics are listed in Table~\ref{tab:params}.

\subsection{Virialization}\label{sec:energy}

\begin{figure}[t]
	\includegraphics[width=\columnwidth]{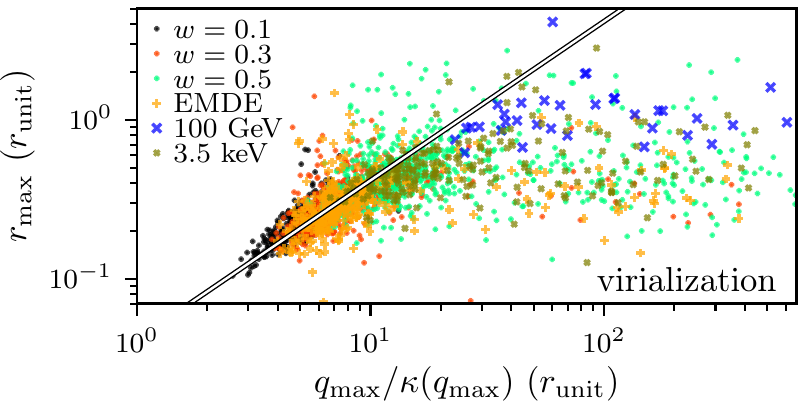}
	\caption{\label{fig:model_rmaxv} For the virialization model, this figure plots the values of $r_\mathrm{max}$ measured in our simulations against their predicted values.  For visual reference, an example proportionality curve is plotted (not a fit).  Like the turnaround model, this model dramatically overpredicts $r_\mathrm{max}$ for a subpopulation of the halos, a discrepancy that can be corrected in a similar fashion (see the text).}
\end{figure}

\begin{table*}[t]
	\caption{\label{tab:params} A summary of the predictive models along with their simulation-tuned coefficients.  In each model, the halo quantity is proportional to the peak quantity with the given coefficient between them.  The mean and rms deviation $\sigma$ in the proportionality coefficient are computed in log space.  We express the spread as a ratio: in particular, the middle 50\% spread is the ratio between the 75th and 25th percentiles, while the 1$\sigma$ spread is the ratio corresponding to the log-space $\sigma$.  We also list the number of halos contributing to the coefficient statistics.}
	\begin{ruledtabular}
		\begin{tabular}{ccccccccc}
			\shortstack{Halo\\quantity} & Peak quantity & Model & \shortstack{Median\\coefficient} & \shortstack{Middle 50\%\\spread} & \shortstack{Mean\\coefficient} & \shortstack{1$\sigma$\\spread} & \shortstack{Sample\\size} & \shortstack{Section\\ref.}\\
			\hline\vspace{-3mm}\\
			$A$ & $\bar\rho_0 \delta^{9/4} |\nabla^2\delta|^{-3/4}$ & Spherical collapse &
			$8.76$ & $1.43$ & $8.50$ & $1.44$ & $2451$ & \ref{sec:sc} \\
			$A$ & $\bar\rho_0 \delta^{9/4} |\nabla^2\delta|^{-3/4}f_\mathrm{ec}^{-3/2}(e,p)$ & Ellipsoidal collapse &
			$12.1$ & $1.31$ & $12.2$ & $1.36$ & $2450$\footnote{One peak is so ellipsoidal that Eq.~(\ref{ec}) has no solution, so this peak is discarded from analysis.} & \ref{sec:ellipsoid} \\
			$r_\mathrm{max}$ & $q_\mathrm{max}/\Delta(q_\mathrm{max})$ & Turnaround, $q_\mathrm{max}<q_\mathrm{v}$ &
			$0.131$ & $1.34$ & $0.132$ & $1.39$ & $1759$ & \ref{sec:frozen} \\
			$r_\mathrm{max}$ & $q_\mathrm{v}/\Delta(q_\mathrm{v})$ & Turnaround, $q_\mathrm{max}>q_\mathrm{v}$ &
			$0.088$ & $2.18$ & $0.085$ & $1.75$ & $712$ & \ref{sec:frozen} \\
			$r_\mathrm{max}$ & $q_\mathrm{max}/[\Delta(q_\mathrm{max})X(q_\mathrm{max})]$ & $s=0$ contraction &
			$0.414$ & $1.40$\footnote{The $s=0$ coefficient has spread $1.33$ and $1.74$ over halos for which the turnaround model predicts $q_\mathrm{max}<q_\mathrm{v}$ and $q_\mathrm{max}>q_\mathrm{v}$, respectively, implying that the contraction model is superior or comparable to the turnaround model in both cases.} & $0.406$ & $1.52$ & $2471$ & \ref{sec:adiabatic} \\
			$r_\mathrm{max}$ & $q_\mathrm{max}/[\Delta(q_\mathrm{max})X(q_\mathrm{max})]$ & $s=1$ contraction &
			$0.846$ & $1.39$\footnote{Similarly, the $s=1$ coefficient has spread $1.33$ and $1.68$ in these two cases.} & $0.848$ & $1.49$ & $2471$ & \ref{sec:adiabatic} \\
			$r_\mathrm{max}$ & $q_\mathrm{max}/\kappa(q_\mathrm{max})$ & Virialization, $q_\mathrm{max}<q_\mathrm{ta}$ &
			$0.042$ & $1.36$ & $0.042$ & $1.37$ & $1578$ & \ref{sec:energy} \\
			$r_\mathrm{max}$ & $q_\mathrm{v}/\Delta(q_\mathrm{v})$ & Virialization, $q_\mathrm{max}>q_\mathrm{ta}$ &
			$0.092$ & $1.95$ & $0.087$ & $1.70$ & $893$ & \ref{sec:energy} \\
			$M(r_\mathrm{max})$ & $(4\pi/3)q_\mathrm{max}^3\bar\rho_0$ & Turnaround, $q_\mathrm{max}<q_\mathrm{v}$ &
			$0.273$ & $1.64$ & $0.258$ & $1.84$ & $1514$ & \ref{sec:frozen} \\
			$M(r_\mathrm{max})$ & $(4\pi/3)q_\mathrm{v}^3\bar\rho_0$ & Turnaround, $q_\mathrm{max}>q_\mathrm{v}$ &
			$0.134$ & $2.97$ & $0.123$ & $3.13$ & $639$ & \ref{sec:frozen} \\
			$M(r_\mathrm{max})$ & $(4\pi/3)q_\mathrm{max}^3\bar\rho_0 X(q_\mathrm{max})$ & $s=0$ contraction &
			$0.441$ & $1.76$ & $0.396$ & $2.30$ & $2153$ & \ref{sec:adiabatic} \\
			$M(r_\mathrm{max})$ & $(4\pi/3)q_\mathrm{max}^3\bar\rho_0 X(q_\mathrm{max})$ & $s=1$ contraction &
			$0.658$ & $1.73$ & $0.619$ & $2.23$ & $2153$ & \ref{sec:adiabatic} \\
			$M(r_\mathrm{max})$ & $(4\pi/3)q_\mathrm{max}^3\bar\rho_0$ & Virialization, $q_\mathrm{max}<q_\mathrm{ta}$ &
			$0.150$ & $1.59$ & $0.145$ & $1.71$ & $1341$ & \ref{sec:energy} \\
			$M(r_\mathrm{max})$ & $(4\pi/3)q_\mathrm{v}^3\bar\rho_0$ & Virialization, $q_\mathrm{max}>q_\mathrm{ta}$ &
			$0.143$ & $2.79$ & $0.129$ & $2.98$ & $812$ & \ref{sec:energy}
		\end{tabular}
	\end{ruledtabular}
\end{table*}

Reference~\cite{salvador2012theoretical} developed a model for halo density profiles in which a mass shell freezes where its enclosed energy is virialized.\footnote{Note that a model in which a shell freezes where \textit{its own} energy is virialized is equivalent to the turnaround model above.}  In this model, the final radius of the $q$ shell is $r(q)=-(3/10)GM(q)^2/E(q)$, where $M(q)$ and $E(q)$ are, respectively, the mass and energy enclosed within the $q$ shell in the linear density field.  This model may be expressed as
\begin{equation}\label{re}
r(q)=\beta\frac{q}{\kappa(q)}, \ \ \kappa(q) \equiv \frac{2}{q^5}\int_0^q q^{\prime 4}\mathrm{d}q^\prime\Delta(q^\prime)
\end{equation}
with $\beta=3/10$ and $M(q)$ defined as in Eq.~(\ref{M_f}).  For this model, $\mathrm{d}\ln M/\mathrm{d}\ln r = 3/[6-2\Delta(q)/\kappa(q)]$.  As before, $q_\mathrm{max}$ is obtained by maximizing $M(q)/r(q)$ or by solving $\mathrm{d}\ln M/\mathrm{d}\ln r = 1$.  We test this model in a similar fashion, allowing $\beta$ to float and plotting in Fig.~\ref{fig:model_rmaxv} the simulated $r_\mathrm{max}$ against the model prediction (using the $N_\mathrm{r}$ halos with $r_\mathrm{max}>r_\mathrm{vir}$).  Evidently, this model performs similarly to the turnaround model.  Again, there is a multitude of halos of which the $q_\mathrm{max}$ shells have not yet accreted, leading to dramatically overpredicted values of $r_\mathrm{max}$.  In this case, the $q_\mathrm{max}<q_\mathrm{v}$ condition turns out to be too restrictive; it eliminates too many halos.  Instead, we require $q_\mathrm{max}<q_\mathrm{ta}$, where $q_\mathrm{ta}$ is Lagrangian radius of the last shell to turn around.  In particular, $q_\mathrm{ta}$ is the smallest $q$ satisfying $a\Delta(q)\leq\delta_\mathrm{ta}$, where $\delta_\mathrm{ta}=1.062$ and $a$ is the scale factor at which we are studying the halo population.  If $q_\mathrm{max}<q_\mathrm{ta}$, we use $q_\mathrm{max}$ to predict $r_\mathrm{max}$ using Eq.~(\ref{re}); otherwise, we use $q_\mathrm{v}$ to predict $r_\mathrm{max}$ using Eq.~(\ref{rmax_c}).  We also use this model to predict $M(r_\mathrm{max})$ in a similar fashion (again using the $N_\mathrm{r}^{<3\mathrm{MM}}$ halos that also underwent fewer than three major mergers), allowing $\beta_M$ to float.  The scatter and statistics of these predictions\footnote{We remark that the optimal $\beta\simeq 0.042$ is much smaller than the exact value $3/10$ claimed in Ref.~\cite{salvador2012theoretical}.} are shown in Fig.~\ref{fig:hist_rmax} and Table~\ref{tab:params}, and we find that this model's scatter is comparable to that of the turnaround model.

\subsection{Discussion}\label{sec:rdis}

Table~\ref{tab:params} summarizes the $r_\mathrm{max}$ and $M(r_\mathrm{max})$ models and the statistics of their simulation-tuned parameters.  Evidently, all of these models exhibit similar scatter.  For instance, when $q_\mathrm{max}<q_\mathrm{v}$ or $q_\mathrm{ta}$, each model's predicted $r_\mathrm{max}$ exhibits about 30\% scatter in the middle half of its predictions.  These similarities suggest there may be a statistical floor limiting the precision of all three models equally.  We propose two sources of such a floor.  The first is artificial discreteness noise in our simulations.  As noted in Paper II, there is discreteness noise in each radial bin of the density profile that is significantly larger than Poisson noise and may be associated with the accretion of artificial fragments.  Our procedure of averaging radial profiles over a finite time interval (see Appendix~\ref{sec:halos}) mitigates but does not eliminate this noise.\footnote{There are other sources of discreteness noise in our analysis, but they are minor.  The binning of our density profiles in factors of $1.1$ introduces artificial scatter, but this scatter does not exceed 5\% in $r_\mathrm{max}$ and is further reduced by our use of interpolation (see Appendix~\ref{sec:halos}).  There is also noise in the predicted values of $r_\mathrm{max}$ because they arise from a finite density grid, but since $q_\mathrm{max}$ is typically much larger than a grid cell [so $\delta(q_\mathrm{max})$ averages over a large number of cells], the resulting scatter is small.}

The other probable source of a statistical floor is physical and owes to a simplifying assumption common to all three models: spherical symmetry.  Observe in Fig.~\ref{fig:collapse_init} that at the initial radius of the $q_\mathrm{max}$ shell the density profile of the peak that is depicted---a typical peak drawn from the $w=0.3$ power spectrum---is highly ellipsoidal.  It is likely that a substantial fraction of the approximately $30\%$ scatter in $r_\mathrm{max}$ results from deviations from spherical symmetry in the initial density peaks.  To correct this, models must be employed that move beyond the assumption of spherically symmetric initial conditions, perhaps employing ellipsoidal collapse arguments \cite{bond1996peak,sheth2001ellipsoidal,sheth2002excursion} or drawing from nonspherical self-similar infall theory \cite{ryden1993self,lithwick2011self}.  The principal difficulty in moving beyond spherical symmetry is that the three-dimensional shape of a peak is different at each radius, an effect for which no model has accounted (to our knowledge).  However, the spherical models presented here exhibit just approximately $30\%$ scatter in $r_\mathrm{max}$ and approximately $60\%$ scatter in $M(r_\mathrm{max})$ over the full range of cosmologies we simulated, a success that should not be understated.  As we will see in Sec.~\ref{sec:summary}, halo mergers present a larger source of error.

\begin{figure}[t]
	\includegraphics[width=\columnwidth]{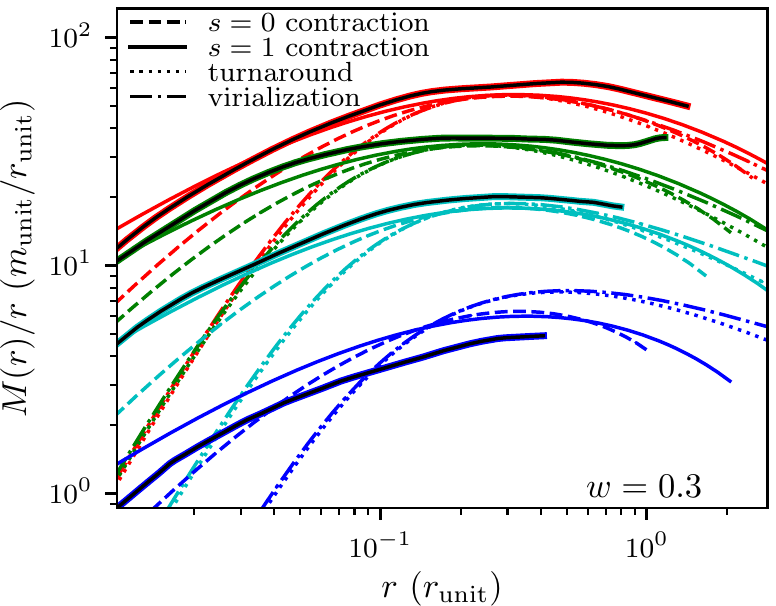} \\
	\includegraphics[width=\columnwidth]{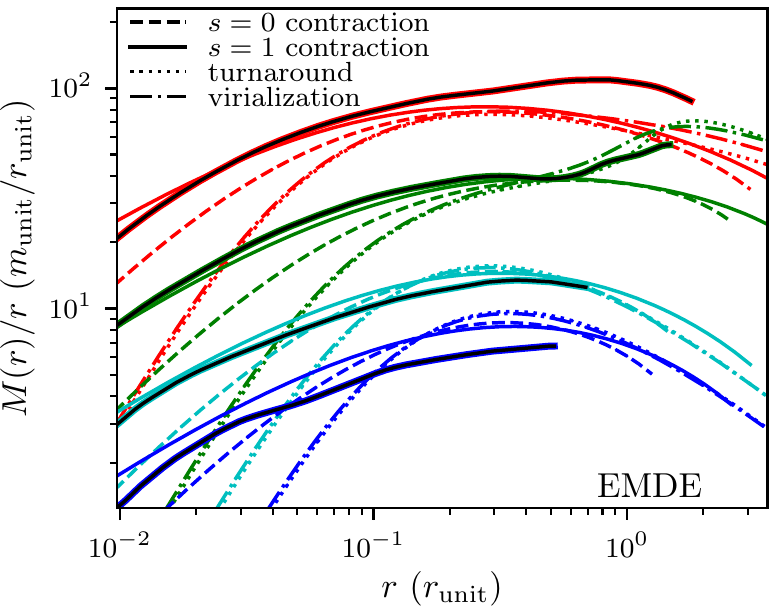}
	\caption{\label{fig:model_profiles} The mass profiles (thick, dark lines) of a few halos in the $w=0.3$ (top) and EMDE (bottom) simulations compared to their model predictions (thin lines).  This figure is intended to depict how the shapes of the profiles compare, so the sample was selected to have little discrepancy in the overall normalization of $M/r$.  The simulated profiles are plotted out to their virial radii.  The turnaround and virialization model predictions largely overlap.}
\end{figure}

Beyond predicting $r_\mathrm{max}$ and $M(r_\mathrm{max})$ (or equivalently $v_\mathrm{max}$), we may ask how well the spherical infall models can predict the full density profiles of the first halos.  No model successfully predicts $\rho\propto r^{-3/2}$ at small radii\footnote{An $s=3/2$ contraction model predicts $\rho\propto r^{-3/2}$ at small radii, but only in a contrived way; the adiabatic approximation is not valid when this part of the profile develops.}, and this is not surprising because in the moments after halo collapse the accretion of new material is not adiabatic with respect to the orbits of the already bound material.  Nevertheless, we may explore the radial range over which the spherical infall models do accurately predict the density profiles.  Figure~\ref{fig:model_profiles} compares the mass profiles of a small halo sample to their predicted profiles using the median proportionality coefficients in Table~\ref{tab:params}.  Evidently, the turnaround and virialization models do not accurately predict the density profile below $r_\mathrm{max}$, although they may succeed at larger radii.  The $s=0$ and $s=1$ contraction models, and especially the latter, more accurately capture the shape of the profile at smaller radii.  However, as shown in Fig.~\ref{fig:shells}, the true density profile of a mass shell is more complicated than $\rho\propto r^{-s}$, so a more sophisticated model for a shell's density profile should yield still better results.

In this section, we presented models that can predict the outer portions of the density profiles of the first halos in any cosmology.  While the models themselves are not new, our calibrations enable their use as predictive tools.  Moreover, our simulations demonstrate the universality of these models by showing that they succeed across wildly dissimilar cosmological scenarios. 

\section{Predicting halo populations}\label{sec:summary}

So far, we have studied the relationship between a density peak and its resulting halo.  However, our ultimate goal is to study the larger populations.  Recall from Sec.~\ref{sec:sims} that not all peaks matched to halos in our simulations and not all halos matched to peaks.  These discrepancies are reflected in Table~\ref{tab:halo_count} and may arise from physical processes, such as halo mergers, or numerical artifacts in our simulations.  We have shown that if we know that a peak developed into a halo that persisted to some later redshift then we can predict the properties of that halo from the properties of the peak.  Ultimately, however, we wish to predict an entire population of halos directly from a population of peaks.  In light of the halo- and peak-count discrepancies, can our models proceed in this way?

\subsection{Population comparisons}

\begin{figure}[t]
	\includegraphics[width=\columnwidth]{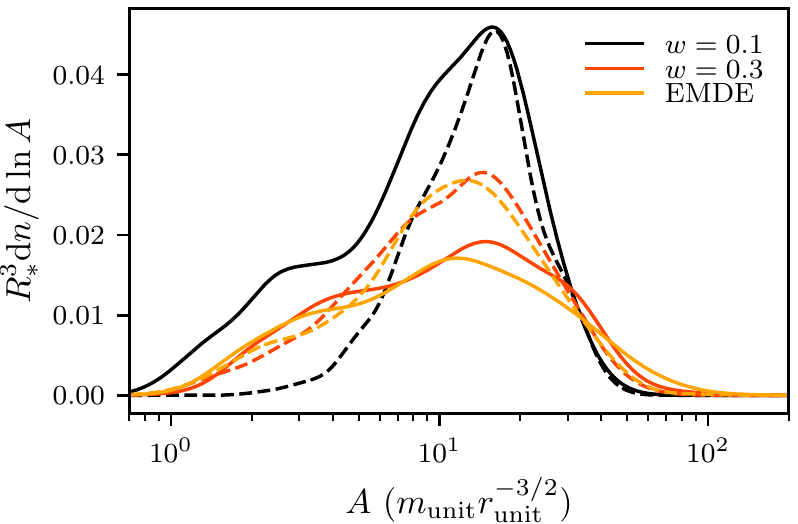}\\
	\includegraphics[width=\columnwidth]{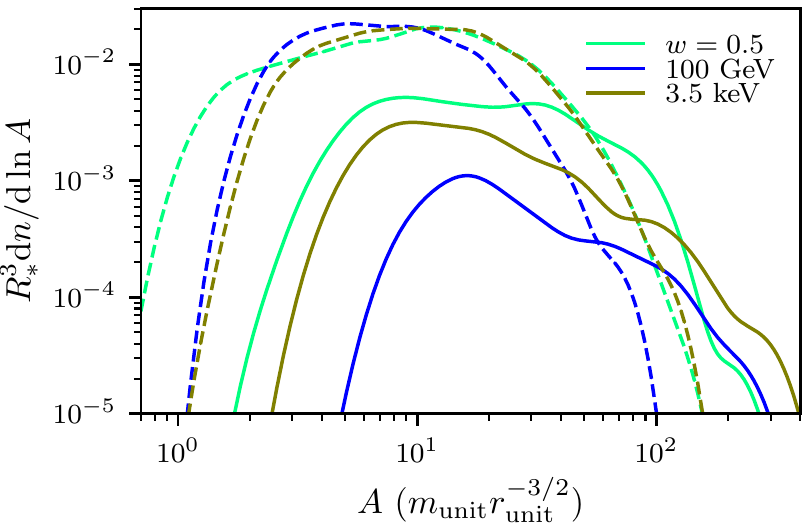}
	\caption{\label{fig:pop_A} A comparison between the simulated halo population, as solid lines, and the population predicted from the peaks in the linear density field, as dashed lines, distributed in the inner asymptotic coefficient $A$.  We use the ellipsoidal collapse model to predict $A$.  Top: The three simulations with narrower power spectra.  Bottom: The three simulations with broader power spectra.  The vertical axis is logarithmic here to accommodate the differences in scale.  We find that our model can capture halo populations arising from more narrowly supported power spectra, but it does not describe well the populations arising from broader power spectra because of the predominance of halo mergers in those scenarios.}
\end{figure}

We first study the population of halos distributed in the asymptote $A$.  In Fig.~\ref{fig:pop_A}, we compare the entire halo population found in our simulations to the population predicted by accounting for every peak in the initial density fields that would have collapsed by simulation termination.  We use the ellipsoidal collapse model.  Generally, we see that for the narrower $w=0.1$, $w=0.3$, and EMDE power spectra the predicted population matches the simulated population reasonably well.  In these scenarios, our model underpredicts halos at the low-density end, a surprising result that may be due to artificial fragmentation.  In simulations with a small-scale power-spectrum cutoff (like ours; see Fig.~\ref{fig:power}), discreteness noise causes filamentary structures to fragment into halos even below the scales of the smallest density fluctuations.  These halos have been shown to be unphysical simulation artifacts \cite{angulo2013warm,lovell2014properties}, but they could contribute to the excess of less-dense halos in our simulations relative to model predictions in the cases of the narrower power spectra.

Otherwise, Fig.~\ref{fig:pop_A} shows a tendency for the model to overpredict halos at the middle-density range and underpredict at the highest-density range.  This discrepancy can be attributed to mergers, for which our model does not account.  Halo mergers reduce the number of halos while raising the central density of the merger remnants.  For the broader $w=0.5$, 100 GeV, and 3.5 keV power spectra (bottom panel of Fig.~\ref{fig:pop_A}), the merger-based discrepancy is amplified: the model underpredicts the densest halos and dramatically overpredicts the rest.  Evidently, while our model can predict the density profiles of individual halos, an understanding of halo mergers is necessary to accurately predict halo populations in scenarios with more broadly supported power spectra.  We will return to this point.

\begin{figure}[t]
	\includegraphics[width=\columnwidth]{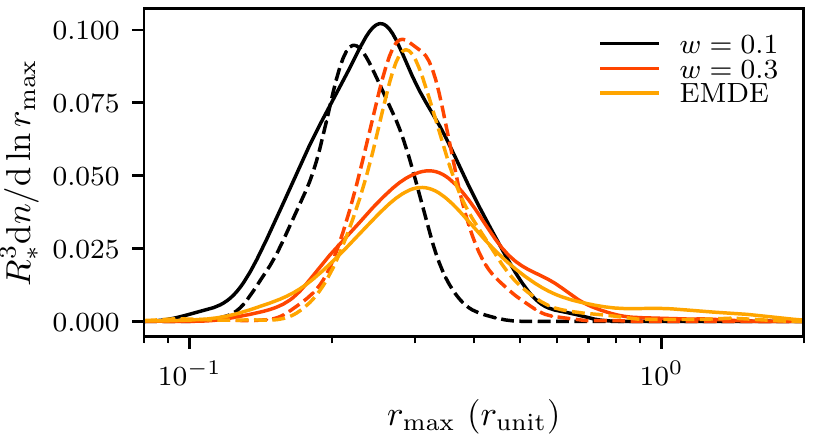}\\
	\includegraphics[width=\columnwidth]{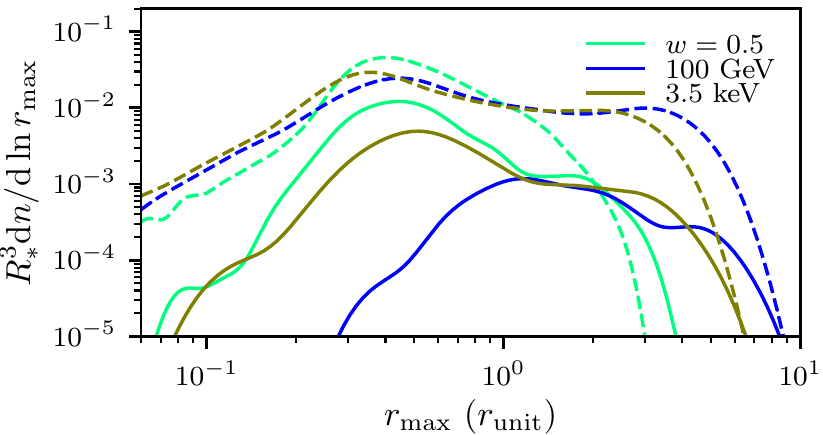}
	\caption{\label{fig:pop_r} Similar to Fig.~\ref{fig:pop_A}, but studying the halo population distributed in the radius $r_\mathrm{max}$ instead.  We use the $s=0$ contraction model.  As in Fig.~\ref{fig:pop_A}, we find that our model describes the halo populations arising from more narrowly supported power spectra with reasonable success, although the predicted distributions tend to be more sharply peaked.  However, halo mergers greatly alter the halo populations in scenarios with broader power spectra.}
\end{figure}

We next study the distribution in the radius $r_\mathrm{max}$.  Many of the simulated halos had $r_\mathrm{max}>r_\mathrm{vir}$, and while we discarded those halos from previous analyses, doing so now would alter the populations.  Instead, we account for this problem here by substituting $r_\mathrm{vir}$ for $r_\mathrm{max}$ in those cases.  As Fig.~\ref{fig:pop_r} shows, the $s=0$ contraction model predicts the $r_\mathrm{max}$ distribution reasonably well for the narrow power spectra, with the bulk of the discrepancy arising from overprediction of the total halo count due to halo mergers.  The predicted distributions are also more sharply peaked, an effect that may owe to the model's neglect of spherical asymmetry (see Sec.~\ref{sec:rdis}).  For the broader power spectra, the halo count discrepancy is magnified due to the much larger frequency of mergers.  However, unlike in the case of the asymptote $A$, we will soon see that the radius $r_\mathrm{max}$ does not significantly increase due to mergers beyond what the models already predict.  Hence, there is no significant underprediction of the largest-radius halos.

We can also ask how well our models predict aggregate observational signals.  If we assume that all halos have $\rho\propto r^{-3/2}$ inner density profiles, then the total dark matter annihilation rate in halos is proportional to the sum $\sum A^2$ over all halos.\footnote{There is a logarithmic sensitivity to $r_\mathrm{max}$, which we neglect.  Also, note that there is a density cap imposed by annihilations \cite{berezinsky1992distribution}, so the signal converges even for $\rho\propto r^{-3/2}$.}  In Table~\ref{tab:pop}, we show how accurately the ellipsoidal collapse model predicts the aggregate annihilation signal.  Remarkably, despite not accounting for halo mergers, we successfully predict the annihilation signal to within a factor of 1.3 for all simulations.  The increased central density within merger remnants has compensated for the drop in halo count.  One caveat is that in this calculation, we did not account for changes in the slope $\gamma$ of the $\rho\propto r^{-\gamma}$ inner profile resulting from mergers.  This change must be accounted for separately and will significantly reduce the annihilation signal (see Paper II).

\begin{table}[t]
	\caption{\label{tab:pop} This table shows how accurately our models predict aggregate halo signals.  We consider simplified aggregate signals $\sum A^2$ for annihilation and $\sum M(r_\mathrm{max})r_\mathrm{max}$ for lensing, each summed over the population of predicted or simulated halos.  In each case, we list the ratio of the signal predicted from the linear density field to the signal aggregated over halos in the simulation box.  We use the ellipsoidal collapse and $s=0$ contraction models.}
	\begin{ruledtabular}
		\begin{tabular}{ccc}
			Simulation & $\sum A^2$ & $\sum M(r_\mathrm{max})r_\mathrm{max}$\\
			\hline\vspace{-3mm}\\
			$w=0.1$ & 0.91 & 0.56 \\
			$w=0.3$ & 0.89 & 0.69 \\
			$w=0.5$ & 0.91 & 2.2 \\
			EMDE    & 0.78 & 0.66 \\
			100 GeV & 1.15 & 33 \\
			3.5 keV & 1.16 & 14 \\
		\end{tabular}
	\end{ruledtabular}
\end{table}

We also consider an aggregate microlensing signal.  The apparent image of a background star is deflected by an angle proportional to $M_\mathrm{2D}(\xi)/\xi$, where $M_\mathrm{2D}(\xi)$ is the projected mass enclosed within impact parameter $\xi$.  As an approximation, we claim that the deflection due to a halo is proportional to $M(r_\mathrm{max})/r_\mathrm{max}$ and a function related to the impact parameter.  Integrating over impact parameters introduces a factor $r_\mathrm{max}^2$, the characteristic area of the halo.  Hence, for a halo population, the aggregate lensing signal, considered as the expected deflection of a given image, is proportional to $\sum M(r_\mathrm{max})r_\mathrm{max}$ summed over all halos.

In Table~\ref{tab:pop}, we show how accurately the $s=0$ contraction model predicts this signal.  We find that our models predict the lensing signal significantly worse than the annihilation signal, although in four of the simulations, the prediction still agrees to about a factor of 2.  For the simulations drawn from the 3.5 keV and 100 GeV power spectra, however, the model overpredicts the lensing signal by an order of magnitude.  These power spectra yield the greatest prevalence of halo mergers, and unlike in the case of annihilations, mergers do not sufficiently boost the remnant's lensing signal relative to its model prediction to compensate for the loss of halo count in mergers.  Hence, the lensing signal in the simulations is much smaller than that predicted from the linear density field.  We remark, however, that the story may change if instrument sensitivities are taken into account.  Mergers predominantly destroy smaller halos, and their signals may have been beyond sensitivity limits regardless.

It is clear from these results that an accurate accounting of halo mergers is necessary to predict a halo population in any generality.  However, for narrower power spectra such as our $w=0.1$, $w=0.3$, or EMDE cases, mergers are subdominant, and our models can predict the populations reasonably well.  Moreover, when considering aggregate annihilation signals, halo mergers may not have a significant impact beyond altering the slope $\gamma$ of the $\rho\propto r^{-\gamma}$ inner density profiles.  For these uses, we describe in Appendix~\ref{sec:stats} a method to sample the halo population directly from the matter power spectrum $\mathcal{P}(k)$, bypassing the step of drawing a density field $\delta(\vec x)$.  Halo populations predicted using this method are slightly different from those predicted using our density fields, but we propose that they are more accurate; the method we describe can easily sample a much larger number of peaks, and it is not subject to errors related to the finite size and finite grid spacing of a sampled density field.\footnote{If sampling directly from the power spectrum is more accurate, one may wonder why we chose to compare our simulation results to those predicted from the less-accurate finite density fields (Figs. \ref{fig:pop_A} and~\ref{fig:pop_r}).  The reason for this choice is that it leads to a more explicit test of our model.  Since we compare a simulation's results to model predictions using the same simulation's initial density field, any discrepancy that is not a simulation artifact can be attributed to the model.}

\subsection{Halo mergers}\label{sec:merge}

A full treatment of mergers is beyond the scope of this work, but we are poised to make some observations.  Our procedure for matching a halo to its predecessor density peak involved tracking each halo backward through time.  During this process, we counted the number of major mergers this halo underwent, which we define to be a merger between two halos with mass ratio smaller than 3.  Over half of our halos with well-resolved asymptotes experienced at least one major merger, and 12\% experienced at least three.

In an effort to find a simple way to cull halos that end up merging, we explored cutting out density peaks that were too close to an earlier-collapsing density peak.  We considered two characteristic comoving length scales below which to make these cuts: the scale $q_\mathrm{pk}$ associated with the density peak (see Sec.~\ref{sec:asymptote}) and the scale $q_\mathrm{max}$ where $\epsilon(q_\mathrm{max})=2/3$ (see Sec.~\ref{sec:frozen}).  Unfortunately, neither of these cuts produced sensible results.  The $q_\mathrm{pk}$ cut culled far too few halos, while the $q_\mathrm{max}$ cut culled far too many.  Ultimately, we expect that a more sophisticated accounting of mergers will be necessary, possibly following along the lines of extended Press-Schechter theory \cite{lacey1993merger} or the \textsc{Peak Patch} algorithm \cite{bond1996peak,bond1996peak2,stein2018mass}.

\begin{figure}[t]
	\includegraphics[width=\linewidth]{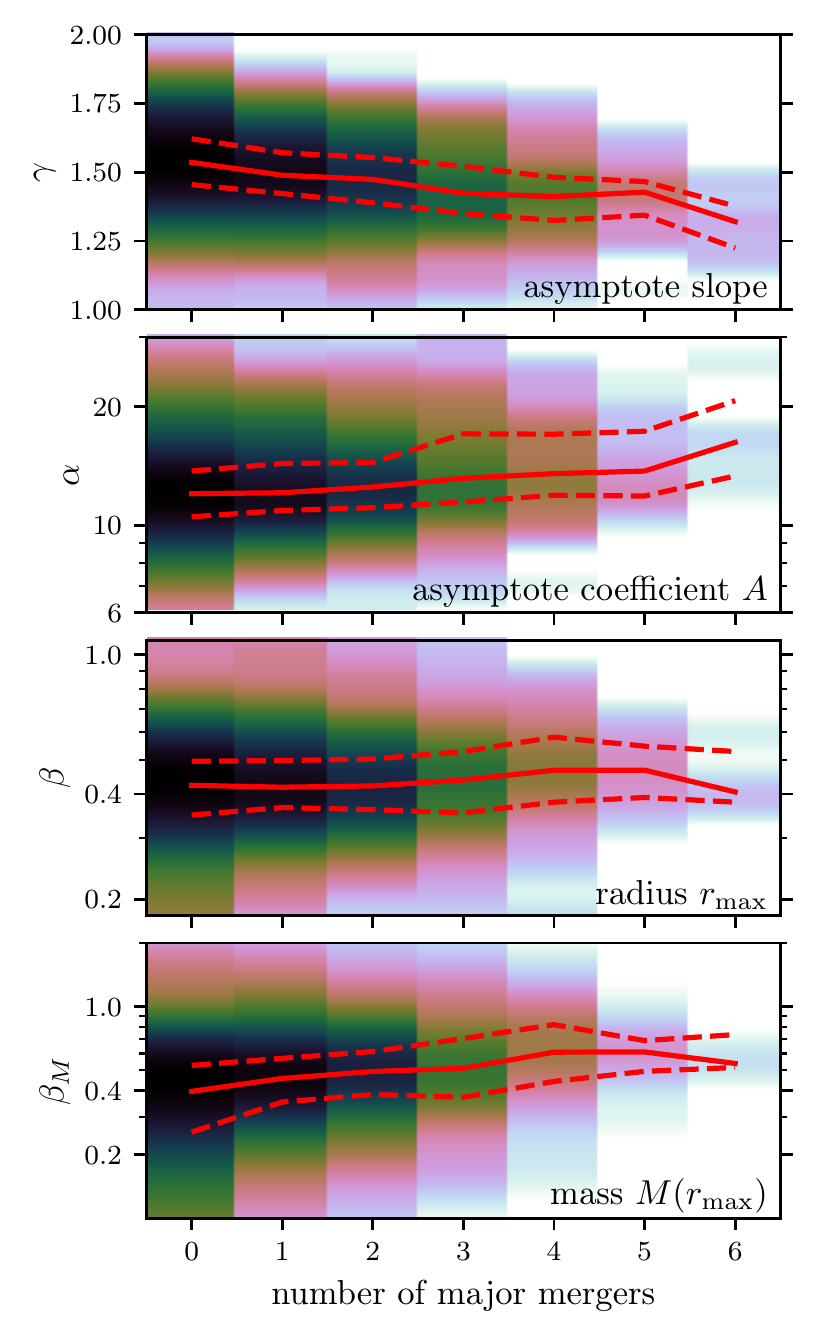}
	\caption{\label{fig:merge} The impact of major mergers on the density profiles in our simulations.  Top: The effect on the power-law index $\gamma$ of the $\rho\propto r^{-\gamma}$ inner asymptote.  Since we do not resolve these radii well, much of the scatter likely arises from numerical noise.  Middle to bottom: The effects on the asymptote $A$, radius $r_\mathrm{max}$, and mass $M(r_\mathrm{max})$, respectively, plotted as the ratio between the measured value and the prediction using our models.  For the asymptote, we use the ellipsoidal collapse model, while for $r_\mathrm{max}$ and $M(r_\mathrm{max})$, we use the $s=0$ contraction model.  The color scale, which is logarithmic, represents a density estimate (in log space for $\alpha$, $\beta$, and $\beta_M$; darker is denser), while the lines mark the median and the 25th and 75th percentiles at each merger count.}
\end{figure}

Another question is how halo density profiles change due to mergers.  Reference~\cite{ogiya2016dynamical} found that successive mergers cause the slope $\gamma$ of the small-radius asymptote of the density profile, $\rho\propto r^{-\gamma}$, to become shallower than its initial value of $\gamma=3/2$.  The same work also found that successive mergers increase the central density.  Figure~\ref{fig:merge} shows that both of these results are borne out in our own simulations as well.  There is a clear trend wherein more major mergers lead to successively denser\footnote{While the coefficient $A$ is only well defined if $\gamma=3/2$, our procedure for measuring $A$ (see Appendix~\ref{sec:halos}) obtains a logarithmically averaged value of $\rho r^{3/2}$ over the inner density profile regardless.} but shallower inner structures.  On the other hand, we also see in Fig.~\ref{fig:merge} that major mergers do not significantly increase $r_\mathrm{max}$ beyond what our models already predict, although they do increase $M(r_\mathrm{max})$.

\section{Conclusion}\label{sec:conclusion}

The first halos form by direct collapse of peaks in the primordial density field.  If some of these halos survive the subsequent hierarchical clustering process, as evidence suggests \cite{polisensky2015fingerprints,angulo2017earth,ogiya2016dynamical,drakos2018major,berezinsky2008remnants,hayashi2003structural,penarrubia2010impact,ogiya2019dash}, then they would be the densest dark matter objects in the Universe.  In this work, we presented models that predict the density profiles of these halos directly from the properties of the density peaks that formed them, and we used high-resolution cosmological simulations in a variety of different scenarios to tune and validate these models.  The models are described in Secs. \ref{sec:asymptote} and~\ref{sec:rmax} and summarized in Table~\ref{tab:params}.  They have simulation-tuned parameters, but these parameters are independent of cosmology and do not need to be retuned to accommodate different scenarios.

We treated small and large radii separately, with the understanding that these regimes form under radically different circumstances.  At small radii, the first halos develop $\rho=A r^{-3/2}$ density profiles \cite{ishiyama2010gamma,anderhalden2013density,*anderhalden2013erratum,ishiyama2014hierarchical,polisensky2015fingerprints,angulo2017earth,ogiya2017sets,delos2018ultracompact,delos2018density}, and our model predicts the coefficient $A$ with remarkable success.  Over the full range of cosmologies we explored, the scatter from our model is only in the vicinity of 30\%-40\%, and it is likely that a significant fraction of that scatter arises from numerical noise in our simulations.  The density profiles at large radii are more varied, but we parametrize them using the radius $r_\mathrm{max}$ at which the circular velocity is maximized along with the mass $M(r_\mathrm{max})$ enclosed within that radius.  For this regime, we employ models already present in the literature \cite{gunn1972infall,dalal2010origin,salvador2012theoretical}, but we supply calibrations that enable their use as predictive tools applicable to any cosmology.  The scatter from our models is roughly 40\%-50\% in $r_\mathrm{max}$ and a factor of 2 in $M(r_\mathrm{max})$, and, similarly, not all of this scatter is physical.

In this way, the models we presented can predict the halo arising from a given density peak.  Our goal, however, is to predict populations of halos.  For a power spectrum of density fluctuations that is narrowly supported, such as the spectrum imprinted by an EMDE \cite{erickcek2011reheating,barenboim2014structure,fan2014nonthermal,erickcek2015dark} or certain inflationary models \cite{salopek1989designing,starobinsky1992,ivanov1994inflation,randall1996supernatural,starobinsky1998beyond,martin2000nonvacuum,chung2000probing,barnaby2009particle,barnaby2010features,bugaev2011curvature}, our models replicate the entire halo population reasonably well.  Thus, our models can serve as a tool to predict the observational signals of such cosmologies.  However, for more broadly supported power spectra, such as those arising from a scale-invariant initial spectrum, halo mergers dramatically alter the halo population, causing the populations that arise in our simulations to differ substantially from the populations our models predict.  Interestingly, modulo changes in the slopes $\gamma$ of the $\rho\propto r^{-\gamma}$ inner density profiles, dark matter annihilation signals seem to be sufficiently close to additive in halo mergers that our models still predict the aggregate annihilation signal to within 30\% in every cosmology we tested.

Nevertheless, it is not clear that this additivity in the annihilation signal should extend beyond the timescales spanned by our simulations.  More broadly, our models predict the initial halo population, but a proper understanding of halo mergers is needed in order to robustly connect it to the population today.  It is necessary to understand both how mergers are distributed across halos and time and how they impact halo density profiles.  Methods exist that can predict the distribution of mergers, the most prominent of which is the extended Press-Schechter theory \cite{lacey1993merger}.  The \textsc{Peak Patch} algorithm \cite{stein2018mass} represents a method that may be easier to adapt, among other advantages, if more computationally expensive to apply.  Meanwhile, the larger challenge is to predict how halo mergers alter density profiles.  References \cite{ogiya2016dynamical,angulo2017earth,drakos2018major} (for major mergers) and~\cite{hayashi2003structural,penarrubia2010impact,ogiya2019dash} (for minor mergers) represent steps toward this goal, but there is not, as yet, a sufficiently general model.

Moreover, there is room for improvement in our models themselves.  In predicting the density profiles at large radii, we assume that each mass shell contributes density in a profile that is a single power law up to a maximum radius.  In reality, shell profiles follow more complicated forms that are sensitive to the total density profile; see Fig.~\ref{fig:shells} and Refs~\cite{lithwick2011self,dalal2010origin}.  Utilizing more accurate shell profiles would likely improve the model predictions for the density profile, especially in predicting its broader shape rather than only $r_\mathrm{max}$ and $M(r_\mathrm{max})$.  Also, our model for the density profile at large radii completely discounts any deviations from sphericity in the initial peak.  Another avenue for improvement may be to incorporate ellipsoidal collapse.  At small radii, the reason the $\rho\propto r^{-3/2}$ profile arises is not well understood, and its accuracy has only been confirmed down to the resolution limits of $N$-body simulations.  A physical understanding must be developed of the mechanism by which this profile arises in order to confirm whether the profile truly extends to arbitrarily small radii.

The primordial density field and its power spectrum of density fluctuations comprise a valuable window into the early Universe and the nature of dark matter.  Our work in this paper was carried out as part of an effort to use the observational signatures of dark matter halos to probe these fluctuations.  Further research is still needed to understand the impact of mergers on halo populations before these signatures, or their nonobservation, can be employed to robustly constrain cosmology.  Nevertheless, the models presented in this work, which predict the initial halo population, represent a step forward in our capacity to use this probe.

\begin{acknowledgments}
The authors thank Dragan Huterer for bringing Ref.~\cite{dalal2010origin} to their attention.  The simulations for this work were carried out on the KillDevil and Dogwood computing clusters at the University of North Carolina at Chapel Hill.  M.\,S.\,D. and A.\,L.\,E. were partially supported by NSF Grant No. PHY-1752752.  Several key figures in this work employ the cube-helix color scheme developed by Ref.~\cite{green2011colour}.
\end{acknowledgments}

\appendix

\section{Collecting halo data}\label{sec:halos}

In Sec.~\ref{sec:sims}, we carried out six simulations and catalogued all halos present at the final redshift of each.  The models we present in Secs. \ref{sec:asymptote} and~\ref{sec:rmax} require data on the density profiles of these halos, and in this section, we detail our methods for collecting these data.

\begin{figure}[t]
	\centering
	\includegraphics[width=\columnwidth]{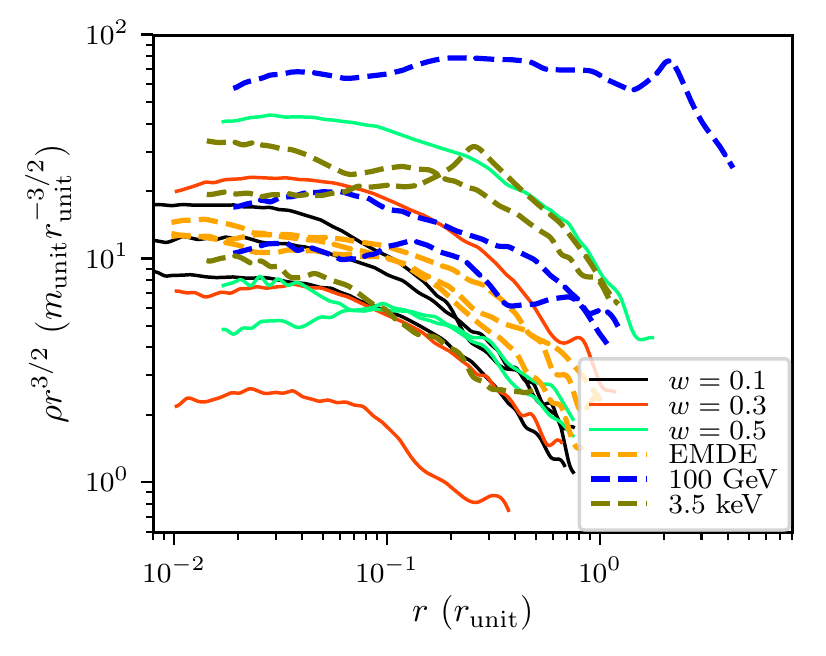}
	\caption{\label{fig:densityprofiles} The density profiles of three random halos from each simulation's final box.}
\end{figure}

We first need to obtain the density profile and enclosed mass profile of each halo at the final redshift of each simulation.  To reduce random noise, we use the procedure described in Paper II, wherein the profiles are averaged over a time interval.  The profiles are binned at successive factors of $1.1$ in radius, but to mitigate noise associated with the binning scheme, we use a cubic spline to smoothly interpolate them.  These profiles are only valid down to the radius $r_\mathrm{soft}$ corresponding to the separation below which simulation forces become non-Newtonian; for \textsc{Gadget-2}, $r_\mathrm{soft}$ is 2.8 times the force softening length.  At large radii, we cut off the density profile at the radius $r_\mathrm{vir}$ inside which the mean enclosed density is 200 times the background density.  Figure~\ref{fig:densityprofiles} shows a random sample of halo density profiles from each simulation.

The model developed in Sec.~\ref{sec:asymptote} predicts the coefficient $A$ of the $\rho=Ar^{-3/2}$ asymptote of the density profile at small $r$.  Thus, we wish to extract $A$ from each density profile.  Starting at $r_\mathrm{soft}$, we find the cumulative (logarithmic) average of $\rho r^{3/2}$ across radial bins, and we set $A$ to be the maximum of this cumulative average.  The averaging procedure is intended to minimize noise resulting from employing a narrow range of radii while at the same time minimizing the influence of any bend in the density profile at larger radii.  However, we also tested the alternative procedure of simply finding the average of $\rho r^{3/2}$ within a factor of 3 in radius above $r_\mathrm{soft}$ and found it to yield similar results.

We also wish to accommodate deviations from ${\rho\propto r^{-3/2}}$ at small radii.  Unlike those of Paper II, our simulations do not have the resolution to clearly resolve the power-law index of the small-radius asymptote, but we may still hope to see statistical correlations in our large sample.  For each halo, we measure the slope $\gamma$ of the small-radius asymptote $\rho\propto r^{-\gamma}$ by considering the radius range from $r_\mathrm{soft}$ to $3r_\mathrm{soft}$ and fitting a line in log space to the density profile within this range.

To test the models presented in Sec.~\ref{sec:rmax}, we also need to compute the radius $r_\mathrm{max}$ at which the circular velocity is maximized along with the mass $M(r_\mathrm{max})$.  Since we already obtained the enclosed mass profile $M(r)$ for each halo, we simply find the maximum value of $M(r)/r$.  The parameters $r_\mathrm{max}$ and $M(r_\mathrm{max})$ are then defined to be the radius and enclosed mass at which $M(r)/r$ is maximized.

\section{Collecting peak parameters}\label{sec:peaks}

The models we present in Secs. \ref{sec:asymptote} and~\ref{sec:rmax} employ data on the peaks in the linear density field ${\delta(\vec x)\equiv\delta(\vec x,a)/a}$ (evaluated during matter domination).  We obtain these data using Fourier methods, defining
\begin{equation}
\delta(\vec k) \equiv \int\!\mathrm d^3\vec x\, \mathrm e^{-i \vec k \cdot \vec x} \delta(\vec x).
\end{equation}
The derivatives of $\delta$ at a peak located at $\vec x$ immediately follow as
\begin{equation}
\partial_i\partial_j\delta = -\int\!\frac{\mathrm d^3\vec k}{(2\pi)^3} \mathrm e^{i \vec k \cdot \vec x} k_i k_j\delta(\vec k).
\end{equation}
Similarly, using Poisson's equation, the derivatives of the (peculiar) gravitational potential are
\begin{equation}
\partial_i \partial_j \phi = 4\pi G\bar\rho_0\int\!\frac{\mathrm d^3\vec k}{(2\pi)^3} \mathrm e^{i \vec k \cdot \vec x} \frac{k_i k_j}{\vec k^2}\delta(\vec k).
\end{equation}
The ellipsoidal refinement described in Sec.~\ref{sec:ellipsoid} requires the three-dimensional shape parameters $e_\phi$ and $p_\phi$ for the potential $\phi$ about the peak.  Taking $\lambda_1\geq\lambda_2\geq\lambda_3$ to be the eigenvalues of $\partial_i\partial_j\phi$, these parameters are defined
\begin{equation}
e_\phi\equiv \frac{\lambda_1-\lambda_3}{2(\lambda_1+\lambda_2+\lambda_3)}
\ \ \mathrm{and} \ \ 
p_\phi\equiv \frac{\lambda_1+\lambda_3-2\lambda_2}{2(\lambda_1+\lambda_2+\lambda_3)}.
\end{equation}

Our model in Sec.~\ref{sec:rmax} requires the density profile $\delta(r)$ and mass profile $\Delta(r)$ about the peak.  For a peak centered at $\vec x$, these profiles are computed as
\begin{equation}\label{key}
\left\{
\begin{array} {c}
\delta(r) \\
\Delta(r) \\
\zeta(r) \\
\end{array}
\right\}
= \int\!\frac{\mathrm d^3\vec k}{(2\pi)^3} \mathrm e^{i \vec k \cdot \vec x} \delta(\vec k)
\left\{
\begin{array} {c}
\sinc(k r) \\
W(k r) \\
\cos(k r) \\
\end{array}
\right\},
\end{equation}
where $\sinc(x)\equiv \sin(x)/x$ and $W$ is the top-hat window function, $W(x)\equiv (3/x^3)(\sin x - x\cos x)$.  The third profile $\zeta(r)$ is useful because it is related to the derivative of $\delta(r)$.  In particular, to numerically integrate Eq.~(\ref{dX}), we interpolate $\delta(r)$ and $\Delta(r)$ with piecewise polynomials, using the relations
\begin{equation}
\frac{\mathrm{d}\delta}{\mathrm{d}\ln r} = \zeta(r)-\delta(r)
\ \ 
\text{and}
\ \ 
\frac{\mathrm{d}\Delta}{\mathrm{d}\ln r} = 3\left[\delta(r)-\Delta(r)\right]
\end{equation}
to fix their derivatives.

\begin{figure}[t]
	\centering
	\includegraphics[width=\columnwidth]{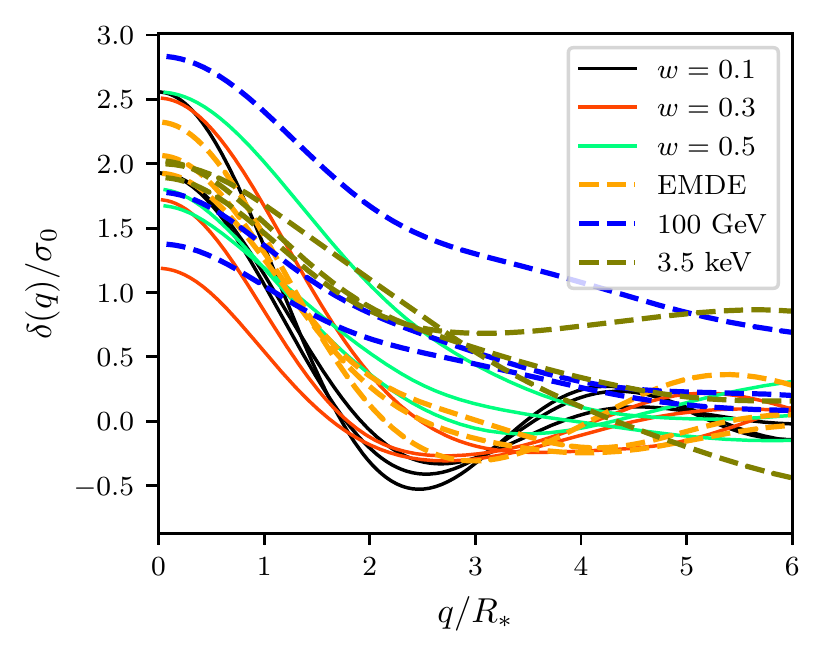}
	\caption{\label{fig:deltaprofiles} The density profiles of three peaks from each simulation's initial box.  These peaks are chosen to match the halos of which the density profiles are depicted in Fig.~\ref{fig:densityprofiles}.}
\end{figure}

Figure~\ref{fig:deltaprofiles} shows the density profiles $\delta(r)$ of three peaks from each of the six simulations.  The peaks displayed are those that later collapse into the halos depicted in Fig.~\ref{fig:densityprofiles}.

\section{Predicting the halo population from the power spectrum}\label{sec:stats}

A primary goal of this work is to enable a prediction of the halo population given a power spectrum $\mathcal{P}(k)$ of density fluctuations.  One option is to sample a density field from the power spectrum, use the methods described in Appendix~\ref{sec:peaks} to characterize the peaks, and then apply the models developed in this work.  However, it is possible to exploit the statistics of a Gaussian random field to compute the halo distribution more directly. In this section, we outline a practical procedure to perform this computation by sampling from the peak distribution.  Similarly to earlier sections, we define $\mathcal{P}(k)\equiv\mathcal{P}(k,a)/a^2$, where $\mathcal{P}(k,a)$ is the dimensionless matter power spectrum evaluated using linear theory during matter domination.  All quantities derived therefrom, such as $\sigma_j$ and $\delta(\vec x)$, inherit similar scaling.

\subsection{Number density of peaks}

The first step is to find the total number density $n$ of peaks.  As derived in Ref.~\cite{bardeen1986statistics}, the differential number density of peaks in a Gaussian random field, in terms of parameters $\nu\equiv \delta/\sigma_0>0$ and $x\equiv -\nabla^2\delta/\sigma_2>0$, is
\begin{equation}\label{dndnudx}
\frac{\mathrm{d}^2n}{\mathrm{d}\nu\mathrm{d}x}
=
\frac{\mathrm e^{-\nu^2/2}}{(2\pi)^2R_*^3}f(x)\frac{\exp\!\left[-\frac{1}{2}(x-\gamma\nu)^2/(1-\gamma^2)\right]}{[2\pi(1-\gamma^2)]^{1/2}},
\end{equation}
where $\gamma\equiv \sigma_1^2/(\sigma_0\sigma_2)$, $\sigma_j$ and $R_*$ are defined in Eqs. (\ref{sigmaj}) and~(\ref{Rstar}), and
\begin{align}\label{bbks_f}
f(x)
\equiv&
\frac{x^3-3x}{2}\left[\erf\!\left(\sqrt{5/2}x\right)+\erf\!\left(\sqrt{5/8}x\right)\right]
\nonumber\\
+&
\sqrt{\frac{2}{5\pi}}\left[\left(\frac{31}{4}x^2+\frac{8}{5}\right)\mathrm e^{-\frac{5}{8}x^2}+\left(\frac{x^2}{2}-\frac{8}{5}\right)\mathrm e^{-\frac{5}{2}x^2}\right].
\end{align}
The $\nu$ integral can be carried out analytically, leading to 
\begin{equation}\label{dndx}
\frac{\mathrm{d}n}{\mathrm{d}x}=
\frac{f(x)}{8\pi^2R_*^3}\mathrm e^{-x^2/2}\left[1+\erf\!\left(\frac{x\gamma}{\sqrt{2(1-\gamma^2)}}\right)\right],
\end{equation}
and this equation can be integrated numerically over ${x\geq 0}$ to obtain $n$.

\subsection{Asymptote $A$}

Next, we use Monte Carlo methods to compute the distribution of coefficients $A$ of the $\rho = A r^{-3/2}$ small-radius asymptote.  The model described in Sec.~\ref{sec:asymptote} predicts this asymptote from the amplitude $\delta$ and curvature $|\nabla^2\delta|$ of the density peak along with the shape parameters\footnote{Since these parameters describe the potential, they are different from the parameters $e$ and $p$ in Ref.~\cite{bardeen1986statistics}.} $e$ and $p$ associated with the potential about the peak.  Equation~(\ref{dndnudx}) supplies the peak distribution in $\nu\equiv \delta/\sigma_0$ and $x\equiv -\nabla^2\delta/\sigma_2$.  Meanwhile, the conditional distribution of $e$ and $p$ for a peak of height $\nu$, derived in Ref.~\cite{sheth2001ellipsoidal}, is
\begin{equation}\label{key}
f(e,p|\nu) = \frac{1125}{\sqrt{10\pi}}e(e^2-p^2)\nu^5\exp\left[-\frac{5}{2}\nu^2(3e^2+p^2)\right].
\end{equation}

To compute the distribution of $A$, we now employ a Monte Carlo procedure.  We use rejection methods to sample $x$ from Eq.~(\ref{dndx}) and then sample $\nu$ from Eq.~(\ref{dndnudx}).  Next, we employ the cumulative distributions of $e$ and $p$,
\begin{equation}\label{Fe}
\begin{split}
F(e|\nu) =&\ 
\mathrm e^{-\frac{15}{2}e^2\nu^2} \left(1-15e^2\nu^2\right) \erf\!\left(\sqrt{5/2}e\nu\right)
\\&
-3\sqrt{10/\pi}e\nu\mathrm e^{-10e^2\nu^2}+\erf\!\left(\sqrt{10}e\nu\right)
\end{split}
\end{equation}
and
\begin{equation}\label{Fp}
\begin{split}
F(p|e,\nu) = \frac{1}{2}\Bigg\{\!\!&
\left(5 e^2 \nu ^2-1\right)\!\! \left[\erf\!\left(\!\sqrt{\frac{5}{2}} e \nu \right)+\erf\!\left(\!\sqrt{\frac{5}{2}}p\nu\right)\right]
\\&
+\sqrt{10/\pi}\nu\left(p \mathrm e^{-\frac{5}{2}p^2\nu^2}+e \mathrm e^{-\frac{5}{2}e^2\nu^2} \right)
\Bigg\}
\\&
\!\!\!\!\!\!\!\!\!\!\!\!\!\!\!\!\!\!\!\!\!\!\!\!\!\!\times
\left[\left(5 e^2 \nu ^2-1\right) \erf\!\left(\!\sqrt{\frac{5}{2}} e \nu \right)+\sqrt{\frac{10}{\pi}} e \nu \mathrm e^{-\frac{5}{2}e^2\nu^2} \right]^{-1},
\end{split}
\end{equation}
numerically inverting them to inverse transform sample $e$ and $p$.  This procedure yields a sample of peaks with parameters $\nu$, $x$, $e$, and $p$.  Finally, we use Eqs. (\ref{ec}) and~(\ref{asymp_e}) to convert this sample into a halo sample distributed in the asymptote $A$, and we multiply the distribution by the total number density $n$ to obtain the differential number density $\mathrm{d}n/\mathrm{d}\ln A$.

\begin{figure}[t]
	\centering
	\includegraphics[width=\columnwidth]{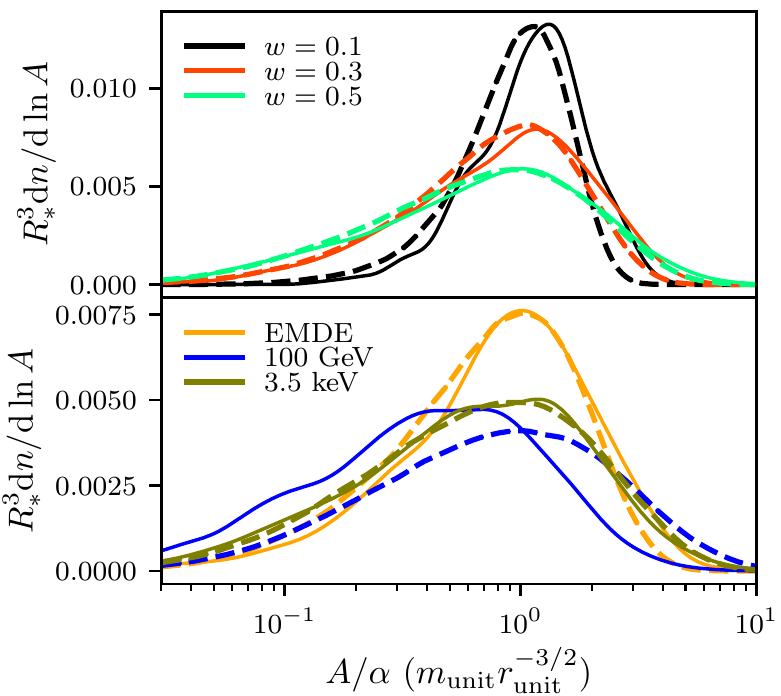}
	\caption{\label{fig:dist_A} A test of the Monte Carlo halo-sampling method of Appendix~\ref{sec:stats}.  This figure plots the peaks drawn from our six power spectra distributed in the predictions for $A$ from Sec.~\ref{sec:ellipsoid} (with simulation-tuned parameter $\alpha$ factored out).  The solid lines show the distributions in the initial density field used for our simulations, while the dashed lines show the distributions computed using the Monte Carlo method.}
\end{figure}

As a test, we sample 400,000 density peaks from each of our six power spectra, and for each peak, we compute its predicted asymptote $A$ (without the proportionality constant) using the ellipsoidal collapse model of Sec.~\ref{sec:ellipsoid}.  We plot the resulting distributions in Fig.~\ref{fig:dist_A} superposed with the distributions we find in the randomly generated simulation boxes.  We find that the two distributions match well for all of our power spectra except for the 100 GeV spectrum.  For this power spectrum, directly sampling the power spectrum yields halos of higher predicted density that we find in the box.  This discrepancy is explained by noting that the simulation box only samples fluctuation modes up to the size of the box.  The 100 GeV power spectrum has sufficient power at larger scales that neglecting it significantly reduces the amplitudes of fluctuations and therefore the density of the resulting halos.  In this way, sampling peaks directly from the power spectrum is more accurate than using the intermediate step of sampling a density field.

\begin{figure}[t]
\centering
\includegraphics[width=\columnwidth]{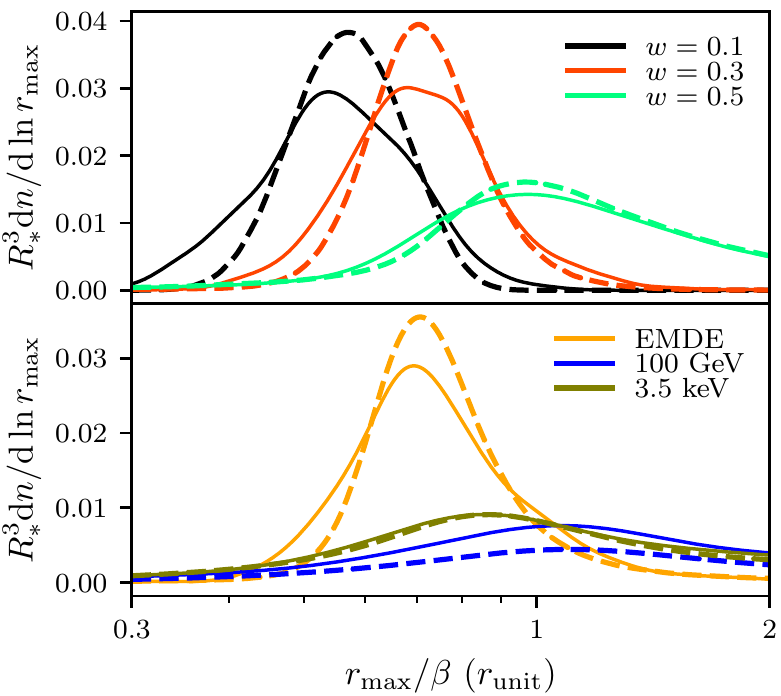}
\caption{\label{fig:dist_rmax} Same as Fig.~\ref{fig:dist_A}, but distributed in the predictions for $r_\mathrm{max}$ from Sec.~\ref{sec:adiabatic} using $s=0$.  Here, the simulation-tuned parameter $\beta$ is factored out.  The solid lines show the distributions in the initial density field used for our simulations, while the dashed lines show the distributions computed using the Monte Carlo method.}
\end{figure}

\begin{figure}[t]
\centering
\includegraphics[width=\columnwidth]{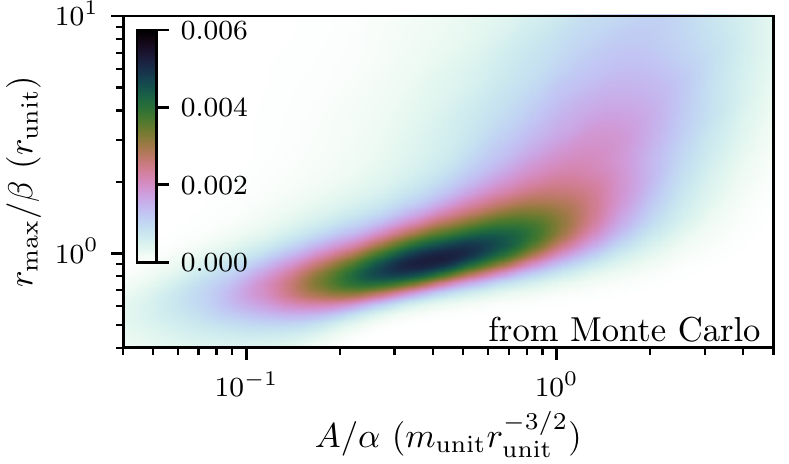}\\
\includegraphics[width=\columnwidth]{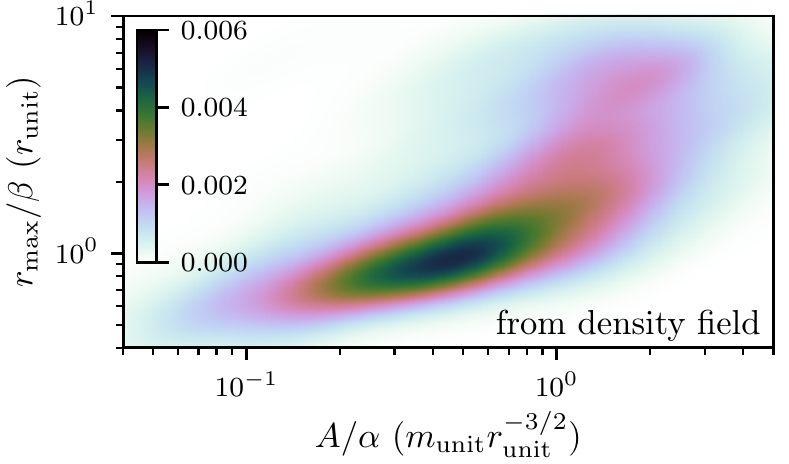}
\caption{\label{fig:dist_2d} The peaks drawn from the 3.5 keV power spectrum distributed in the predictions for $A$ and $r_\mathrm{max}$ from Secs. \ref{sec:ellipsoid} and~\ref{sec:adiabatic}, respectively (with simulation-tuned parameters $\alpha$ and $\beta$ factored out).  Top: The distribution computed using the Monte Carlo method of Appendix~\ref{sec:stats}.  Bottom: The distribution in the initial density field used for our simulation.  This figure shows that the Monte Carlo method accurately reproduces the correlations between $A$ and $r_\mathrm{max}$ found in the density field.  The color scale indicates the differential number density $\mathrm{d}^2n/(\mathrm{d}\ln A\, \mathrm{d}\ln r_\mathrm{max})$ in units of $R_*^{-3}$.}
\end{figure}

\subsection{Outer profile: $r_\mathrm{max}$ and $M(r_\mathrm{max})$}

So far, we have obtained the halo population distributed in the small-radius asymptote $A$.  The next step is to extend this computation to find the multivariate distribution in $A$ and the outer profile parameters $r_\mathrm{max}$ and $M(r_\mathrm{max})$ using the models discussed in Sec.~\ref{sec:rmax}.  This calculation is more difficult because, instead of the finite number of Gaussian variables relevant to the neighborhood of each peak, we must now handle a Gaussian distribution in the infinitely many variables corresponding to the density profile about the peak at each radius $q$.  Nevertheless, a numerical computation is tractable.

The idea is to Monte Carlo sample the full density profile $\delta(q)$ about the peak.  To accomplish this, we must discretize the radial coordinate such that
\begin{equation}\label{r_discrete}
0\leq q_1 < q_2 < ... < q_N
\end{equation}
for some large $N$.  The maximum radius may be initially set at some $q_N\gtrsim R_*$ and raised as needed.  For shorthand, we will write $\delta_i=\delta(q_i)$ and use $\vec\delta$ to represent the full vector of $\delta_i$s.  We now seek the distribution of $\vec\delta$ conditioned on the data we already used to find the asymptote $A$.  In fact, $\vec\delta$ has vanishing covariance with the three-dimensional shape of the peak, so we need only the conditional distribution $f(\vec\delta | \nu,x)$.  This distribution is Gaussian with mean
\begin{equation}\label{delta_mean}
\bar\delta_i=\frac{
	\left(\left\langle\delta_i\nu\right\rangle-\gamma\left\langle\delta_i x\right\rangle\right)\nu
	+\left(\left\langle\delta_i x\right\rangle-\gamma\left\langle\delta_i\nu\right\rangle\right)x
}{1-\gamma^2}
\end{equation}
(recalling $\gamma\equiv \sigma_1^2/[\sigma_0\sigma_2]$) and covariance matrix
\begin{align}\label{delta_cov}
C_{ij}=
\left\langle\delta_i\delta_j\right\rangle
-\frac{1}{1-\gamma^2}\big[&
\left\langle\delta_i\nu\right\rangle\left\langle\delta_j\nu\right\rangle
+\left\langle\delta_i x\right\rangle\left\langle\delta_j x\right\rangle
\nonumber\\
&-\gamma\left(
\left\langle\delta_i\nu\right\rangle\left\langle\delta_j x\right\rangle
+\left\langle\delta_i x\right\rangle\left\langle\delta_j\nu\right\rangle
\right)
\big]
\end{align}
with the necessary covariances given by
\begin{equation}\label{cov}
\left\{
\begin{array} {c}
\sigma_0\left\langle \delta_i\nu \right\rangle \\
\sigma_2\left\langle \delta_i x \right\rangle \\
\left\langle\delta_i\delta_j\right\rangle \\
\end{array}
\right\}
= \int_0^\infty \frac{\mathrm d k}{k} \mathcal{P}(k)
\left\{
\begin{array} {c}
\sinc(k q_i) \\
k^2 \sinc(k q_i) \\
\sinc(k q_i) \sinc(k q_j) \\
\end{array}
\right\}.
\end{equation}


It is helpful to diagonalize $C = P D P^T$ so that $P$ is an orthogonal matrix and $D=\mathrm{diag}(\lambda_1,...,\lambda_N)$ is diagonal.  If we define a new vector ${\vec\kappa \equiv P^T(\vec\delta-\bar{\vec\delta})}$, then $\vec\kappa$ is distributed as
\begin{equation}\label{fdiag}
f(\vec\kappa | \nu,x) = \prod_{i=1}^N \frac{1}{(2\pi\lambda_i)^{1/2}}\exp\!\left(-\frac{\kappa_i^2}{2\lambda_i}\right).
\end{equation}
We then sample each $\kappa_i$ from its respective univariate Gaussian distribution,
and the density profile $\vec\delta$ immediately follows using $\vec\delta=\bar{\vec\delta}+P\vec\kappa$.  From the density profile $\vec\delta\rightarrow\delta(q)$, we can find the mass profile $\Delta(q)$ using Eq.~(\ref{Delta_int}).  With these profiles in hand, it is now a simple matter to apply any of the models detailed in Sec.~\ref{sec:rmax}.

To test this procedure, we sample density profiles for the same 400,000 peaks from each power spectrum for which we already sampled the peak parameters.  Using the $s=0$ contraction model in Sec.~\ref{sec:adiabatic}, Fig.~\ref{fig:dist_rmax} compares the $r_\mathrm{max}$ distributions binned from our initial density fields to those computed using the Monte Carlo method.  As before, the distributions from the 100 GeV power spectrum are discrepant, likely owing again to the finite size of the random density fields.  Also, there is a tendency for the random density fields to produce less sharply peaked distributions than the Monte Carlo method, especially for the $w=0.1$, $w=0.3$, and EMDE power spectra.  It is unclear where this discrepancy arises, but it could be connected to the finite grid resolution of the random density fields.  For a second test, we also plot the combined $A$-$r_\mathrm{max}$ distribution for the 3.5 keV power spectrum in Fig.~\ref{fig:dist_2d}.  The Monte Carlo method proposed in this section accurately reproduces the correlations between $A$ and $r_\mathrm{max}$ found in the density field.

\clearpage

\bibliography{references}

 \newcommand{\noop}[1]{}
\begin{thebibliography}{147}%
\makeatletter
\providecommand \@ifxundefined [1]{%
 \@ifx{#1\undefined}
}%
\providecommand \@ifnum [1]{%
 \ifnum #1\expandafter \@firstoftwo
 \else \expandafter \@secondoftwo
 \fi
}%
\providecommand \@ifx [1]{%
 \ifx #1\expandafter \@firstoftwo
 \else \expandafter \@secondoftwo
 \fi
}%
\providecommand \natexlab [1]{#1}%
\providecommand \enquote  [1]{``#1''}%
\providecommand \bibnamefont  [1]{#1}%
\providecommand \bibfnamefont [1]{#1}%
\providecommand \citenamefont [1]{#1}%
\providecommand \href@noop [0]{\@secondoftwo}%
\providecommand \href [0]{\begingroup \@sanitize@url \@href}%
\providecommand \@href[1]{\@@startlink{#1}\@@href}%
\providecommand \@@href[1]{\endgroup#1\@@endlink}%
\providecommand \@sanitize@url [0]{\catcode `\\12\catcode `\$12\catcode
  `\&12\catcode `\#12\catcode `\^12\catcode `\_12\catcode `\%12\relax}%
\providecommand \@@startlink[1]{}%
\providecommand \@@endlink[0]{}%
\providecommand \url  [0]{\begingroup\@sanitize@url \@url }%
\providecommand \@url [1]{\endgroup\@href {#1}{\urlprefix }}%
\providecommand \urlprefix  [0]{URL }%
\providecommand \Eprint [0]{\href }%
\providecommand \doibase [0]{http://dx.doi.org/}%
\providecommand \selectlanguage [0]{\@gobble}%
\providecommand \bibinfo  [0]{\@secondoftwo}%
\providecommand \bibfield  [0]{\@secondoftwo}%
\providecommand \translation [1]{[#1]}%
\providecommand \BibitemOpen [0]{}%
\providecommand \bibitemStop [0]{}%
\providecommand \bibitemNoStop [0]{.\EOS\space}%
\providecommand \EOS [0]{\spacefactor3000\relax}%
\providecommand \BibitemShut  [1]{\csname bibitem#1\endcsname}%
\let\auto@bib@innerbib\@empty
\bibitem [{\citenamefont {Navarro}\ \emph {et~al.}(1996)\citenamefont
  {Navarro}, \citenamefont {Frenk},\ and\ \citenamefont
  {White}}]{navarro1996structure}%
  \BibitemOpen
  \bibfield  {author} {\bibinfo {author} {\bibfnamefont {J.~F.}\ \bibnamefont
  {Navarro}}, \bibinfo {author} {\bibfnamefont {C.~S.}\ \bibnamefont {Frenk}},
  \ and\ \bibinfo {author} {\bibfnamefont {S.~D.~M.}\ \bibnamefont {White}},\
  }\href@noop {} {\bibfield  {journal} {\bibinfo  {journal} {Astrophys. J.}\
  }\textbf {\bibinfo {volume} {462}},\ \bibinfo {pages} {563} (\bibinfo {year}
  {1996})},\ \Eprint {http://arxiv.org/abs/astro-ph/9508025} {astro-ph/9508025}
  \BibitemShut {NoStop}%
\bibitem [{\citenamefont {Navarro}\ \emph {et~al.}(1997)\citenamefont
  {Navarro}, \citenamefont {Frenk},\ and\ \citenamefont
  {White}}]{navarro1997universal}%
  \BibitemOpen
  \bibfield  {author} {\bibinfo {author} {\bibfnamefont {J.~F.}\ \bibnamefont
  {Navarro}}, \bibinfo {author} {\bibfnamefont {C.~S.}\ \bibnamefont {Frenk}},
  \ and\ \bibinfo {author} {\bibfnamefont {S.~D.~M.}\ \bibnamefont {White}},\
  }\href@noop {} {\bibfield  {journal} {\bibinfo  {journal} {Astrophys. J.}\
  }\textbf {\bibinfo {volume} {490}},\ \bibinfo {pages} {493} (\bibinfo {year}
  {1997})},\ \Eprint {http://arxiv.org/abs/astro-ph/9611107} {astro-ph/9611107}
  \BibitemShut {NoStop}%
\bibitem [{\citenamefont {Navarro}\ \emph {et~al.}(2010)\citenamefont
  {Navarro}, \citenamefont {Ludlow}, \citenamefont {Springel}, \citenamefont
  {Wang}, \citenamefont {Vogelsberger}, \citenamefont {White}, \citenamefont
  {Jenkins}, \citenamefont {Frenk},\ and\ \citenamefont
  {Helmi}}]{navarro2010diversity}%
  \BibitemOpen
  \bibfield  {author} {\bibinfo {author} {\bibfnamefont {J.~F.}\ \bibnamefont
  {Navarro}}, \bibinfo {author} {\bibfnamefont {A.}~\bibnamefont {Ludlow}},
  \bibinfo {author} {\bibfnamefont {V.}~\bibnamefont {Springel}}, \bibinfo
  {author} {\bibfnamefont {J.}~\bibnamefont {Wang}}, \bibinfo {author}
  {\bibfnamefont {M.}~\bibnamefont {Vogelsberger}}, \bibinfo {author}
  {\bibfnamefont {S.~D.~M.}\ \bibnamefont {White}}, \bibinfo {author}
  {\bibfnamefont {A.}~\bibnamefont {Jenkins}}, \bibinfo {author} {\bibfnamefont
  {C.~S.}\ \bibnamefont {Frenk}}, \ and\ \bibinfo {author} {\bibfnamefont
  {A.}~\bibnamefont {Helmi}},\ }\href@noop {} {\bibfield  {journal} {\bibinfo
  {journal} {Mon. Not. R. Astron. Soc.}\ }\textbf {\bibinfo {volume} {402}},\
  \bibinfo {pages} {21} (\bibinfo {year} {2010})}\BibitemShut {NoStop}%
\bibitem [{\citenamefont {Ishiyama}\ \emph {et~al.}(2010)\citenamefont
  {Ishiyama}, \citenamefont {Makino},\ and\ \citenamefont
  {Ebisuzaki}}]{ishiyama2010gamma}%
  \BibitemOpen
  \bibfield  {author} {\bibinfo {author} {\bibfnamefont {T.}~\bibnamefont
  {Ishiyama}}, \bibinfo {author} {\bibfnamefont {J.}~\bibnamefont {Makino}}, \
  and\ \bibinfo {author} {\bibfnamefont {T.}~\bibnamefont {Ebisuzaki}},\
  }\href@noop {} {\bibfield  {journal} {\bibinfo  {journal} {Astrophys. J.
  Lett.}\ }\textbf {\bibinfo {volume} {723}},\ \bibinfo {pages} {L195}
  (\bibinfo {year} {2010})},\ \Eprint {http://arxiv.org/abs/1006.3392}
  {1006.3392} \BibitemShut {NoStop}%
\bibitem [{\citenamefont {Anderhalden}\ and\ \citenamefont
  {Diemand}(2013{\natexlab{a}})}]{anderhalden2013density}%
  \BibitemOpen
  \bibfield  {author} {\bibinfo {author} {\bibfnamefont {D.}~\bibnamefont
  {Anderhalden}}\ and\ \bibinfo {author} {\bibfnamefont {J.}~\bibnamefont
  {Diemand}},\ }\href@noop {} {\bibfield  {journal} {\bibinfo  {journal} {J.
  Cosmol. Astropart. Phys.}\ }\textbf {\bibinfo {volume} {04}},\ \bibinfo
  {pages} {009} (\bibinfo {year} {2013}{\natexlab{a}})},\ \Eprint
  {http://arxiv.org/abs/1302.0003} {1302.0003} \BibitemShut {NoStop}%
\bibitem [{\citenamefont {Anderhalden}\ and\ \citenamefont
  {Diemand}(2013{\natexlab{b}})}]{anderhalden2013erratum}%
  \BibitemOpen
  \bibfield  {author} {\bibinfo {author} {\bibfnamefont {D.}~\bibnamefont
  {Anderhalden}}\ and\ \bibinfo {author} {\bibfnamefont {J.}~\bibnamefont
  {Diemand}},\ }\href@noop {} {\bibfield  {journal} {\bibinfo  {journal} {J.
  Cosmol. Astropart. Phys.}\ }\textbf {\bibinfo {volume} {08}},\ \bibinfo
  {pages} {E02} (\bibinfo {year} {2013}{\natexlab{b}})}\BibitemShut {NoStop}%
\bibitem [{\citenamefont {Ishiyama}(2014)}]{ishiyama2014hierarchical}%
  \BibitemOpen
  \bibfield  {author} {\bibinfo {author} {\bibfnamefont {T.}~\bibnamefont
  {Ishiyama}},\ }\href@noop {} {\bibfield  {journal} {\bibinfo  {journal}
  {Astrophys. J.}\ }\textbf {\bibinfo {volume} {788}},\ \bibinfo {pages} {27}
  (\bibinfo {year} {2014})},\ \Eprint {http://arxiv.org/abs/1404.1650}
  {1404.1650} \BibitemShut {NoStop}%
\bibitem [{\citenamefont {Polisensky}\ and\ \citenamefont
  {Ricotti}(2015)}]{polisensky2015fingerprints}%
  \BibitemOpen
  \bibfield  {author} {\bibinfo {author} {\bibfnamefont {E.}~\bibnamefont
  {Polisensky}}\ and\ \bibinfo {author} {\bibfnamefont {M.}~\bibnamefont
  {Ricotti}},\ }\href@noop {} {\bibfield  {journal} {\bibinfo  {journal} {Mon.
  Not. R. Astron. Soc.}\ }\textbf {\bibinfo {volume} {450}},\ \bibinfo {pages}
  {2172} (\bibinfo {year} {2015})},\ \Eprint {http://arxiv.org/abs/1504.02126}
  {1504.02126} \BibitemShut {NoStop}%
\bibitem [{\citenamefont {Ogiya}\ and\ \citenamefont
  {Hahn}(2018)}]{ogiya2017sets}%
  \BibitemOpen
  \bibfield  {author} {\bibinfo {author} {\bibfnamefont {G.}~\bibnamefont
  {Ogiya}}\ and\ \bibinfo {author} {\bibfnamefont {O.}~\bibnamefont {Hahn}},\
  }\href@noop {} {\bibfield  {journal} {\bibinfo  {journal} {Mon. Not. R.
  Astron. Soc.}\ }\textbf {\bibinfo {volume} {473}},\ \bibinfo {pages} {4339}
  (\bibinfo {year} {2018})},\ \Eprint {http://arxiv.org/abs/1707.07693}
  {1707.07693} \BibitemShut {NoStop}%
\bibitem [{\citenamefont {Angulo}\ \emph {et~al.}(2017)\citenamefont {Angulo},
  \citenamefont {Hahn}, \citenamefont {Ludlow},\ and\ \citenamefont
  {Bonoli}}]{angulo2017earth}%
  \BibitemOpen
  \bibfield  {author} {\bibinfo {author} {\bibfnamefont {R.~E.}\ \bibnamefont
  {Angulo}}, \bibinfo {author} {\bibfnamefont {O.}~\bibnamefont {Hahn}},
  \bibinfo {author} {\bibfnamefont {A.~D.}\ \bibnamefont {Ludlow}}, \ and\
  \bibinfo {author} {\bibfnamefont {S.}~\bibnamefont {Bonoli}},\ }\href@noop {}
  {\bibfield  {journal} {\bibinfo  {journal} {Mon. Not. R. Astron. Soc.}\
  }\textbf {\bibinfo {volume} {471}},\ \bibinfo {pages} {4687} (\bibinfo {year}
  {2017})},\ \Eprint {http://arxiv.org/abs/1604.03131} {1604.03131}
  \BibitemShut {NoStop}%
\bibitem [{\citenamefont {Ogiya}\ \emph {et~al.}(2016)\citenamefont {Ogiya},
  \citenamefont {Nagai},\ and\ \citenamefont {Ishiyama}}]{ogiya2016dynamical}%
  \BibitemOpen
  \bibfield  {author} {\bibinfo {author} {\bibfnamefont {G.}~\bibnamefont
  {Ogiya}}, \bibinfo {author} {\bibfnamefont {D.}~\bibnamefont {Nagai}}, \ and\
  \bibinfo {author} {\bibfnamefont {T.}~\bibnamefont {Ishiyama}},\ }\href@noop
  {} {\bibfield  {journal} {\bibinfo  {journal} {Mon. Not. R. Astron. Soc.}\
  }\textbf {\bibinfo {volume} {461}},\ \bibinfo {pages} {3385} (\bibinfo {year}
  {2016})},\ \Eprint {http://arxiv.org/abs/1604.02866} {1604.02866}
  \BibitemShut {NoStop}%
\bibitem [{\citenamefont {Lynden-Bell}(1967)}]{lynden1967statistical}%
  \BibitemOpen
  \bibfield  {author} {\bibinfo {author} {\bibfnamefont {D.}~\bibnamefont
  {Lynden-Bell}},\ }\href@noop {} {\bibfield  {journal} {\bibinfo  {journal}
  {Mon. Not. R. Astron. Soc.}\ }\textbf {\bibinfo {volume} {136}},\ \bibinfo
  {pages} {101} (\bibinfo {year} {1967})}\BibitemShut {NoStop}%
\bibitem [{\citenamefont {Drakos}\ \emph {et~al.}()\citenamefont {Drakos},
  \citenamefont {Taylor}, \citenamefont {Berrouet}, \citenamefont {Robotham},\
  and\ \citenamefont {Power}}]{drakos2018major}%
  \BibitemOpen
  \bibfield  {author} {\bibinfo {author} {\bibfnamefont {N.~E.}\ \bibnamefont
  {Drakos}}, \bibinfo {author} {\bibfnamefont {J.~E.}\ \bibnamefont {Taylor}},
  \bibinfo {author} {\bibfnamefont {A.}~\bibnamefont {Berrouet}}, \bibinfo
  {author} {\bibfnamefont {A.~S.}\ \bibnamefont {Robotham}}, \ and\ \bibinfo
  {author} {\bibfnamefont {C.}~\bibnamefont {Power}},\ }\href@noop {} {\
  }\Eprint {http://arxiv.org/abs/1811.12844} {1811.12844} \BibitemShut
  {NoStop}%
\bibitem [{\citenamefont {Berezinsky}\ \emph {et~al.}(2008)\citenamefont
  {Berezinsky}, \citenamefont {Dokuchaev},\ and\ \citenamefont
  {Eroshenko}}]{berezinsky2008remnants}%
  \BibitemOpen
  \bibfield  {author} {\bibinfo {author} {\bibfnamefont {V.}~\bibnamefont
  {Berezinsky}}, \bibinfo {author} {\bibfnamefont {V.}~\bibnamefont
  {Dokuchaev}}, \ and\ \bibinfo {author} {\bibfnamefont {Y.}~\bibnamefont
  {Eroshenko}},\ }\href@noop {} {\bibfield  {journal} {\bibinfo  {journal}
  {Phys. Rev. D}\ }\textbf {\bibinfo {volume} {77}},\ \bibinfo {pages} {083519}
  (\bibinfo {year} {2008})},\ \Eprint {http://arxiv.org/abs/0712.3499}
  {0712.3499} \BibitemShut {NoStop}%
\bibitem [{\citenamefont {Hayashi}\ \emph {et~al.}(2003)\citenamefont
  {Hayashi}, \citenamefont {Navarro}, \citenamefont {Taylor}, \citenamefont
  {Stadel},\ and\ \citenamefont {Quinn}}]{hayashi2003structural}%
  \BibitemOpen
  \bibfield  {author} {\bibinfo {author} {\bibfnamefont {E.}~\bibnamefont
  {Hayashi}}, \bibinfo {author} {\bibfnamefont {J.~F.}\ \bibnamefont
  {Navarro}}, \bibinfo {author} {\bibfnamefont {J.~E.}\ \bibnamefont {Taylor}},
  \bibinfo {author} {\bibfnamefont {J.}~\bibnamefont {Stadel}}, \ and\ \bibinfo
  {author} {\bibfnamefont {T.}~\bibnamefont {Quinn}},\ }\href@noop {}
  {\bibfield  {journal} {\bibinfo  {journal} {Astrophys. J.}\ }\textbf
  {\bibinfo {volume} {584}},\ \bibinfo {pages} {541} (\bibinfo {year}
  {2003})},\ \Eprint {http://arxiv.org/abs/astro-ph/0203004} {astro-ph/0203004}
  \BibitemShut {NoStop}%
\bibitem [{\citenamefont {Penarrubia}\ \emph {et~al.}(2010)\citenamefont
  {Penarrubia}, \citenamefont {Benson}, \citenamefont {Walker}, \citenamefont
  {Gilmore}, \citenamefont {McConnachie},\ and\ \citenamefont
  {Mayer}}]{penarrubia2010impact}%
  \BibitemOpen
  \bibfield  {author} {\bibinfo {author} {\bibfnamefont {J.}~\bibnamefont
  {Penarrubia}}, \bibinfo {author} {\bibfnamefont {A.~J.}\ \bibnamefont
  {Benson}}, \bibinfo {author} {\bibfnamefont {M.~G.}\ \bibnamefont {Walker}},
  \bibinfo {author} {\bibfnamefont {G.}~\bibnamefont {Gilmore}}, \bibinfo
  {author} {\bibfnamefont {A.~W.}\ \bibnamefont {McConnachie}}, \ and\ \bibinfo
  {author} {\bibfnamefont {L.}~\bibnamefont {Mayer}},\ }\href@noop {}
  {\bibfield  {journal} {\bibinfo  {journal} {Mon. Not. R. Astron. Soc.}\
  }\textbf {\bibinfo {volume} {406}},\ \bibinfo {pages} {1290} (\bibinfo {year}
  {2010})},\ \Eprint {http://arxiv.org/abs/1002.3376} {1002.3376} \BibitemShut
  {NoStop}%
\bibitem [{\citenamefont {Ogiya}\ \emph {et~al.}(2019)\citenamefont {Ogiya},
  \citenamefont {Van~den Bosch}, \citenamefont {Hahn}, \citenamefont {Green},
  \citenamefont {Miller},\ and\ \citenamefont {Burkert}}]{ogiya2019dash}%
  \BibitemOpen
  \bibfield  {author} {\bibinfo {author} {\bibfnamefont {G.}~\bibnamefont
  {Ogiya}}, \bibinfo {author} {\bibfnamefont {F.~C.}\ \bibnamefont {Van~den
  Bosch}}, \bibinfo {author} {\bibfnamefont {O.}~\bibnamefont {Hahn}}, \bibinfo
  {author} {\bibfnamefont {S.~B.}\ \bibnamefont {Green}}, \bibinfo {author}
  {\bibfnamefont {T.~B.}\ \bibnamefont {Miller}}, \ and\ \bibinfo {author}
  {\bibfnamefont {A.}~\bibnamefont {Burkert}},\ }\href@noop {} {\bibfield
  {journal} {\bibinfo  {journal} {Mon. Not. R. Astron. Soc.}\ }\textbf
  {\bibinfo {volume} {485}},\ \bibinfo {pages} {189} (\bibinfo {year}
  {2019})}\BibitemShut {NoStop}%
\bibitem [{\citenamefont {Diemand}\ \emph {et~al.}(2007)\citenamefont
  {Diemand}, \citenamefont {Kuhlen},\ and\ \citenamefont
  {Madau}}]{diemand2007dark}%
  \BibitemOpen
  \bibfield  {author} {\bibinfo {author} {\bibfnamefont {J.}~\bibnamefont
  {Diemand}}, \bibinfo {author} {\bibfnamefont {M.}~\bibnamefont {Kuhlen}}, \
  and\ \bibinfo {author} {\bibfnamefont {P.}~\bibnamefont {Madau}},\
  }\href@noop {} {\bibfield  {journal} {\bibinfo  {journal} {Astrophys. J.}\
  }\textbf {\bibinfo {volume} {657}},\ \bibinfo {pages} {262} (\bibinfo {year}
  {2007})},\ \Eprint {http://arxiv.org/abs/1507.08656} {1507.08656}
  \BibitemShut {NoStop}%
\bibitem [{\citenamefont {Strigari}\ \emph {et~al.}(2007)\citenamefont
  {Strigari}, \citenamefont {Koushiappas}, \citenamefont {Bullock},\ and\
  \citenamefont {Kaplinghat}}]{strigari2007precise}%
  \BibitemOpen
  \bibfield  {author} {\bibinfo {author} {\bibfnamefont {L.~E.}\ \bibnamefont
  {Strigari}}, \bibinfo {author} {\bibfnamefont {S.~M.}\ \bibnamefont
  {Koushiappas}}, \bibinfo {author} {\bibfnamefont {J.~S.}\ \bibnamefont
  {Bullock}}, \ and\ \bibinfo {author} {\bibfnamefont {M.}~\bibnamefont
  {Kaplinghat}},\ }\href@noop {} {\bibfield  {journal} {\bibinfo  {journal}
  {Phys. Rev. D}\ }\textbf {\bibinfo {volume} {75}},\ \bibinfo {pages} {083526}
  (\bibinfo {year} {2007})},\ \Eprint {http://arxiv.org/abs/astro-ph/0611925}
  {astro-ph/0611925} \BibitemShut {NoStop}%
\bibitem [{\citenamefont {Pieri}\ \emph {et~al.}(2008)\citenamefont {Pieri},
  \citenamefont {Bertone},\ and\ \citenamefont {Branchini}}]{pieri2008dark}%
  \BibitemOpen
  \bibfield  {author} {\bibinfo {author} {\bibfnamefont {L.}~\bibnamefont
  {Pieri}}, \bibinfo {author} {\bibfnamefont {G.}~\bibnamefont {Bertone}}, \
  and\ \bibinfo {author} {\bibfnamefont {E.}~\bibnamefont {Branchini}},\
  }\href@noop {} {\bibfield  {journal} {\bibinfo  {journal} {Mon. Not. R.
  Astron. Soc.}\ }\textbf {\bibinfo {volume} {384}},\ \bibinfo {pages} {1627}
  (\bibinfo {year} {2008})},\ \Eprint {http://arxiv.org/abs/0706.2101}
  {0706.2101} \BibitemShut {NoStop}%
\bibitem [{\citenamefont {Jeltema}\ and\ \citenamefont
  {Profumo}(2008)}]{jeltema2008searching}%
  \BibitemOpen
  \bibfield  {author} {\bibinfo {author} {\bibfnamefont {T.~E.}\ \bibnamefont
  {Jeltema}}\ and\ \bibinfo {author} {\bibfnamefont {S.}~\bibnamefont
  {Profumo}},\ }\href@noop {} {\bibfield  {journal} {\bibinfo  {journal}
  {Astrophys. J.}\ }\textbf {\bibinfo {volume} {686}},\ \bibinfo {pages} {1045}
  (\bibinfo {year} {2008})},\ \Eprint {http://arxiv.org/abs/0805.1054}
  {0805.1054} \BibitemShut {NoStop}%
\bibitem [{\citenamefont {Bartels}\ and\ \citenamefont
  {Ando}(2015)}]{bartels2015boosting}%
  \BibitemOpen
  \bibfield  {author} {\bibinfo {author} {\bibfnamefont {R.}~\bibnamefont
  {Bartels}}\ and\ \bibinfo {author} {\bibfnamefont {S.}~\bibnamefont {Ando}},\
  }\href@noop {} {\bibfield  {journal} {\bibinfo  {journal} {Phys. Rev. D}\
  }\textbf {\bibinfo {volume} {92}},\ \bibinfo {pages} {123508} (\bibinfo
  {year} {2015})},\ \Eprint {http://arxiv.org/abs/1507.08656} {1507.08656}
  \BibitemShut {NoStop}%
\bibitem [{\citenamefont {Ando}\ \emph {et~al.}()\citenamefont {Ando},
  \citenamefont {Ishiyama},\ and\ \citenamefont {Hiroshima}}]{ando2019halo}%
  \BibitemOpen
  \bibfield  {author} {\bibinfo {author} {\bibfnamefont {S.}~\bibnamefont
  {Ando}}, \bibinfo {author} {\bibfnamefont {T.}~\bibnamefont {Ishiyama}}, \
  and\ \bibinfo {author} {\bibfnamefont {N.}~\bibnamefont {Hiroshima}},\
  }\href@noop {} {\ }\Eprint {http://arxiv.org/abs/1903.11427} {1903.11427}
  \BibitemShut {NoStop}%
\bibitem [{\citenamefont {Boden}\ \emph {et~al.}(1998)\citenamefont {Boden},
  \citenamefont {Shao},\ and\ \citenamefont
  {Van~Buren}}]{boden1998astrometric}%
  \BibitemOpen
  \bibfield  {author} {\bibinfo {author} {\bibfnamefont {A.}~\bibnamefont
  {Boden}}, \bibinfo {author} {\bibfnamefont {M.}~\bibnamefont {Shao}}, \ and\
  \bibinfo {author} {\bibfnamefont {D.}~\bibnamefont {Van~Buren}},\ }\href@noop
  {} {\bibfield  {journal} {\bibinfo  {journal} {Astrophys. J.}\ }\textbf
  {\bibinfo {volume} {502}},\ \bibinfo {pages} {538} (\bibinfo {year}
  {1998})},\ \Eprint {http://arxiv.org/abs/astro-ph/9802179} {astro-ph/9802179}
  \BibitemShut {NoStop}%
\bibitem [{\citenamefont {Dominik}\ and\ \citenamefont
  {Sahu}(2000)}]{dominik2000astrometric}%
  \BibitemOpen
  \bibfield  {author} {\bibinfo {author} {\bibfnamefont {M.}~\bibnamefont
  {Dominik}}\ and\ \bibinfo {author} {\bibfnamefont {K.~C.}\ \bibnamefont
  {Sahu}},\ }\href@noop {} {\bibfield  {journal} {\bibinfo  {journal}
  {Astrophys. J.}\ }\textbf {\bibinfo {volume} {534}},\ \bibinfo {pages} {213}
  (\bibinfo {year} {2000})},\ \Eprint {http://arxiv.org/abs/astro-ph/9805360}
  {astro-ph/9805360} \BibitemShut {NoStop}%
\bibitem [{\citenamefont {Erickcek}\ and\ \citenamefont
  {Law}(2011)}]{erickcek2011astrometric}%
  \BibitemOpen
  \bibfield  {author} {\bibinfo {author} {\bibfnamefont {A.~L.}\ \bibnamefont
  {Erickcek}}\ and\ \bibinfo {author} {\bibfnamefont {N.~M.}\ \bibnamefont
  {Law}},\ }\href@noop {} {\bibfield  {journal} {\bibinfo  {journal}
  {Astrophys. J.}\ }\textbf {\bibinfo {volume} {729}},\ \bibinfo {pages} {49}
  (\bibinfo {year} {2011})},\ \Eprint {http://arxiv.org/abs/1007.4228}
  {1007.4228} \BibitemShut {NoStop}%
\bibitem [{\citenamefont {Van~Tilburg}\ \emph {et~al.}(2018)\citenamefont
  {Van~Tilburg}, \citenamefont {Taki},\ and\ \citenamefont
  {Weiner}}]{van2018halometry}%
  \BibitemOpen
  \bibfield  {author} {\bibinfo {author} {\bibfnamefont {K.}~\bibnamefont
  {Van~Tilburg}}, \bibinfo {author} {\bibfnamefont {A.-M.}\ \bibnamefont
  {Taki}}, \ and\ \bibinfo {author} {\bibfnamefont {N.}~\bibnamefont
  {Weiner}},\ }\href@noop {} {\bibfield  {journal} {\bibinfo  {journal} {J.
  Cosmol. Astropart. Phys.}\ }\textbf {\bibinfo {volume} {2018}},\ \bibinfo
  {pages} {041} (\bibinfo {year} {2018})},\ \Eprint
  {http://arxiv.org/abs/1804.01991} {1804.01991} \BibitemShut {NoStop}%
\bibitem [{\citenamefont {Siegel}\ \emph {et~al.}(2007)\citenamefont {Siegel},
  \citenamefont {Hertzberg},\ and\ \citenamefont {Fry}}]{siegel2007probing}%
  \BibitemOpen
  \bibfield  {author} {\bibinfo {author} {\bibfnamefont {E.~R.}\ \bibnamefont
  {Siegel}}, \bibinfo {author} {\bibfnamefont {M.}~\bibnamefont {Hertzberg}}, \
  and\ \bibinfo {author} {\bibfnamefont {J.}~\bibnamefont {Fry}},\ }\href@noop
  {} {\bibfield  {journal} {\bibinfo  {journal} {Mon. Not. R. Astron. Soc.}\
  }\textbf {\bibinfo {volume} {382}},\ \bibinfo {pages} {879} (\bibinfo {year}
  {2007})},\ \Eprint {http://arxiv.org/abs/astro-ph/0702546} {astro-ph/0702546}
  \BibitemShut {NoStop}%
\bibitem [{\citenamefont {Erkal}\ and\ \citenamefont
  {Belokurov}(2015)}]{erkal2015forensics}%
  \BibitemOpen
  \bibfield  {author} {\bibinfo {author} {\bibfnamefont {D.}~\bibnamefont
  {Erkal}}\ and\ \bibinfo {author} {\bibfnamefont {V.}~\bibnamefont
  {Belokurov}},\ }\href@noop {} {\bibfield  {journal} {\bibinfo  {journal}
  {Mon. Not. R. Astron. Soc.}\ }\textbf {\bibinfo {volume} {450}},\ \bibinfo
  {pages} {1136} (\bibinfo {year} {2015})},\ \Eprint
  {http://arxiv.org/abs/1412.6035} {1412.6035} \BibitemShut {NoStop}%
\bibitem [{\citenamefont {Erkal}\ \emph {et~al.}(2016)\citenamefont {Erkal},
  \citenamefont {Belokurov}, \citenamefont {Bovy},\ and\ \citenamefont
  {Sanders}}]{erkal2016number}%
  \BibitemOpen
  \bibfield  {author} {\bibinfo {author} {\bibfnamefont {D.}~\bibnamefont
  {Erkal}}, \bibinfo {author} {\bibfnamefont {V.}~\bibnamefont {Belokurov}},
  \bibinfo {author} {\bibfnamefont {J.}~\bibnamefont {Bovy}}, \ and\ \bibinfo
  {author} {\bibfnamefont {J.~L.}\ \bibnamefont {Sanders}},\ }\href@noop {}
  {\bibfield  {journal} {\bibinfo  {journal} {Mon. Not. R. Astron. Soc.}\
  }\textbf {\bibinfo {volume} {463}},\ \bibinfo {pages} {102} (\bibinfo {year}
  {2016})},\ \Eprint {http://arxiv.org/abs/1606.04946} {1606.04946}
  \BibitemShut {NoStop}%
\bibitem [{\citenamefont {Buschmann}\ \emph {et~al.}(2018)\citenamefont
  {Buschmann}, \citenamefont {Kopp}, \citenamefont {Safdi},\ and\ \citenamefont
  {Wu}}]{buschmann2018stellar}%
  \BibitemOpen
  \bibfield  {author} {\bibinfo {author} {\bibfnamefont {M.}~\bibnamefont
  {Buschmann}}, \bibinfo {author} {\bibfnamefont {J.}~\bibnamefont {Kopp}},
  \bibinfo {author} {\bibfnamefont {B.~R.}\ \bibnamefont {Safdi}}, \ and\
  \bibinfo {author} {\bibfnamefont {C.-L.}\ \bibnamefont {Wu}},\ }\href@noop {}
  {\bibfield  {journal} {\bibinfo  {journal} {Phys. Rev. Lett.}\ }\textbf
  {\bibinfo {volume} {120}},\ \bibinfo {pages} {211101} (\bibinfo {year}
  {2018})},\ \Eprint {http://arxiv.org/abs/1711.03554} {1711.03554}
  \BibitemShut {NoStop}%
\bibitem [{\citenamefont {Hlozek}\ \emph {et~al.}(2012)\citenamefont {Hlozek}
  \emph {et~al.}}]{hlozek2012atacama}%
  \BibitemOpen
  \bibfield  {author} {\bibinfo {author} {\bibfnamefont {R.}~\bibnamefont
  {Hlozek}} \emph {et~al.},\ }\href@noop {} {\bibfield  {journal} {\bibinfo
  {journal} {Astrophys. J.}\ }\textbf {\bibinfo {volume} {749}},\ \bibinfo
  {pages} {90} (\bibinfo {year} {2012})},\ \Eprint
  {http://arxiv.org/abs/1105.4887} {1105.4887} \BibitemShut {NoStop}%
\bibitem [{\citenamefont {Bird}\ \emph {et~al.}(2011)\citenamefont {Bird},
  \citenamefont {Peiris}, \citenamefont {Viel},\ and\ \citenamefont
  {Verde}}]{bird2011minimally}%
  \BibitemOpen
  \bibfield  {author} {\bibinfo {author} {\bibfnamefont {S.}~\bibnamefont
  {Bird}}, \bibinfo {author} {\bibfnamefont {H.~V.}\ \bibnamefont {Peiris}},
  \bibinfo {author} {\bibfnamefont {M.}~\bibnamefont {Viel}}, \ and\ \bibinfo
  {author} {\bibfnamefont {L.}~\bibnamefont {Verde}},\ }\href@noop {}
  {\bibfield  {journal} {\bibinfo  {journal} {Mon. Not. R. Astron. Soc.}\
  }\textbf {\bibinfo {volume} {413}},\ \bibinfo {pages} {1717} (\bibinfo {year}
  {2011})},\ \Eprint {http://arxiv.org/abs/1010.1519} {1010.1519} \BibitemShut
  {NoStop}%
\bibitem [{\citenamefont {Josan}\ \emph {et~al.}(2009)\citenamefont {Josan},
  \citenamefont {Green},\ and\ \citenamefont {Malik}}]{josan2009generalized}%
  \BibitemOpen
  \bibfield  {author} {\bibinfo {author} {\bibfnamefont {A.~S.}\ \bibnamefont
  {Josan}}, \bibinfo {author} {\bibfnamefont {A.~M.}\ \bibnamefont {Green}}, \
  and\ \bibinfo {author} {\bibfnamefont {K.~A.}\ \bibnamefont {Malik}},\
  }\href@noop {} {\bibfield  {journal} {\bibinfo  {journal} {Phys. Rev. D}\
  }\textbf {\bibinfo {volume} {79}},\ \bibinfo {pages} {103520} (\bibinfo
  {year} {2009})},\ \Eprint {http://arxiv.org/abs/0903.3184} {0903.3184}
  \BibitemShut {NoStop}%
\bibitem [{\citenamefont {Chluba}\ \emph {et~al.}(2012)\citenamefont {Chluba},
  \citenamefont {Erickcek},\ and\ \citenamefont
  {Ben-Dayan}}]{chluba2012probing}%
  \BibitemOpen
  \bibfield  {author} {\bibinfo {author} {\bibfnamefont {J.}~\bibnamefont
  {Chluba}}, \bibinfo {author} {\bibfnamefont {A.~L.}\ \bibnamefont
  {Erickcek}}, \ and\ \bibinfo {author} {\bibfnamefont {I.}~\bibnamefont
  {Ben-Dayan}},\ }\href@noop {} {\bibfield  {journal} {\bibinfo  {journal}
  {Astrophys. J.}\ }\textbf {\bibinfo {volume} {758}},\ \bibinfo {pages} {76}
  (\bibinfo {year} {2012})},\ \Eprint {http://arxiv.org/abs/1203.2681}
  {1203.2681} \BibitemShut {NoStop}%
\bibitem [{\citenamefont {Lidsey}\ \emph {et~al.}(1997)\citenamefont {Lidsey},
  \citenamefont {Liddle}, \citenamefont {Kolb}, \citenamefont {Copeland},
  \citenamefont {Barreiro},\ and\ \citenamefont
  {Abney}}]{lidsey1997reconstructing}%
  \BibitemOpen
  \bibfield  {author} {\bibinfo {author} {\bibfnamefont {J.~E.}\ \bibnamefont
  {Lidsey}}, \bibinfo {author} {\bibfnamefont {A.~R.}\ \bibnamefont {Liddle}},
  \bibinfo {author} {\bibfnamefont {E.~W.}\ \bibnamefont {Kolb}}, \bibinfo
  {author} {\bibfnamefont {E.~J.}\ \bibnamefont {Copeland}}, \bibinfo {author}
  {\bibfnamefont {T.}~\bibnamefont {Barreiro}}, \ and\ \bibinfo {author}
  {\bibfnamefont {M.}~\bibnamefont {Abney}},\ }\href@noop {} {\bibfield
  {journal} {\bibinfo  {journal} {Rev. Mod. Phys.}\ }\textbf {\bibinfo {volume}
  {69}},\ \bibinfo {pages} {373} (\bibinfo {year} {1997})},\ \Eprint
  {http://arxiv.org/abs/astro-ph/9508078} {astro-ph/9508078} \BibitemShut
  {NoStop}%
\bibitem [{\citenamefont {Erickcek}\ and\ \citenamefont
  {Sigurdson}(2011)}]{erickcek2011reheating}%
  \BibitemOpen
  \bibfield  {author} {\bibinfo {author} {\bibfnamefont {A.~L.}\ \bibnamefont
  {Erickcek}}\ and\ \bibinfo {author} {\bibfnamefont {K.}~\bibnamefont
  {Sigurdson}},\ }\href@noop {} {\bibfield  {journal} {\bibinfo  {journal}
  {Phys. Rev. D}\ }\textbf {\bibinfo {volume} {84}},\ \bibinfo {pages} {083503}
  (\bibinfo {year} {2011})},\ \Eprint {http://arxiv.org/abs/1106.0536}
  {1106.0536} \BibitemShut {NoStop}%
\bibitem [{\citenamefont {Barenboim}\ and\ \citenamefont
  {Rasero}(2014)}]{barenboim2014structure}%
  \BibitemOpen
  \bibfield  {author} {\bibinfo {author} {\bibfnamefont {G.}~\bibnamefont
  {Barenboim}}\ and\ \bibinfo {author} {\bibfnamefont {J.}~\bibnamefont
  {Rasero}},\ }\href@noop {} {\bibfield  {journal} {\bibinfo  {journal} {J.
  High Energy Phys.}\ }\textbf {\bibinfo {volume} {2014}},\ \bibinfo {pages}
  {138} (\bibinfo {year} {2014})},\ \Eprint {http://arxiv.org/abs/1311.4034}
  {1311.4034} \BibitemShut {NoStop}%
\bibitem [{\citenamefont {Fan}\ \emph {et~al.}(2014)\citenamefont {Fan},
  \citenamefont {{\"O}zsoy},\ and\ \citenamefont {Watson}}]{fan2014nonthermal}%
  \BibitemOpen
  \bibfield  {author} {\bibinfo {author} {\bibfnamefont {J.~J.}\ \bibnamefont
  {Fan}}, \bibinfo {author} {\bibfnamefont {O.}~\bibnamefont {{\"O}zsoy}}, \
  and\ \bibinfo {author} {\bibfnamefont {S.}~\bibnamefont {Watson}},\
  }\href@noop {} {\bibfield  {journal} {\bibinfo  {journal} {Phys. Rev. D}\
  }\textbf {\bibinfo {volume} {90}},\ \bibinfo {pages} {043536} (\bibinfo
  {year} {2014})},\ \Eprint {http://arxiv.org/abs/1405.7373} {1405.7373}
  \BibitemShut {NoStop}%
\bibitem [{\citenamefont {Erickcek}(2015)}]{erickcek2015dark}%
  \BibitemOpen
  \bibfield  {author} {\bibinfo {author} {\bibfnamefont {A.~L.}\ \bibnamefont
  {Erickcek}},\ }\href@noop {} {\bibfield  {journal} {\bibinfo  {journal}
  {Phys. Rev. D}\ }\textbf {\bibinfo {volume} {92}},\ \bibinfo {pages} {103505}
  (\bibinfo {year} {2015})},\ \Eprint {http://arxiv.org/abs/1504.03335}
  {1504.03335} \BibitemShut {NoStop}%
\bibitem [{\citenamefont {Redmond}\ \emph {et~al.}(2018)\citenamefont
  {Redmond}, \citenamefont {Trezza},\ and\ \citenamefont
  {Erickcek}}]{redmond2018growth}%
  \BibitemOpen
  \bibfield  {author} {\bibinfo {author} {\bibfnamefont {K.}~\bibnamefont
  {Redmond}}, \bibinfo {author} {\bibfnamefont {A.}~\bibnamefont {Trezza}}, \
  and\ \bibinfo {author} {\bibfnamefont {A.~L.}\ \bibnamefont {Erickcek}},\
  }\href@noop {} {\bibfield  {journal} {\bibinfo  {journal} {Phys. Rev. D}\
  }\textbf {\bibinfo {volume} {98}},\ \bibinfo {pages} {063504} (\bibinfo
  {year} {2018})},\ \Eprint {http://arxiv.org/abs/1807.01327} {1807.01327}
  \BibitemShut {NoStop}%
\bibitem [{\citenamefont {Josan}\ and\ \citenamefont
  {Green}(2010)}]{josan2010gamma}%
  \BibitemOpen
  \bibfield  {author} {\bibinfo {author} {\bibfnamefont {A.~S.}\ \bibnamefont
  {Josan}}\ and\ \bibinfo {author} {\bibfnamefont {A.~M.}\ \bibnamefont
  {Green}},\ }\href@noop {} {\bibfield  {journal} {\bibinfo  {journal} {Phys.
  Rev. D}\ }\textbf {\bibinfo {volume} {82}},\ \bibinfo {pages} {083527}
  (\bibinfo {year} {2010})},\ \Eprint {http://arxiv.org/abs/1006.4970}
  {1006.4970} \BibitemShut {NoStop}%
\bibitem [{\citenamefont {Bringmann}\ \emph {et~al.}(2012)\citenamefont
  {Bringmann}, \citenamefont {Scott},\ and\ \citenamefont
  {Akrami}}]{bringmann2012improved}%
  \BibitemOpen
  \bibfield  {author} {\bibinfo {author} {\bibfnamefont {T.}~\bibnamefont
  {Bringmann}}, \bibinfo {author} {\bibfnamefont {P.}~\bibnamefont {Scott}}, \
  and\ \bibinfo {author} {\bibfnamefont {Y.}~\bibnamefont {Akrami}},\
  }\href@noop {} {\bibfield  {journal} {\bibinfo  {journal} {Phys. Rev. D}\
  }\textbf {\bibinfo {volume} {85}},\ \bibinfo {pages} {125027} (\bibinfo
  {year} {2012})},\ \Eprint {http://arxiv.org/abs/1110.2484} {1110.2484}
  \BibitemShut {NoStop}%
\bibitem [{\citenamefont {Li}\ \emph {et~al.}(2012)\citenamefont {Li},
  \citenamefont {Erickcek},\ and\ \citenamefont {Law}}]{li2012new}%
  \BibitemOpen
  \bibfield  {author} {\bibinfo {author} {\bibfnamefont {F.}~\bibnamefont
  {Li}}, \bibinfo {author} {\bibfnamefont {A.~L.}\ \bibnamefont {Erickcek}}, \
  and\ \bibinfo {author} {\bibfnamefont {N.~M.}\ \bibnamefont {Law}},\
  }\href@noop {} {\bibfield  {journal} {\bibinfo  {journal} {Phys. Rev. D}\
  }\textbf {\bibinfo {volume} {86}},\ \bibinfo {pages} {043519} (\bibinfo
  {year} {2012})},\ \Eprint {http://arxiv.org/abs/1202.1284} {1202.1284}
  \BibitemShut {NoStop}%
\bibitem [{\citenamefont {Yang}\ \emph
  {et~al.}(2013{\natexlab{a}})\citenamefont {Yang}, \citenamefont {Yang},\ and\
  \citenamefont {Zong}}]{yang2013neutrino}%
  \BibitemOpen
  \bibfield  {author} {\bibinfo {author} {\bibfnamefont {Y.}~\bibnamefont
  {Yang}}, \bibinfo {author} {\bibfnamefont {G.}~\bibnamefont {Yang}}, \ and\
  \bibinfo {author} {\bibfnamefont {H.}~\bibnamefont {Zong}},\ }\href@noop {}
  {\bibfield  {journal} {\bibinfo  {journal} {Phys. Rev. D}\ }\textbf {\bibinfo
  {volume} {87}},\ \bibinfo {pages} {103525} (\bibinfo {year}
  {2013}{\natexlab{a}})},\ \Eprint {http://arxiv.org/abs/1305.4213} {1305.4213}
  \BibitemShut {NoStop}%
\bibitem [{\citenamefont {Yang}\ \emph
  {et~al.}(2013{\natexlab{b}})\citenamefont {Yang}, \citenamefont {Yang},\ and\
  \citenamefont {Zong}}]{yang2013dark}%
  \BibitemOpen
  \bibfield  {author} {\bibinfo {author} {\bibfnamefont {Y.-P.}\ \bibnamefont
  {Yang}}, \bibinfo {author} {\bibfnamefont {G.-L.}\ \bibnamefont {Yang}}, \
  and\ \bibinfo {author} {\bibfnamefont {H.-S.}\ \bibnamefont {Zong}},\
  }\href@noop {} {\bibfield  {journal} {\bibinfo  {journal} {Europhys. Lett.}\
  }\textbf {\bibinfo {volume} {101}},\ \bibinfo {pages} {69001} (\bibinfo
  {year} {2013}{\natexlab{b}})},\ \Eprint {http://arxiv.org/abs/1210.1409}
  {1210.1409} \BibitemShut {NoStop}%
\bibitem [{\citenamefont {Yang}(2014)}]{yang2014constraints}%
  \BibitemOpen
  \bibfield  {author} {\bibinfo {author} {\bibfnamefont {Y.}~\bibnamefont
  {Yang}},\ }\href@noop {} {\bibfield  {journal} {\bibinfo  {journal} {Int. J.
  Mod. Phys. A}\ }\textbf {\bibinfo {volume} {29}},\ \bibinfo {pages} {1450194}
  (\bibinfo {year} {2014})},\ \Eprint {http://arxiv.org/abs/1501.00789}
  {1501.00789} \BibitemShut {NoStop}%
\bibitem [{\citenamefont {Clark}\ \emph {et~al.}(2016)\citenamefont {Clark},
  \citenamefont {Lewis},\ and\ \citenamefont
  {Scott}}]{clark2015investigatingII}%
  \BibitemOpen
  \bibfield  {author} {\bibinfo {author} {\bibfnamefont {H.~A.}\ \bibnamefont
  {Clark}}, \bibinfo {author} {\bibfnamefont {G.~F.}\ \bibnamefont {Lewis}}, \
  and\ \bibinfo {author} {\bibfnamefont {P.}~\bibnamefont {Scott}},\
  }\href@noop {} {\bibfield  {journal} {\bibinfo  {journal} {Mon. Not. R.
  Astron. Soc.}\ }\textbf {\bibinfo {volume} {456}},\ \bibinfo {pages} {1402}
  (\bibinfo {year} {2016})},\ \Eprint {http://arxiv.org/abs/1509.02941}
  {1509.02941} \BibitemShut {NoStop}%
\bibitem [{\citenamefont {Clark}\ \emph {et~al.}(2017)\citenamefont {Clark},
  \citenamefont {Lewis},\ and\ \citenamefont {Scott}}]{clark2017erratumII}%
  \BibitemOpen
  \bibfield  {author} {\bibinfo {author} {\bibfnamefont {H.~A.}\ \bibnamefont
  {Clark}}, \bibinfo {author} {\bibfnamefont {G.~F.}\ \bibnamefont {Lewis}}, \
  and\ \bibinfo {author} {\bibfnamefont {P.}~\bibnamefont {Scott}},\
  }\href@noop {} {\bibfield  {journal} {\bibinfo  {journal} {Mon. Not. R.
  Astron. Soc.}\ }\textbf {\bibinfo {volume} {464}},\ \bibinfo {pages} {955}
  (\bibinfo {year} {2017})}\BibitemShut {NoStop}%
\bibitem [{\citenamefont {Yang}\ and\ \citenamefont {Qin}(2017)}]{yang2017tau}%
  \BibitemOpen
  \bibfield  {author} {\bibinfo {author} {\bibfnamefont {Y.}~\bibnamefont
  {Yang}}\ and\ \bibinfo {author} {\bibfnamefont {Y.}~\bibnamefont {Qin}},\
  }\href@noop {} {\bibfield  {journal} {\bibinfo  {journal} {Phys. Rev. D}\
  }\textbf {\bibinfo {volume} {96}},\ \bibinfo {pages} {103509} (\bibinfo
  {year} {2017})},\ \Eprint {http://arxiv.org/abs/1711.00993} {1711.00993}
  \BibitemShut {NoStop}%
\bibitem [{\citenamefont {Nakama}\ \emph {et~al.}(2018)\citenamefont {Nakama},
  \citenamefont {Suyama}, \citenamefont {Kohri},\ and\ \citenamefont
  {Hiroshima}}]{nakama2018constraints}%
  \BibitemOpen
  \bibfield  {author} {\bibinfo {author} {\bibfnamefont {T.}~\bibnamefont
  {Nakama}}, \bibinfo {author} {\bibfnamefont {T.}~\bibnamefont {Suyama}},
  \bibinfo {author} {\bibfnamefont {K.}~\bibnamefont {Kohri}}, \ and\ \bibinfo
  {author} {\bibfnamefont {N.}~\bibnamefont {Hiroshima}},\ }\href@noop {}
  {\bibfield  {journal} {\bibinfo  {journal} {Phys. Rev. D}\ }\textbf {\bibinfo
  {volume} {97}},\ \bibinfo {pages} {023539} (\bibinfo {year} {2018})},\
  \Eprint {http://arxiv.org/abs/1712.08820} {1712.08820} \BibitemShut {NoStop}%
\bibitem [{\citenamefont {Ricotti}\ and\ \citenamefont
  {Gould}(2009)}]{ricotti2009new}%
  \BibitemOpen
  \bibfield  {author} {\bibinfo {author} {\bibfnamefont {M.}~\bibnamefont
  {Ricotti}}\ and\ \bibinfo {author} {\bibfnamefont {A.}~\bibnamefont
  {Gould}},\ }\href@noop {} {\bibfield  {journal} {\bibinfo  {journal}
  {Astrophys. J.}\ }\textbf {\bibinfo {volume} {707}},\ \bibinfo {pages} {979}
  (\bibinfo {year} {2009})},\ \Eprint {http://arxiv.org/abs/0908.0735}
  {0908.0735} \BibitemShut {NoStop}%
\bibitem [{\citenamefont {Bertschinger}(1985)}]{bertschinger1985self}%
  \BibitemOpen
  \bibfield  {author} {\bibinfo {author} {\bibfnamefont {E.}~\bibnamefont
  {Bertschinger}},\ }\href@noop {} {\bibfield  {journal} {\bibinfo  {journal}
  {Astrophys. J. Suppl. Ser.}\ }\textbf {\bibinfo {volume} {58}},\ \bibinfo
  {pages} {39} (\bibinfo {year} {1985})}\BibitemShut {NoStop}%
\bibitem [{\citenamefont {Gosenca}\ \emph {et~al.}(2017)\citenamefont
  {Gosenca}, \citenamefont {Adamek}, \citenamefont {Byrnes},\ and\
  \citenamefont {Hotchkiss}}]{gosenca20173d}%
  \BibitemOpen
  \bibfield  {author} {\bibinfo {author} {\bibfnamefont {M.}~\bibnamefont
  {Gosenca}}, \bibinfo {author} {\bibfnamefont {J.}~\bibnamefont {Adamek}},
  \bibinfo {author} {\bibfnamefont {C.~T.}\ \bibnamefont {Byrnes}}, \ and\
  \bibinfo {author} {\bibfnamefont {S.}~\bibnamefont {Hotchkiss}},\ }\href@noop
  {} {\bibfield  {journal} {\bibinfo  {journal} {Phys. Rev. D}\ }\textbf
  {\bibinfo {volume} {96}},\ \bibinfo {pages} {123519} (\bibinfo {year}
  {2017})},\ \Eprint {http://arxiv.org/abs/1710.02055} {1710.02055}
  \BibitemShut {NoStop}%
\bibitem [{\citenamefont {Delos}\ \emph
  {et~al.}(2018{\natexlab{a}})\citenamefont {Delos}, \citenamefont {Erickcek},
  \citenamefont {Bailey},\ and\ \citenamefont
  {Alvarez}}]{delos2018ultracompact}%
  \BibitemOpen
  \bibfield  {author} {\bibinfo {author} {\bibfnamefont {M.~S.}\ \bibnamefont
  {Delos}}, \bibinfo {author} {\bibfnamefont {A.~L.}\ \bibnamefont {Erickcek}},
  \bibinfo {author} {\bibfnamefont {A.~P.}\ \bibnamefont {Bailey}}, \ and\
  \bibinfo {author} {\bibfnamefont {M.~A.}\ \bibnamefont {Alvarez}},\
  }\href@noop {} {\bibfield  {journal} {\bibinfo  {journal} {Phys. Rev. D}\
  }\textbf {\bibinfo {volume} {97}},\ \bibinfo {pages} {041303(R)} (\bibinfo
  {year} {2018}{\natexlab{a}})},\ \Eprint {http://arxiv.org/abs/1712.05421}
  {1712.05421} \BibitemShut {NoStop}%
\bibitem [{\citenamefont {Delos}\ \emph
  {et~al.}(2018{\natexlab{b}})\citenamefont {Delos}, \citenamefont {Erickcek},
  \citenamefont {Bailey},\ and\ \citenamefont {Alvarez}}]{delos2018density}%
  \BibitemOpen
  \bibfield  {author} {\bibinfo {author} {\bibfnamefont {M.~S.}\ \bibnamefont
  {Delos}}, \bibinfo {author} {\bibfnamefont {A.~L.}\ \bibnamefont {Erickcek}},
  \bibinfo {author} {\bibfnamefont {A.~P.}\ \bibnamefont {Bailey}}, \ and\
  \bibinfo {author} {\bibfnamefont {M.~A.}\ \bibnamefont {Alvarez}},\
  }\href@noop {} {\bibfield  {journal} {\bibinfo  {journal} {Phys. Rev. D}\
  }\textbf {\bibinfo {volume} {98}},\ \bibinfo {pages} {063527} (\bibinfo
  {year} {2018}{\natexlab{b}})},\ \Eprint {http://arxiv.org/abs/1806.07389}
  {1806.07389} \BibitemShut {NoStop}%
\bibitem [{\citenamefont {Gunn}\ and\ \citenamefont
  {Gott~III}(1972)}]{gunn1972infall}%
  \BibitemOpen
  \bibfield  {author} {\bibinfo {author} {\bibfnamefont {J.~E.}\ \bibnamefont
  {Gunn}}\ and\ \bibinfo {author} {\bibfnamefont {J.~R.}\ \bibnamefont
  {Gott~III}},\ }\href@noop {} {\bibfield  {journal} {\bibinfo  {journal}
  {Astrophys. J.}\ }\textbf {\bibinfo {volume} {176}},\ \bibinfo {pages} {1}
  (\bibinfo {year} {1972})}\BibitemShut {NoStop}%
\bibitem [{\citenamefont {Gott}(1975)}]{gott1975formation}%
  \BibitemOpen
  \bibfield  {author} {\bibinfo {author} {\bibfnamefont {J.~R.}\ \bibnamefont
  {Gott}},\ }\href@noop {} {\bibfield  {journal} {\bibinfo  {journal}
  {Astrophys. J.}\ }\textbf {\bibinfo {volume} {201}},\ \bibinfo {pages} {296}
  (\bibinfo {year} {1975})}\BibitemShut {NoStop}%
\bibitem [{\citenamefont {Gunn}(1977)}]{gunn1977massive}%
  \BibitemOpen
  \bibfield  {author} {\bibinfo {author} {\bibfnamefont {J.}~\bibnamefont
  {Gunn}},\ }\href@noop {} {\bibfield  {journal} {\bibinfo  {journal}
  {Astrophys. J.}\ }\textbf {\bibinfo {volume} {218}},\ \bibinfo {pages} {592}
  (\bibinfo {year} {1977})}\BibitemShut {NoStop}%
\bibitem [{\citenamefont {Fillmore}\ and\ \citenamefont
  {Goldreich}(1984)}]{fillmore1984self}%
  \BibitemOpen
  \bibfield  {author} {\bibinfo {author} {\bibfnamefont {J.~A.}\ \bibnamefont
  {Fillmore}}\ and\ \bibinfo {author} {\bibfnamefont {P.}~\bibnamefont
  {Goldreich}},\ }\href@noop {} {\bibfield  {journal} {\bibinfo  {journal}
  {Astrophys. J.}\ }\textbf {\bibinfo {volume} {281}},\ \bibinfo {pages} {1}
  (\bibinfo {year} {1984})}\BibitemShut {NoStop}%
\bibitem [{\citenamefont {Hoffman}\ and\ \citenamefont
  {Shaham}(1985)}]{hoffman1985local}%
  \BibitemOpen
  \bibfield  {author} {\bibinfo {author} {\bibfnamefont {Y.}~\bibnamefont
  {Hoffman}}\ and\ \bibinfo {author} {\bibfnamefont {J.}~\bibnamefont
  {Shaham}},\ }\href@noop {} {\bibfield  {journal} {\bibinfo  {journal}
  {Astrophys. J.}\ }\textbf {\bibinfo {volume} {297}},\ \bibinfo {pages} {16}
  (\bibinfo {year} {1985})}\BibitemShut {NoStop}%
\bibitem [{\citenamefont {Ryden}\ and\ \citenamefont
  {Gunn}(1987)}]{ryden1987galaxy}%
  \BibitemOpen
  \bibfield  {author} {\bibinfo {author} {\bibfnamefont {B.~S.}\ \bibnamefont
  {Ryden}}\ and\ \bibinfo {author} {\bibfnamefont {J.~E.}\ \bibnamefont
  {Gunn}},\ }\href@noop {} {\bibfield  {journal} {\bibinfo  {journal}
  {Astrophys. J.}\ }\textbf {\bibinfo {volume} {318}},\ \bibinfo {pages} {15}
  (\bibinfo {year} {1987})}\BibitemShut {NoStop}%
\bibitem [{\citenamefont {Zaroubi}\ and\ \citenamefont
  {Hoffman}(1993)}]{zaroubi1993gravitational}%
  \BibitemOpen
  \bibfield  {author} {\bibinfo {author} {\bibfnamefont {S.}~\bibnamefont
  {Zaroubi}}\ and\ \bibinfo {author} {\bibfnamefont {Y.}~\bibnamefont
  {Hoffman}},\ }\href@noop {} {\bibfield  {journal} {\bibinfo  {journal}
  {Astrophys. J.}\ }\textbf {\bibinfo {volume} {416}},\ \bibinfo {pages} {410}
  (\bibinfo {year} {1993})}\BibitemShut {NoStop}%
\bibitem [{\citenamefont {{\L}okas}\ and\ \citenamefont
  {Hoffman}(2000)}]{lokas2000formation}%
  \BibitemOpen
  \bibfield  {author} {\bibinfo {author} {\bibfnamefont {E.~L.}\ \bibnamefont
  {{\L}okas}}\ and\ \bibinfo {author} {\bibfnamefont {Y.}~\bibnamefont
  {Hoffman}},\ }\href@noop {} {\bibfield  {journal} {\bibinfo  {journal}
  {Astrophys. J. Lett.}\ }\textbf {\bibinfo {volume} {542}},\ \bibinfo {pages}
  {L139} (\bibinfo {year} {2000})},\ \Eprint
  {http://arxiv.org/abs/astro-ph/0005566} {astro-ph/0005566} \BibitemShut
  {NoStop}%
\bibitem [{\citenamefont {Dalal}\ \emph {et~al.}()\citenamefont {Dalal},
  \citenamefont {Lithwick},\ and\ \citenamefont {Kuhlen}}]{dalal2010origin}%
  \BibitemOpen
  \bibfield  {author} {\bibinfo {author} {\bibfnamefont {N.}~\bibnamefont
  {Dalal}}, \bibinfo {author} {\bibfnamefont {Y.}~\bibnamefont {Lithwick}}, \
  and\ \bibinfo {author} {\bibfnamefont {M.}~\bibnamefont {Kuhlen}},\
  }\href@noop {} {\ }\Eprint {http://arxiv.org/abs/1010.2539} {1010.2539}
  \BibitemShut {NoStop}%
\bibitem [{\citenamefont {Ryden}(1993)}]{ryden1993self}%
  \BibitemOpen
  \bibfield  {author} {\bibinfo {author} {\bibfnamefont {B.~S.}\ \bibnamefont
  {Ryden}},\ }\href@noop {} {\bibfield  {journal} {\bibinfo  {journal}
  {Astrophys. J.}\ }\textbf {\bibinfo {volume} {418}},\ \bibinfo {pages} {4}
  (\bibinfo {year} {1993})}\BibitemShut {NoStop}%
\bibitem [{\citenamefont {Lithwick}\ and\ \citenamefont
  {Dalal}(2011)}]{lithwick2011self}%
  \BibitemOpen
  \bibfield  {author} {\bibinfo {author} {\bibfnamefont {Y.}~\bibnamefont
  {Lithwick}}\ and\ \bibinfo {author} {\bibfnamefont {N.}~\bibnamefont
  {Dalal}},\ }\href@noop {} {\bibfield  {journal} {\bibinfo  {journal}
  {Astrophys. J.}\ }\textbf {\bibinfo {volume} {734}},\ \bibinfo {pages} {100}
  (\bibinfo {year} {2011})},\ \Eprint {http://arxiv.org/abs/1010.3723}
  {1010.3723} \BibitemShut {NoStop}%
\bibitem [{\citenamefont {White}\ and\ \citenamefont
  {Zaritsky}(1992)}]{white1992models}%
  \BibitemOpen
  \bibfield  {author} {\bibinfo {author} {\bibfnamefont {S.~D.}\ \bibnamefont
  {White}}\ and\ \bibinfo {author} {\bibfnamefont {D.}~\bibnamefont
  {Zaritsky}},\ }\href@noop {} {\bibfield  {journal} {\bibinfo  {journal}
  {Astrophys. J.}\ }\textbf {\bibinfo {volume} {394}},\ \bibinfo {pages} {1}
  (\bibinfo {year} {1992})}\BibitemShut {NoStop}%
\bibitem [{\citenamefont {Nusser}(2001)}]{nusser2001self}%
  \BibitemOpen
  \bibfield  {author} {\bibinfo {author} {\bibfnamefont {A.}~\bibnamefont
  {Nusser}},\ }\href@noop {} {\bibfield  {journal} {\bibinfo  {journal} {Mon.
  Not. R. Astron. Soc.}\ }\textbf {\bibinfo {volume} {325}},\ \bibinfo {pages}
  {1397} (\bibinfo {year} {2001})}\BibitemShut {NoStop}%
\bibitem [{\citenamefont {Ascasibar}\ \emph {et~al.}(2004)\citenamefont
  {Ascasibar}, \citenamefont {Yepes}, \citenamefont {Gottl{\"o}ber},\ and\
  \citenamefont {M{\"u}ller}}]{ascasibar2004physical}%
  \BibitemOpen
  \bibfield  {author} {\bibinfo {author} {\bibfnamefont {Y.}~\bibnamefont
  {Ascasibar}}, \bibinfo {author} {\bibfnamefont {G.}~\bibnamefont {Yepes}},
  \bibinfo {author} {\bibfnamefont {S.}~\bibnamefont {Gottl{\"o}ber}}, \ and\
  \bibinfo {author} {\bibfnamefont {V.}~\bibnamefont {M{\"u}ller}},\
  }\href@noop {} {\bibfield  {journal} {\bibinfo  {journal} {Mon. Not. R.
  Astron. Soc.}\ }\textbf {\bibinfo {volume} {352}},\ \bibinfo {pages} {1109}
  (\bibinfo {year} {2004})},\ \Eprint {http://arxiv.org/abs/astro-ph/0312221}
  {astro-ph/0312221} \BibitemShut {NoStop}%
\bibitem [{\citenamefont {Lu}\ \emph {et~al.}(2006)\citenamefont {Lu},
  \citenamefont {Mo}, \citenamefont {Katz},\ and\ \citenamefont
  {Weinberg}}]{lu2006origin}%
  \BibitemOpen
  \bibfield  {author} {\bibinfo {author} {\bibfnamefont {Y.}~\bibnamefont
  {Lu}}, \bibinfo {author} {\bibfnamefont {H.~J.}\ \bibnamefont {Mo}}, \bibinfo
  {author} {\bibfnamefont {N.}~\bibnamefont {Katz}}, \ and\ \bibinfo {author}
  {\bibfnamefont {M.~D.}\ \bibnamefont {Weinberg}},\ }\href@noop {} {\bibfield
  {journal} {\bibinfo  {journal} {Mon. Not. R. Astron. Soc.}\ }\textbf
  {\bibinfo {volume} {368}},\ \bibinfo {pages} {1931} (\bibinfo {year}
  {2006})},\ \Eprint {http://arxiv.org/abs/astro-ph/0508624} {astro-ph/0508624}
  \BibitemShut {NoStop}%
\bibitem [{\citenamefont {Ascasibar}\ \emph {et~al.}(2007)\citenamefont
  {Ascasibar}, \citenamefont {Hoffman},\ and\ \citenamefont
  {Gottl{\"o}ber}}]{ascasibar2007secondary}%
  \BibitemOpen
  \bibfield  {author} {\bibinfo {author} {\bibfnamefont {Y.}~\bibnamefont
  {Ascasibar}}, \bibinfo {author} {\bibfnamefont {Y.}~\bibnamefont {Hoffman}},
  \ and\ \bibinfo {author} {\bibfnamefont {S.}~\bibnamefont {Gottl{\"o}ber}},\
  }\href@noop {} {\bibfield  {journal} {\bibinfo  {journal} {Mon. Not. R.
  Astron. Soc.}\ }\textbf {\bibinfo {volume} {376}},\ \bibinfo {pages} {393}
  (\bibinfo {year} {2007})},\ \Eprint {http://arxiv.org/abs/astro-ph/0609713}
  {astro-ph/0609713} \BibitemShut {NoStop}%
\bibitem [{\citenamefont {Zukin}\ and\ \citenamefont
  {Bertschinger}(2010{\natexlab{a}})}]{zukin2010self}%
  \BibitemOpen
  \bibfield  {author} {\bibinfo {author} {\bibfnamefont {P.}~\bibnamefont
  {Zukin}}\ and\ \bibinfo {author} {\bibfnamefont {E.}~\bibnamefont
  {Bertschinger}},\ }\href@noop {} {\bibfield  {journal} {\bibinfo  {journal}
  {Phys. Rev. D}\ }\textbf {\bibinfo {volume} {82}},\ \bibinfo {pages} {104044}
  (\bibinfo {year} {2010}{\natexlab{a}})},\ \Eprint
  {http://arxiv.org/abs/1008.0639} {1008.0639} \BibitemShut {NoStop}%
\bibitem [{\citenamefont {Zukin}\ and\ \citenamefont
  {Bertschinger}(2010{\natexlab{b}})}]{zukin2010velocity}%
  \BibitemOpen
  \bibfield  {author} {\bibinfo {author} {\bibfnamefont {P.}~\bibnamefont
  {Zukin}}\ and\ \bibinfo {author} {\bibfnamefont {E.}~\bibnamefont
  {Bertschinger}},\ }\href@noop {} {\bibfield  {journal} {\bibinfo  {journal}
  {Phys. Rev. D}\ }\textbf {\bibinfo {volume} {82}},\ \bibinfo {pages} {104045}
  (\bibinfo {year} {2010}{\natexlab{b}})},\ \Eprint
  {http://arxiv.org/abs/1008.1980} {1008.1980} \BibitemShut {NoStop}%
\bibitem [{\citenamefont {Salvador-Sol{\'e}}\ \emph {et~al.}(2012)\citenamefont
  {Salvador-Sol{\'e}}, \citenamefont {Vi{\~n}as}, \citenamefont {Manrique},\
  and\ \citenamefont {Serra}}]{salvador2012theoretical}%
  \BibitemOpen
  \bibfield  {author} {\bibinfo {author} {\bibfnamefont {E.}~\bibnamefont
  {Salvador-Sol{\'e}}}, \bibinfo {author} {\bibfnamefont {J.}~\bibnamefont
  {Vi{\~n}as}}, \bibinfo {author} {\bibfnamefont {A.}~\bibnamefont {Manrique}},
  \ and\ \bibinfo {author} {\bibfnamefont {S.}~\bibnamefont {Serra}},\
  }\href@noop {} {\bibfield  {journal} {\bibinfo  {journal} {Mon. Not. R.
  Astron. Soc.}\ }\textbf {\bibinfo {volume} {423}},\ \bibinfo {pages} {2190}
  (\bibinfo {year} {2012})},\ \Eprint {http://arxiv.org/abs/1104.2334}
  {1104.2334} \BibitemShut {NoStop}%
\bibitem [{\citenamefont {Juan}\ \emph {et~al.}(2014)\citenamefont {Juan},
  \citenamefont {Salvador-Sol{\'e}}, \citenamefont {Dom{\`e}nech},\ and\
  \citenamefont {Manrique}}]{juan2014fixing}%
  \BibitemOpen
  \bibfield  {author} {\bibinfo {author} {\bibfnamefont {E.}~\bibnamefont
  {Juan}}, \bibinfo {author} {\bibfnamefont {E.}~\bibnamefont
  {Salvador-Sol{\'e}}}, \bibinfo {author} {\bibfnamefont {G.}~\bibnamefont
  {Dom{\`e}nech}}, \ and\ \bibinfo {author} {\bibfnamefont {A.}~\bibnamefont
  {Manrique}},\ }\href@noop {} {\bibfield  {journal} {\bibinfo  {journal} {Mon.
  Not. R. Astron. Soc.}\ }\textbf {\bibinfo {volume} {439}},\ \bibinfo {pages}
  {719} (\bibinfo {year} {2014})},\ \Eprint {http://arxiv.org/abs/1401.7335}
  {1401.7335} \BibitemShut {NoStop}%
\bibitem [{\citenamefont {Press}\ and\ \citenamefont
  {Schechter}(1974)}]{press1974formation}%
  \BibitemOpen
  \bibfield  {author} {\bibinfo {author} {\bibfnamefont {W.~H.}\ \bibnamefont
  {Press}}\ and\ \bibinfo {author} {\bibfnamefont {P.}~\bibnamefont
  {Schechter}},\ }\href@noop {} {\bibfield  {journal} {\bibinfo  {journal}
  {Astrophys. J.}\ }\textbf {\bibinfo {volume} {187}},\ \bibinfo {pages} {425}
  (\bibinfo {year} {1974})}\BibitemShut {NoStop}%
\bibitem [{\citenamefont {Sheth}\ \emph {et~al.}(2001)\citenamefont {Sheth},
  \citenamefont {Mo},\ and\ \citenamefont {Tormen}}]{sheth2001ellipsoidal}%
  \BibitemOpen
  \bibfield  {author} {\bibinfo {author} {\bibfnamefont {R.~K.}\ \bibnamefont
  {Sheth}}, \bibinfo {author} {\bibfnamefont {H.}~\bibnamefont {Mo}}, \ and\
  \bibinfo {author} {\bibfnamefont {G.}~\bibnamefont {Tormen}},\ }\href@noop {}
  {\bibfield  {journal} {\bibinfo  {journal} {Mon. Not. R. Astron. Soc.}\
  }\textbf {\bibinfo {volume} {323}},\ \bibinfo {pages} {1} (\bibinfo {year}
  {2001})},\ \Eprint {http://arxiv.org/abs/astro-ph/9907024} {astro-ph/9907024}
  \BibitemShut {NoStop}%
\bibitem [{\citenamefont {Sheth}\ and\ \citenamefont
  {Tormen}(2002)}]{sheth2002excursion}%
  \BibitemOpen
  \bibfield  {author} {\bibinfo {author} {\bibfnamefont {R.~K.}\ \bibnamefont
  {Sheth}}\ and\ \bibinfo {author} {\bibfnamefont {G.}~\bibnamefont {Tormen}},\
  }\href@noop {} {\bibfield  {journal} {\bibinfo  {journal} {Mon. Not. R.
  Astron. Soc.}\ }\textbf {\bibinfo {volume} {329}},\ \bibinfo {pages} {61}
  (\bibinfo {year} {2002})},\ \Eprint {http://arxiv.org/abs/astro-ph/0105113}
  {astro-ph/0105113} \BibitemShut {NoStop}%
\bibitem [{\citenamefont {Avila-Reese}\ \emph {et~al.}(1999)\citenamefont
  {Avila-Reese}, \citenamefont {Firmani}, \citenamefont {Klypin},\ and\
  \citenamefont {Kravtsov}}]{avila1999density}%
  \BibitemOpen
  \bibfield  {author} {\bibinfo {author} {\bibfnamefont {V.}~\bibnamefont
  {Avila-Reese}}, \bibinfo {author} {\bibfnamefont {C.}~\bibnamefont
  {Firmani}}, \bibinfo {author} {\bibfnamefont {A.}~\bibnamefont {Klypin}}, \
  and\ \bibinfo {author} {\bibfnamefont {A.~V.}\ \bibnamefont {Kravtsov}},\
  }\href@noop {} {\bibfield  {journal} {\bibinfo  {journal} {Mon. Not. R.
  Astron. Soc.}\ }\textbf {\bibinfo {volume} {310}},\ \bibinfo {pages} {527}
  (\bibinfo {year} {1999})},\ \Eprint {http://arxiv.org/abs/astro-ph/9906260}
  {astro-ph/9906260} \BibitemShut {NoStop}%
\bibitem [{\citenamefont {Jing}(2000)}]{jing2000density}%
  \BibitemOpen
  \bibfield  {author} {\bibinfo {author} {\bibfnamefont {Y.}~\bibnamefont
  {Jing}},\ }\href@noop {} {\bibfield  {journal} {\bibinfo  {journal}
  {Astrophys. J.}\ }\textbf {\bibinfo {volume} {535}},\ \bibinfo {pages} {30}
  (\bibinfo {year} {2000})},\ \Eprint {http://arxiv.org/abs/astro-ph/9901340}
  {astro-ph/9901340} \BibitemShut {NoStop}%
\bibitem [{\citenamefont {Colin}\ \emph {et~al.}(2004)\citenamefont {Colin},
  \citenamefont {Klypin}, \citenamefont {Valenzuela},\ and\ \citenamefont
  {Gottl{\"o}ber}}]{colin2004dwarf}%
  \BibitemOpen
  \bibfield  {author} {\bibinfo {author} {\bibfnamefont {P.}~\bibnamefont
  {Colin}}, \bibinfo {author} {\bibfnamefont {A.}~\bibnamefont {Klypin}},
  \bibinfo {author} {\bibfnamefont {O.}~\bibnamefont {Valenzuela}}, \ and\
  \bibinfo {author} {\bibfnamefont {S.}~\bibnamefont {Gottl{\"o}ber}},\
  }\href@noop {} {\bibfield  {journal} {\bibinfo  {journal} {Astrophys. J.}\
  }\textbf {\bibinfo {volume} {612}},\ \bibinfo {pages} {50} (\bibinfo {year}
  {2004})},\ \Eprint {http://arxiv.org/abs/astro-ph/0308348} {astro-ph/0308348}
  \BibitemShut {NoStop}%
\bibitem [{\citenamefont {Dolag}\ \emph {et~al.}(2004)\citenamefont {Dolag},
  \citenamefont {Bartelmann}, \citenamefont {Perrotta}, \citenamefont
  {Baccigalupi}, \citenamefont {Moscardini}, \citenamefont {Meneghetti},\ and\
  \citenamefont {Tormen}}]{dolag2004numerical}%
  \BibitemOpen
  \bibfield  {author} {\bibinfo {author} {\bibfnamefont {K.}~\bibnamefont
  {Dolag}}, \bibinfo {author} {\bibfnamefont {M.}~\bibnamefont {Bartelmann}},
  \bibinfo {author} {\bibfnamefont {F.}~\bibnamefont {Perrotta}}, \bibinfo
  {author} {\bibfnamefont {C.}~\bibnamefont {Baccigalupi}}, \bibinfo {author}
  {\bibfnamefont {L.}~\bibnamefont {Moscardini}}, \bibinfo {author}
  {\bibfnamefont {M.}~\bibnamefont {Meneghetti}}, \ and\ \bibinfo {author}
  {\bibfnamefont {G.}~\bibnamefont {Tormen}},\ }\href@noop {} {\bibfield
  {journal} {\bibinfo  {journal} {Astron. Astrophys.}\ }\textbf {\bibinfo
  {volume} {416}},\ \bibinfo {pages} {853} (\bibinfo {year} {2004})},\ \Eprint
  {http://arxiv.org/abs/astro-ph/0309771} {astro-ph/0309771} \BibitemShut
  {NoStop}%
\bibitem [{\citenamefont {Avila-Reese}\ \emph {et~al.}(2005)\citenamefont
  {Avila-Reese}, \citenamefont {Col{\'\i}n}, \citenamefont {Gottl{\"o}ber},
  \citenamefont {Firmani},\ and\ \citenamefont
  {Maulbetsch}}]{avila2005dependence}%
  \BibitemOpen
  \bibfield  {author} {\bibinfo {author} {\bibfnamefont {V.}~\bibnamefont
  {Avila-Reese}}, \bibinfo {author} {\bibfnamefont {P.}~\bibnamefont
  {Col{\'\i}n}}, \bibinfo {author} {\bibfnamefont {S.}~\bibnamefont
  {Gottl{\"o}ber}}, \bibinfo {author} {\bibfnamefont {C.}~\bibnamefont
  {Firmani}}, \ and\ \bibinfo {author} {\bibfnamefont {C.}~\bibnamefont
  {Maulbetsch}},\ }\href@noop {} {\bibfield  {journal} {\bibinfo  {journal}
  {Astrophys. J.}\ }\textbf {\bibinfo {volume} {634}},\ \bibinfo {pages} {51}
  (\bibinfo {year} {2005})},\ \Eprint {http://arxiv.org/abs/astro-ph/0508053}
  {astro-ph/0508053} \BibitemShut {NoStop}%
\bibitem [{\citenamefont {Neto}\ \emph {et~al.}(2007)\citenamefont {Neto},
  \citenamefont {Gao}, \citenamefont {Bett}, \citenamefont {Cole},
  \citenamefont {Navarro}, \citenamefont {Frenk}, \citenamefont {White},
  \citenamefont {Springel},\ and\ \citenamefont
  {Jenkins}}]{neto2007statistics}%
  \BibitemOpen
  \bibfield  {author} {\bibinfo {author} {\bibfnamefont {A.~F.}\ \bibnamefont
  {Neto}}, \bibinfo {author} {\bibfnamefont {L.}~\bibnamefont {Gao}}, \bibinfo
  {author} {\bibfnamefont {P.}~\bibnamefont {Bett}}, \bibinfo {author}
  {\bibfnamefont {S.}~\bibnamefont {Cole}}, \bibinfo {author} {\bibfnamefont
  {J.~F.}\ \bibnamefont {Navarro}}, \bibinfo {author} {\bibfnamefont {C.~S.}\
  \bibnamefont {Frenk}}, \bibinfo {author} {\bibfnamefont {S.~D.}\ \bibnamefont
  {White}}, \bibinfo {author} {\bibfnamefont {V.}~\bibnamefont {Springel}}, \
  and\ \bibinfo {author} {\bibfnamefont {A.}~\bibnamefont {Jenkins}},\
  }\href@noop {} {\bibfield  {journal} {\bibinfo  {journal} {Mon. Not. R.
  Astron. Soc.}\ }\textbf {\bibinfo {volume} {381}},\ \bibinfo {pages} {1450}
  (\bibinfo {year} {2007})},\ \Eprint {http://arxiv.org/abs/0706.2919}
  {0706.2919} \BibitemShut {NoStop}%
\bibitem [{\citenamefont {Duffy}\ \emph {et~al.}(2008)\citenamefont {Duffy},
  \citenamefont {Schaye}, \citenamefont {Kay},\ and\ \citenamefont
  {Dalla~Vecchia}}]{duffy2008dark}%
  \BibitemOpen
  \bibfield  {author} {\bibinfo {author} {\bibfnamefont {A.~R.}\ \bibnamefont
  {Duffy}}, \bibinfo {author} {\bibfnamefont {J.}~\bibnamefont {Schaye}},
  \bibinfo {author} {\bibfnamefont {S.~T.}\ \bibnamefont {Kay}}, \ and\
  \bibinfo {author} {\bibfnamefont {C.}~\bibnamefont {Dalla~Vecchia}},\
  }\href@noop {} {\bibfield  {journal} {\bibinfo  {journal} {Mon. Not. R.
  Astron. Soc. Lett.}\ }\textbf {\bibinfo {volume} {390}},\ \bibinfo {pages}
  {L64} (\bibinfo {year} {2008})},\ \Eprint {http://arxiv.org/abs/0804.2486}
  {0804.2486} \BibitemShut {NoStop}%
\bibitem [{\citenamefont {Duffy}\ \emph {et~al.}(2011)\citenamefont {Duffy},
  \citenamefont {Schaye}, \citenamefont {Kay},\ and\ \citenamefont
  {Dalla~Vecchia}}]{duffy2011erratum}%
  \BibitemOpen
  \bibfield  {author} {\bibinfo {author} {\bibfnamefont {A.~R.}\ \bibnamefont
  {Duffy}}, \bibinfo {author} {\bibfnamefont {J.}~\bibnamefont {Schaye}},
  \bibinfo {author} {\bibfnamefont {S.~T.}\ \bibnamefont {Kay}}, \ and\
  \bibinfo {author} {\bibfnamefont {C.}~\bibnamefont {Dalla~Vecchia}},\
  }\href@noop {} {\bibfield  {journal} {\bibinfo  {journal} {Mon. Not. R.
  Astron. Soc. Lett.}\ }\textbf {\bibinfo {volume} {415}},\ \bibinfo {pages}
  {L85} (\bibinfo {year} {2011})}\BibitemShut {NoStop}%
\bibitem [{\citenamefont {Gao}\ \emph {et~al.}(2008)\citenamefont {Gao},
  \citenamefont {Navarro}, \citenamefont {Cole}, \citenamefont {Frenk},
  \citenamefont {White}, \citenamefont {Springel}, \citenamefont {Jenkins},\
  and\ \citenamefont {Neto}}]{gao2008redshift}%
  \BibitemOpen
  \bibfield  {author} {\bibinfo {author} {\bibfnamefont {L.}~\bibnamefont
  {Gao}}, \bibinfo {author} {\bibfnamefont {J.~F.}\ \bibnamefont {Navarro}},
  \bibinfo {author} {\bibfnamefont {S.}~\bibnamefont {Cole}}, \bibinfo {author}
  {\bibfnamefont {C.~S.}\ \bibnamefont {Frenk}}, \bibinfo {author}
  {\bibfnamefont {S.~D.}\ \bibnamefont {White}}, \bibinfo {author}
  {\bibfnamefont {V.}~\bibnamefont {Springel}}, \bibinfo {author}
  {\bibfnamefont {A.}~\bibnamefont {Jenkins}}, \ and\ \bibinfo {author}
  {\bibfnamefont {A.~F.}\ \bibnamefont {Neto}},\ }\href@noop {} {\bibfield
  {journal} {\bibinfo  {journal} {Mon. Not. R. Astron. Soc.}\ }\textbf
  {\bibinfo {volume} {387}},\ \bibinfo {pages} {536} (\bibinfo {year}
  {2008})},\ \Eprint {http://arxiv.org/abs/0711.0746} {0711.0746} \BibitemShut
  {NoStop}%
\bibitem [{\citenamefont {Macci{\`o}}\ \emph {et~al.}(2008)\citenamefont
  {Macci{\`o}}, \citenamefont {Dutton},\ and\ \citenamefont {Van
  Den~Bosch}}]{maccio2008concentration}%
  \BibitemOpen
  \bibfield  {author} {\bibinfo {author} {\bibfnamefont {A.~V.}\ \bibnamefont
  {Macci{\`o}}}, \bibinfo {author} {\bibfnamefont {A.~A.}\ \bibnamefont
  {Dutton}}, \ and\ \bibinfo {author} {\bibfnamefont {F.~C.}\ \bibnamefont {Van
  Den~Bosch}},\ }\href@noop {} {\bibfield  {journal} {\bibinfo  {journal} {Mon.
  Not. R. Astron. Soc.}\ }\textbf {\bibinfo {volume} {391}},\ \bibinfo {pages}
  {1940} (\bibinfo {year} {2008})},\ \Eprint {http://arxiv.org/abs/0805.1926}
  {0805.1926} \BibitemShut {NoStop}%
\bibitem [{\citenamefont {Klypin}\ \emph {et~al.}(2011)\citenamefont {Klypin},
  \citenamefont {Trujillo-Gomez},\ and\ \citenamefont
  {Primack}}]{klypin2011dark}%
  \BibitemOpen
  \bibfield  {author} {\bibinfo {author} {\bibfnamefont {A.~A.}\ \bibnamefont
  {Klypin}}, \bibinfo {author} {\bibfnamefont {S.}~\bibnamefont
  {Trujillo-Gomez}}, \ and\ \bibinfo {author} {\bibfnamefont {J.}~\bibnamefont
  {Primack}},\ }\href@noop {} {\bibfield  {journal} {\bibinfo  {journal}
  {Astrophys. J.}\ }\textbf {\bibinfo {volume} {740}},\ \bibinfo {pages} {102}
  (\bibinfo {year} {2011})},\ \Eprint {http://arxiv.org/abs/1002.3660}
  {1002.3660} \BibitemShut {NoStop}%
\bibitem [{\citenamefont {Mu{\~n}oz-Cuartas}\ \emph {et~al.}(2011)\citenamefont
  {Mu{\~n}oz-Cuartas}, \citenamefont {Macci{\`o}}, \citenamefont
  {Gottl{\"o}ber},\ and\ \citenamefont {Dutton}}]{munoz2011redshift}%
  \BibitemOpen
  \bibfield  {author} {\bibinfo {author} {\bibfnamefont {J.}~\bibnamefont
  {Mu{\~n}oz-Cuartas}}, \bibinfo {author} {\bibfnamefont {A.}~\bibnamefont
  {Macci{\`o}}}, \bibinfo {author} {\bibfnamefont {S.}~\bibnamefont
  {Gottl{\"o}ber}}, \ and\ \bibinfo {author} {\bibfnamefont {A.}~\bibnamefont
  {Dutton}},\ }\href@noop {} {\bibfield  {journal} {\bibinfo  {journal} {Mon.
  Not. R. Astron. Soc.}\ }\textbf {\bibinfo {volume} {411}},\ \bibinfo {pages}
  {584} (\bibinfo {year} {2011})},\ \Eprint {http://arxiv.org/abs/1007.0438}
  {1007.0438} \BibitemShut {NoStop}%
\bibitem [{\citenamefont {Bhattacharya}\ \emph {et~al.}(2013)\citenamefont
  {Bhattacharya}, \citenamefont {Habib}, \citenamefont {Heitmann},\ and\
  \citenamefont {Vikhlinin}}]{bhattacharya2013dark}%
  \BibitemOpen
  \bibfield  {author} {\bibinfo {author} {\bibfnamefont {S.}~\bibnamefont
  {Bhattacharya}}, \bibinfo {author} {\bibfnamefont {S.}~\bibnamefont {Habib}},
  \bibinfo {author} {\bibfnamefont {K.}~\bibnamefont {Heitmann}}, \ and\
  \bibinfo {author} {\bibfnamefont {A.}~\bibnamefont {Vikhlinin}},\ }\href@noop
  {} {\bibfield  {journal} {\bibinfo  {journal} {Astrophys. J.}\ }\textbf
  {\bibinfo {volume} {766}},\ \bibinfo {pages} {32} (\bibinfo {year} {2013})},\
  \Eprint {http://arxiv.org/abs/1112.5479} {1112.5479} \BibitemShut {NoStop}%
\bibitem [{\citenamefont {Dutton}\ and\ \citenamefont
  {Macci{\`o}}(2014)}]{dutton2014cold}%
  \BibitemOpen
  \bibfield  {author} {\bibinfo {author} {\bibfnamefont {A.~A.}\ \bibnamefont
  {Dutton}}\ and\ \bibinfo {author} {\bibfnamefont {A.~V.}\ \bibnamefont
  {Macci{\`o}}},\ }\href@noop {} {\bibfield  {journal} {\bibinfo  {journal}
  {Mon. Not. R. Astron. Soc.}\ }\textbf {\bibinfo {volume} {441}},\ \bibinfo
  {pages} {3359} (\bibinfo {year} {2014})},\ \Eprint
  {http://arxiv.org/abs/1402.7073} {1402.7073} \BibitemShut {NoStop}%
\bibitem [{\citenamefont {Heitmann}\ \emph {et~al.}(2015)\citenamefont
  {Heitmann}, \citenamefont {Frontiere}, \citenamefont {Sewell}, \citenamefont
  {Habib}, \citenamefont {Pope}, \citenamefont {Finkel}, \citenamefont {Rizzi},
  \citenamefont {Insley},\ and\ \citenamefont {Bhattacharya}}]{heitmann2015q}%
  \BibitemOpen
  \bibfield  {author} {\bibinfo {author} {\bibfnamefont {K.}~\bibnamefont
  {Heitmann}}, \bibinfo {author} {\bibfnamefont {N.}~\bibnamefont {Frontiere}},
  \bibinfo {author} {\bibfnamefont {C.}~\bibnamefont {Sewell}}, \bibinfo
  {author} {\bibfnamefont {S.}~\bibnamefont {Habib}}, \bibinfo {author}
  {\bibfnamefont {A.}~\bibnamefont {Pope}}, \bibinfo {author} {\bibfnamefont
  {H.}~\bibnamefont {Finkel}}, \bibinfo {author} {\bibfnamefont
  {S.}~\bibnamefont {Rizzi}}, \bibinfo {author} {\bibfnamefont
  {J.}~\bibnamefont {Insley}}, \ and\ \bibinfo {author} {\bibfnamefont
  {S.}~\bibnamefont {Bhattacharya}},\ }\href@noop {} {\bibfield  {journal}
  {\bibinfo  {journal} {Astrophys. J. Suppl. Ser.}\ }\textbf {\bibinfo {volume}
  {219}},\ \bibinfo {pages} {34} (\bibinfo {year} {2015})},\ \Eprint
  {http://arxiv.org/abs/1411.3396} {1411.3396} \BibitemShut {NoStop}%
\bibitem [{\citenamefont {Klypin}\ \emph {et~al.}(2016)\citenamefont {Klypin},
  \citenamefont {Yepes}, \citenamefont {Gottl{\"o}ber}, \citenamefont {Prada},\
  and\ \citenamefont {Hess}}]{klypin2016multidark}%
  \BibitemOpen
  \bibfield  {author} {\bibinfo {author} {\bibfnamefont {A.}~\bibnamefont
  {Klypin}}, \bibinfo {author} {\bibfnamefont {G.}~\bibnamefont {Yepes}},
  \bibinfo {author} {\bibfnamefont {S.}~\bibnamefont {Gottl{\"o}ber}}, \bibinfo
  {author} {\bibfnamefont {F.}~\bibnamefont {Prada}}, \ and\ \bibinfo {author}
  {\bibfnamefont {S.}~\bibnamefont {Hess}},\ }\href@noop {} {\bibfield
  {journal} {\bibinfo  {journal} {Mon. Not. R. Astron. Soc.}\ }\textbf
  {\bibinfo {volume} {457}},\ \bibinfo {pages} {4340} (\bibinfo {year}
  {2016})},\ \Eprint {http://arxiv.org/abs/1411.4001} {1411.4001} \BibitemShut
  {NoStop}%
\bibitem [{\citenamefont {Hellwing}\ \emph {et~al.}(2016)\citenamefont
  {Hellwing}, \citenamefont {Frenk}, \citenamefont {Cautun}, \citenamefont
  {Bose}, \citenamefont {Helly}, \citenamefont {Jenkins}, \citenamefont
  {Sawala},\ and\ \citenamefont {Cytowski}}]{hellwing2016copernicus}%
  \BibitemOpen
  \bibfield  {author} {\bibinfo {author} {\bibfnamefont {W.~A.}\ \bibnamefont
  {Hellwing}}, \bibinfo {author} {\bibfnamefont {C.~S.}\ \bibnamefont {Frenk}},
  \bibinfo {author} {\bibfnamefont {M.}~\bibnamefont {Cautun}}, \bibinfo
  {author} {\bibfnamefont {S.}~\bibnamefont {Bose}}, \bibinfo {author}
  {\bibfnamefont {J.}~\bibnamefont {Helly}}, \bibinfo {author} {\bibfnamefont
  {A.}~\bibnamefont {Jenkins}}, \bibinfo {author} {\bibfnamefont
  {T.}~\bibnamefont {Sawala}}, \ and\ \bibinfo {author} {\bibfnamefont
  {M.}~\bibnamefont {Cytowski}},\ }\href@noop {} {\bibfield  {journal}
  {\bibinfo  {journal} {Mon. Not. R. Astron. Soc.}\ }\textbf {\bibinfo {volume}
  {457}},\ \bibinfo {pages} {3492} (\bibinfo {year} {2016})},\ \Eprint
  {http://arxiv.org/abs/1505.06436} {1505.06436} \BibitemShut {NoStop}%
\bibitem [{\citenamefont {Angel}\ \emph {et~al.}(2016)\citenamefont {Angel},
  \citenamefont {Poole}, \citenamefont {Ludlow}, \citenamefont {Duffy},
  \citenamefont {Geil}, \citenamefont {Mutch}, \citenamefont {Mesinger},\ and\
  \citenamefont {Wyithe}}]{angel2016dark}%
  \BibitemOpen
  \bibfield  {author} {\bibinfo {author} {\bibfnamefont {P.~W.}\ \bibnamefont
  {Angel}}, \bibinfo {author} {\bibfnamefont {G.~B.}\ \bibnamefont {Poole}},
  \bibinfo {author} {\bibfnamefont {A.~D.}\ \bibnamefont {Ludlow}}, \bibinfo
  {author} {\bibfnamefont {A.~R.}\ \bibnamefont {Duffy}}, \bibinfo {author}
  {\bibfnamefont {P.~M.}\ \bibnamefont {Geil}}, \bibinfo {author}
  {\bibfnamefont {S.~J.}\ \bibnamefont {Mutch}}, \bibinfo {author}
  {\bibfnamefont {A.}~\bibnamefont {Mesinger}}, \ and\ \bibinfo {author}
  {\bibfnamefont {J.~S.~B.}\ \bibnamefont {Wyithe}},\ }\href@noop {} {\bibfield
   {journal} {\bibinfo  {journal} {Mon. Not. R. Astron. Soc.}\ }\textbf
  {\bibinfo {volume} {459}},\ \bibinfo {pages} {2106} (\bibinfo {year}
  {2016})},\ \Eprint {http://arxiv.org/abs/1512.00560} {1512.00560}
  \BibitemShut {NoStop}%
\bibitem [{\citenamefont {Child}\ \emph {et~al.}(2018)\citenamefont {Child},
  \citenamefont {Habib}, \citenamefont {Heitmann}, \citenamefont {Frontiere},
  \citenamefont {Finkel}, \citenamefont {Pope},\ and\ \citenamefont
  {Morozov}}]{child2018halo}%
  \BibitemOpen
  \bibfield  {author} {\bibinfo {author} {\bibfnamefont {H.~L.}\ \bibnamefont
  {Child}}, \bibinfo {author} {\bibfnamefont {S.}~\bibnamefont {Habib}},
  \bibinfo {author} {\bibfnamefont {K.}~\bibnamefont {Heitmann}}, \bibinfo
  {author} {\bibfnamefont {N.}~\bibnamefont {Frontiere}}, \bibinfo {author}
  {\bibfnamefont {H.}~\bibnamefont {Finkel}}, \bibinfo {author} {\bibfnamefont
  {A.}~\bibnamefont {Pope}}, \ and\ \bibinfo {author} {\bibfnamefont
  {V.}~\bibnamefont {Morozov}},\ }\href@noop {} {\bibfield  {journal} {\bibinfo
   {journal} {Astrophys. J.}\ }\textbf {\bibinfo {volume} {859}},\ \bibinfo
  {pages} {55} (\bibinfo {year} {2018})},\ \Eprint
  {http://arxiv.org/abs/1804.10199} {1804.10199} \BibitemShut {NoStop}%
\bibitem [{\citenamefont {Bullock}\ \emph {et~al.}(2001)\citenamefont
  {Bullock}, \citenamefont {Kolatt}, \citenamefont {Sigad}, \citenamefont
  {Somerville}, \citenamefont {Kravtsov}, \citenamefont {Klypin}, \citenamefont
  {Primack},\ and\ \citenamefont {Dekel}}]{bullock2001profiles}%
  \BibitemOpen
  \bibfield  {author} {\bibinfo {author} {\bibfnamefont {J.~S.}\ \bibnamefont
  {Bullock}}, \bibinfo {author} {\bibfnamefont {T.~S.}\ \bibnamefont {Kolatt}},
  \bibinfo {author} {\bibfnamefont {Y.}~\bibnamefont {Sigad}}, \bibinfo
  {author} {\bibfnamefont {R.~S.}\ \bibnamefont {Somerville}}, \bibinfo
  {author} {\bibfnamefont {A.~V.}\ \bibnamefont {Kravtsov}}, \bibinfo {author}
  {\bibfnamefont {A.~A.}\ \bibnamefont {Klypin}}, \bibinfo {author}
  {\bibfnamefont {J.~R.}\ \bibnamefont {Primack}}, \ and\ \bibinfo {author}
  {\bibfnamefont {A.}~\bibnamefont {Dekel}},\ }\href@noop {} {\bibfield
  {journal} {\bibinfo  {journal} {Mon. Not. R. Astron. Soc.}\ }\textbf
  {\bibinfo {volume} {321}},\ \bibinfo {pages} {559} (\bibinfo {year}
  {2001})},\ \Eprint {http://arxiv.org/abs/astro-ph/9908159} {astro-ph/9908159}
  \BibitemShut {NoStop}%
\bibitem [{\citenamefont {Eke}\ \emph {et~al.}(2001)\citenamefont {Eke},
  \citenamefont {Navarro},\ and\ \citenamefont {Steinmetz}}]{eke2001power}%
  \BibitemOpen
  \bibfield  {author} {\bibinfo {author} {\bibfnamefont {V.~R.}\ \bibnamefont
  {Eke}}, \bibinfo {author} {\bibfnamefont {J.~F.}\ \bibnamefont {Navarro}}, \
  and\ \bibinfo {author} {\bibfnamefont {M.}~\bibnamefont {Steinmetz}},\
  }\href@noop {} {\bibfield  {journal} {\bibinfo  {journal} {Astrophys. J.}\
  }\textbf {\bibinfo {volume} {554}},\ \bibinfo {pages} {114} (\bibinfo {year}
  {2001})},\ \Eprint {http://arxiv.org/abs/astro-ph/0012337} {astro-ph/0012337}
  \BibitemShut {NoStop}%
\bibitem [{\citenamefont {Wechsler}\ \emph {et~al.}(2002)\citenamefont
  {Wechsler}, \citenamefont {Bullock}, \citenamefont {Primack}, \citenamefont
  {Kravtsov},\ and\ \citenamefont {Dekel}}]{wechsler2002concentrations}%
  \BibitemOpen
  \bibfield  {author} {\bibinfo {author} {\bibfnamefont {R.~H.}\ \bibnamefont
  {Wechsler}}, \bibinfo {author} {\bibfnamefont {J.~S.}\ \bibnamefont
  {Bullock}}, \bibinfo {author} {\bibfnamefont {J.~R.}\ \bibnamefont
  {Primack}}, \bibinfo {author} {\bibfnamefont {A.~V.}\ \bibnamefont
  {Kravtsov}}, \ and\ \bibinfo {author} {\bibfnamefont {A.}~\bibnamefont
  {Dekel}},\ }\href@noop {} {\bibfield  {journal} {\bibinfo  {journal}
  {Astrophys. J.}\ }\textbf {\bibinfo {volume} {568}},\ \bibinfo {pages} {52}
  (\bibinfo {year} {2002})},\ \Eprint {http://arxiv.org/abs/astro-ph/0108151}
  {astro-ph/0108151} \BibitemShut {NoStop}%
\bibitem [{\citenamefont {Zhao}\ \emph {et~al.}(2003)\citenamefont {Zhao},
  \citenamefont {Jing}, \citenamefont {Mo},\ and\ \citenamefont
  {B{\"o}rner}}]{zhao2003mass}%
  \BibitemOpen
  \bibfield  {author} {\bibinfo {author} {\bibfnamefont {D.-H.}\ \bibnamefont
  {Zhao}}, \bibinfo {author} {\bibfnamefont {Y.}~\bibnamefont {Jing}}, \bibinfo
  {author} {\bibfnamefont {H.}~\bibnamefont {Mo}}, \ and\ \bibinfo {author}
  {\bibfnamefont {G.}~\bibnamefont {B{\"o}rner}},\ }\href@noop {} {\bibfield
  {journal} {\bibinfo  {journal} {Astrophys. J. Lett.}\ }\textbf {\bibinfo
  {volume} {597}},\ \bibinfo {pages} {L9} (\bibinfo {year} {2003})},\ \Eprint
  {http://arxiv.org/abs/astro-ph/0309375} {astro-ph/0309375} \BibitemShut
  {NoStop}%
\bibitem [{\citenamefont {Zhao}\ \emph {et~al.}(2009)\citenamefont {Zhao},
  \citenamefont {Jing}, \citenamefont {Mo},\ and\ \citenamefont
  {B{\"o}rner}}]{zhao2009accurate}%
  \BibitemOpen
  \bibfield  {author} {\bibinfo {author} {\bibfnamefont {D.}~\bibnamefont
  {Zhao}}, \bibinfo {author} {\bibfnamefont {Y.}~\bibnamefont {Jing}}, \bibinfo
  {author} {\bibfnamefont {H.}~\bibnamefont {Mo}}, \ and\ \bibinfo {author}
  {\bibfnamefont {G.}~\bibnamefont {B{\"o}rner}},\ }\href@noop {} {\bibfield
  {journal} {\bibinfo  {journal} {Astrophys. J.}\ }\textbf {\bibinfo {volume}
  {707}},\ \bibinfo {pages} {354} (\bibinfo {year} {2009})},\ \Eprint
  {http://arxiv.org/abs/0811.0828} {0811.0828} \BibitemShut {NoStop}%
\bibitem [{\citenamefont {Giocoli}\ \emph {et~al.}(2012)\citenamefont
  {Giocoli}, \citenamefont {Tormen},\ and\ \citenamefont
  {Sheth}}]{giocoli2012formation}%
  \BibitemOpen
  \bibfield  {author} {\bibinfo {author} {\bibfnamefont {C.}~\bibnamefont
  {Giocoli}}, \bibinfo {author} {\bibfnamefont {G.}~\bibnamefont {Tormen}}, \
  and\ \bibinfo {author} {\bibfnamefont {R.~K.}\ \bibnamefont {Sheth}},\
  }\href@noop {} {\bibfield  {journal} {\bibinfo  {journal} {Mon. Not. R.
  Astron. Soc.}\ }\textbf {\bibinfo {volume} {422}},\ \bibinfo {pages} {185}
  (\bibinfo {year} {2012})},\ \Eprint {http://arxiv.org/abs/1111.6977}
  {1111.6977} \BibitemShut {NoStop}%
\bibitem [{\citenamefont {Ludlow}\ \emph {et~al.}(2013)\citenamefont {Ludlow},
  \citenamefont {Navarro}, \citenamefont {Boylan-Kolchin}, \citenamefont
  {Bett}, \citenamefont {Angulo}, \citenamefont {Li}, \citenamefont {White},
  \citenamefont {Frenk},\ and\ \citenamefont {Springel}}]{ludlow2013mass}%
  \BibitemOpen
  \bibfield  {author} {\bibinfo {author} {\bibfnamefont {A.~D.}\ \bibnamefont
  {Ludlow}}, \bibinfo {author} {\bibfnamefont {J.~F.}\ \bibnamefont {Navarro}},
  \bibinfo {author} {\bibfnamefont {M.}~\bibnamefont {Boylan-Kolchin}},
  \bibinfo {author} {\bibfnamefont {P.~E.}\ \bibnamefont {Bett}}, \bibinfo
  {author} {\bibfnamefont {R.~E.}\ \bibnamefont {Angulo}}, \bibinfo {author}
  {\bibfnamefont {M.}~\bibnamefont {Li}}, \bibinfo {author} {\bibfnamefont
  {S.~D.}\ \bibnamefont {White}}, \bibinfo {author} {\bibfnamefont
  {C.}~\bibnamefont {Frenk}}, \ and\ \bibinfo {author} {\bibfnamefont
  {V.}~\bibnamefont {Springel}},\ }\href@noop {} {\bibfield  {journal}
  {\bibinfo  {journal} {Mon. Not. R. Astron. Soc.}\ }\textbf {\bibinfo {volume}
  {432}},\ \bibinfo {pages} {1103} (\bibinfo {year} {2013})},\ \Eprint
  {http://arxiv.org/abs/1302.0288} {1302.0288} \BibitemShut {NoStop}%
\bibitem [{\citenamefont {Ludlow}\ \emph {et~al.}(2014)\citenamefont {Ludlow},
  \citenamefont {Navarro}, \citenamefont {Angulo}, \citenamefont
  {Boylan-Kolchin}, \citenamefont {Springel}, \citenamefont {Frenk},\ and\
  \citenamefont {White}}]{ludlow2014mass}%
  \BibitemOpen
  \bibfield  {author} {\bibinfo {author} {\bibfnamefont {A.~D.}\ \bibnamefont
  {Ludlow}}, \bibinfo {author} {\bibfnamefont {J.~F.}\ \bibnamefont {Navarro}},
  \bibinfo {author} {\bibfnamefont {R.~E.}\ \bibnamefont {Angulo}}, \bibinfo
  {author} {\bibfnamefont {M.}~\bibnamefont {Boylan-Kolchin}}, \bibinfo
  {author} {\bibfnamefont {V.}~\bibnamefont {Springel}}, \bibinfo {author}
  {\bibfnamefont {C.}~\bibnamefont {Frenk}}, \ and\ \bibinfo {author}
  {\bibfnamefont {S.~D.}\ \bibnamefont {White}},\ }\href@noop {} {\bibfield
  {journal} {\bibinfo  {journal} {Mon. Not. R. Astron. Soc.}\ }\textbf
  {\bibinfo {volume} {441}},\ \bibinfo {pages} {378} (\bibinfo {year}
  {2014})},\ \Eprint {http://arxiv.org/abs/1312.0945} {1312.0945} \BibitemShut
  {NoStop}%
\bibitem [{\citenamefont {Bosch}\ \emph {et~al.}(2014)\citenamefont {Bosch},
  \citenamefont {Jiang}, \citenamefont {Hearin}, \citenamefont {Campbell},
  \citenamefont {Watson},\ and\ \citenamefont {Padmanabhan}}]{bosch2014coming}%
  \BibitemOpen
  \bibfield  {author} {\bibinfo {author} {\bibfnamefont {F.~C. v.~d.}\
  \bibnamefont {Bosch}}, \bibinfo {author} {\bibfnamefont {F.}~\bibnamefont
  {Jiang}}, \bibinfo {author} {\bibfnamefont {A.}~\bibnamefont {Hearin}},
  \bibinfo {author} {\bibfnamefont {D.}~\bibnamefont {Campbell}}, \bibinfo
  {author} {\bibfnamefont {D.}~\bibnamefont {Watson}}, \ and\ \bibinfo {author}
  {\bibfnamefont {N.}~\bibnamefont {Padmanabhan}},\ }\href@noop {} {\bibfield
  {journal} {\bibinfo  {journal} {Mon. Not. R. Astron. Soc.}\ }\textbf
  {\bibinfo {volume} {445}},\ \bibinfo {pages} {1713} (\bibinfo {year}
  {2014})},\ \Eprint {http://arxiv.org/abs/1409.2750} {1409.2750} \BibitemShut
  {NoStop}%
\bibitem [{\citenamefont {Correa}\ \emph {et~al.}(2015)\citenamefont {Correa},
  \citenamefont {Wyithe}, \citenamefont {Schaye},\ and\ \citenamefont
  {Duffy}}]{correa2015accretion}%
  \BibitemOpen
  \bibfield  {author} {\bibinfo {author} {\bibfnamefont {C.~A.}\ \bibnamefont
  {Correa}}, \bibinfo {author} {\bibfnamefont {J.~S.~B.}\ \bibnamefont
  {Wyithe}}, \bibinfo {author} {\bibfnamefont {J.}~\bibnamefont {Schaye}}, \
  and\ \bibinfo {author} {\bibfnamefont {A.~R.}\ \bibnamefont {Duffy}},\
  }\href@noop {} {\bibfield  {journal} {\bibinfo  {journal} {Mon. Not. R.
  Astron. Soc.}\ }\textbf {\bibinfo {volume} {452}},\ \bibinfo {pages} {1217}
  (\bibinfo {year} {2015})},\ \Eprint {http://arxiv.org/abs/1502.00391}
  {1502.00391} \BibitemShut {NoStop}%
\bibitem [{\citenamefont {Ludlow}\ \emph {et~al.}(2016)\citenamefont {Ludlow},
  \citenamefont {Bose}, \citenamefont {Angulo}, \citenamefont {Wang},
  \citenamefont {Hellwing}, \citenamefont {Navarro}, \citenamefont {Cole},\
  and\ \citenamefont {Frenk}}]{ludlow2016mass}%
  \BibitemOpen
  \bibfield  {author} {\bibinfo {author} {\bibfnamefont {A.~D.}\ \bibnamefont
  {Ludlow}}, \bibinfo {author} {\bibfnamefont {S.}~\bibnamefont {Bose}},
  \bibinfo {author} {\bibfnamefont {R.~E.}\ \bibnamefont {Angulo}}, \bibinfo
  {author} {\bibfnamefont {L.}~\bibnamefont {Wang}}, \bibinfo {author}
  {\bibfnamefont {W.~A.}\ \bibnamefont {Hellwing}}, \bibinfo {author}
  {\bibfnamefont {J.~F.}\ \bibnamefont {Navarro}}, \bibinfo {author}
  {\bibfnamefont {S.}~\bibnamefont {Cole}}, \ and\ \bibinfo {author}
  {\bibfnamefont {C.~S.}\ \bibnamefont {Frenk}},\ }\href@noop {} {\bibfield
  {journal} {\bibinfo  {journal} {Mon. Not. R. Astron. Soc.}\ }\textbf
  {\bibinfo {volume} {460}},\ \bibinfo {pages} {1214} (\bibinfo {year}
  {2016})},\ \Eprint {http://arxiv.org/abs/1601.02624} {1601.02624}
  \BibitemShut {NoStop}%
\bibitem [{\citenamefont {Prada}\ \emph {et~al.}(2012)\citenamefont {Prada},
  \citenamefont {Klypin}, \citenamefont {Cuesta}, \citenamefont
  {Betancort-Rijo},\ and\ \citenamefont {Primack}}]{prada2012halo}%
  \BibitemOpen
  \bibfield  {author} {\bibinfo {author} {\bibfnamefont {F.}~\bibnamefont
  {Prada}}, \bibinfo {author} {\bibfnamefont {A.~A.}\ \bibnamefont {Klypin}},
  \bibinfo {author} {\bibfnamefont {A.~J.}\ \bibnamefont {Cuesta}}, \bibinfo
  {author} {\bibfnamefont {J.~E.}\ \bibnamefont {Betancort-Rijo}}, \ and\
  \bibinfo {author} {\bibfnamefont {J.}~\bibnamefont {Primack}},\ }\href@noop
  {} {\bibfield  {journal} {\bibinfo  {journal} {Mon. Not. R. Astron. Soc.}\
  }\textbf {\bibinfo {volume} {423}},\ \bibinfo {pages} {3018} (\bibinfo {year}
  {2012})},\ \Eprint {http://arxiv.org/abs/1104.5130} {1104.5130} \BibitemShut
  {NoStop}%
\bibitem [{\citenamefont {Okoli}\ and\ \citenamefont
  {Afshordi}(2016)}]{okoli2015concentration}%
  \BibitemOpen
  \bibfield  {author} {\bibinfo {author} {\bibfnamefont {C.}~\bibnamefont
  {Okoli}}\ and\ \bibinfo {author} {\bibfnamefont {N.}~\bibnamefont
  {Afshordi}},\ }\href@noop {} {\bibfield  {journal} {\bibinfo  {journal} {Mon.
  Not. R. Astron. Soc.}\ }\textbf {\bibinfo {volume} {456}},\ \bibinfo {pages}
  {3068} (\bibinfo {year} {2016})},\ \Eprint {http://arxiv.org/abs/1510.03868}
  {1510.03868} \BibitemShut {NoStop}%
\bibitem [{\citenamefont {Diemer}\ and\ \citenamefont
  {Kravtsov}(2015)}]{diemer2015universal}%
  \BibitemOpen
  \bibfield  {author} {\bibinfo {author} {\bibfnamefont {B.}~\bibnamefont
  {Diemer}}\ and\ \bibinfo {author} {\bibfnamefont {A.~V.}\ \bibnamefont
  {Kravtsov}},\ }\href@noop {} {\bibfield  {journal} {\bibinfo  {journal}
  {Astrophys. J.}\ }\textbf {\bibinfo {volume} {799}},\ \bibinfo {pages} {108}
  (\bibinfo {year} {2015})},\ \Eprint {http://arxiv.org/abs/1407.4730}
  {1407.4730} \BibitemShut {NoStop}%
\bibitem [{\citenamefont {Diemer}\ and\ \citenamefont
  {Joyce}(2019)}]{diemer2019accurate}%
  \BibitemOpen
  \bibfield  {author} {\bibinfo {author} {\bibfnamefont {B.}~\bibnamefont
  {Diemer}}\ and\ \bibinfo {author} {\bibfnamefont {M.}~\bibnamefont {Joyce}},\
  }\href@noop {} {\bibfield  {journal} {\bibinfo  {journal} {Astrophys. J.}\
  }\textbf {\bibinfo {volume} {871}},\ \bibinfo {pages} {168} (\bibinfo {year}
  {2019})},\ \Eprint {http://arxiv.org/abs/1809.07326} {1809.07326}
  \BibitemShut {NoStop}%
\bibitem [{\citenamefont {Bond}\ and\ \citenamefont
  {Myers}(1996{\natexlab{a}})}]{bond1996peak}%
  \BibitemOpen
  \bibfield  {author} {\bibinfo {author} {\bibfnamefont {J.}~\bibnamefont
  {Bond}}\ and\ \bibinfo {author} {\bibfnamefont {S.}~\bibnamefont {Myers}},\
  }\href@noop {} {\bibfield  {journal} {\bibinfo  {journal} {Astrophys. J.
  Suppl. Ser.}\ }\textbf {\bibinfo {volume} {103}},\ \bibinfo {pages} {1}
  (\bibinfo {year} {1996}{\natexlab{a}})}\BibitemShut {NoStop}%
\bibitem [{\citenamefont {Ascasibar}\ and\ \citenamefont
  {Gottl{\"o}ber}(2008)}]{ascasibar2008dynamical}%
  \BibitemOpen
  \bibfield  {author} {\bibinfo {author} {\bibfnamefont {Y.}~\bibnamefont
  {Ascasibar}}\ and\ \bibinfo {author} {\bibfnamefont {S.}~\bibnamefont
  {Gottl{\"o}ber}},\ }\href@noop {} {\bibfield  {journal} {\bibinfo  {journal}
  {Mon. Not. R. Astron. Soc.}\ }\textbf {\bibinfo {volume} {386}},\ \bibinfo
  {pages} {2022} (\bibinfo {year} {2008})},\ \Eprint
  {http://arxiv.org/abs/0802.4348} {0802.4348} \BibitemShut {NoStop}%
\bibitem [{\citenamefont {Bardeen}\ \emph {et~al.}(1986)\citenamefont
  {Bardeen}, \citenamefont {Bond}, \citenamefont {Kaiser},\ and\ \citenamefont
  {Szalay}}]{bardeen1986statistics}%
  \BibitemOpen
  \bibfield  {author} {\bibinfo {author} {\bibfnamefont {J.~M.}\ \bibnamefont
  {Bardeen}}, \bibinfo {author} {\bibfnamefont {J.~R.}\ \bibnamefont {Bond}},
  \bibinfo {author} {\bibfnamefont {N.}~\bibnamefont {Kaiser}}, \ and\ \bibinfo
  {author} {\bibfnamefont {A.~S.}\ \bibnamefont {Szalay}},\ }\href@noop {}
  {\bibfield  {journal} {\bibinfo  {journal} {Astrophys. J.}\ }\textbf
  {\bibinfo {volume} {304}},\ \bibinfo {pages} {15} (\bibinfo {year}
  {1986})}\BibitemShut {NoStop}%
\bibitem [{\citenamefont {Salopek}\ \emph {et~al.}(1989)\citenamefont
  {Salopek}, \citenamefont {Bond},\ and\ \citenamefont
  {Bardeen}}]{salopek1989designing}%
  \BibitemOpen
  \bibfield  {author} {\bibinfo {author} {\bibfnamefont {D.~S.}\ \bibnamefont
  {Salopek}}, \bibinfo {author} {\bibfnamefont {J.~R.}\ \bibnamefont {Bond}}, \
  and\ \bibinfo {author} {\bibfnamefont {J.~M.}\ \bibnamefont {Bardeen}},\
  }\href@noop {} {\bibfield  {journal} {\bibinfo  {journal} {Phys. Rev. D}\
  }\textbf {\bibinfo {volume} {40}},\ \bibinfo {pages} {1753} (\bibinfo {year}
  {1989})}\BibitemShut {NoStop}%
\bibitem [{\citenamefont {Starobinskij}(1992)}]{starobinsky1992}%
  \BibitemOpen
  \bibfield  {author} {\bibinfo {author} {\bibfnamefont {A.~A.}\ \bibnamefont
  {Starobinskij}},\ }\href@noop {} {\bibfield  {journal} {\bibinfo  {journal}
  {JETP Lett.}\ }\textbf {\bibinfo {volume} {55}},\ \bibinfo {pages} {489}
  (\bibinfo {year} {1992})}\BibitemShut {NoStop}%
\bibitem [{\citenamefont {Ivanov}\ \emph {et~al.}(1994)\citenamefont {Ivanov},
  \citenamefont {Naselsky},\ and\ \citenamefont
  {Novikov}}]{ivanov1994inflation}%
  \BibitemOpen
  \bibfield  {author} {\bibinfo {author} {\bibfnamefont {P.}~\bibnamefont
  {Ivanov}}, \bibinfo {author} {\bibfnamefont {P.}~\bibnamefont {Naselsky}}, \
  and\ \bibinfo {author} {\bibfnamefont {I.}~\bibnamefont {Novikov}},\
  }\href@noop {} {\bibfield  {journal} {\bibinfo  {journal} {Phys. Rev. D}\
  }\textbf {\bibinfo {volume} {50}},\ \bibinfo {pages} {7173} (\bibinfo {year}
  {1994})}\BibitemShut {NoStop}%
\bibitem [{\citenamefont {Randall}\ \emph {et~al.}(1996)\citenamefont
  {Randall}, \citenamefont {Solja{\v{c}}i{\'c}},\ and\ \citenamefont
  {Guth}}]{randall1996supernatural}%
  \BibitemOpen
  \bibfield  {author} {\bibinfo {author} {\bibfnamefont {L.}~\bibnamefont
  {Randall}}, \bibinfo {author} {\bibfnamefont {M.}~\bibnamefont
  {Solja{\v{c}}i{\'c}}}, \ and\ \bibinfo {author} {\bibfnamefont {A.~H.}\
  \bibnamefont {Guth}},\ }\href@noop {} {\bibfield  {journal} {\bibinfo
  {journal} {Nucl. Phys.}\ }\textbf {\bibinfo {volume} {B472}},\ \bibinfo
  {pages} {377} (\bibinfo {year} {1996})},\ \Eprint
  {http://arxiv.org/abs/hep-ph/9512439} {hep-ph/9512439} \BibitemShut {NoStop}%
\bibitem [{\citenamefont {Starobinsky}(1998)}]{starobinsky1998beyond}%
  \BibitemOpen
  \bibfield  {author} {\bibinfo {author} {\bibfnamefont {A.~A.}\ \bibnamefont
  {Starobinsky}},\ }\href@noop {} {\bibfield  {journal} {\bibinfo  {journal}
  {Grav. Cosmol.}\ }\textbf {\bibinfo {volume} {4}},\ \bibinfo {pages} {489}
  (\bibinfo {year} {1998})},\ \Eprint {http://arxiv.org/abs/astro-ph/9811360}
  {astro-ph/9811360} \BibitemShut {NoStop}%
\bibitem [{\citenamefont {Martin}\ \emph {et~al.}(2000)\citenamefont {Martin},
  \citenamefont {Riazuelo},\ and\ \citenamefont
  {Sakellariadou}}]{martin2000nonvacuum}%
  \BibitemOpen
  \bibfield  {author} {\bibinfo {author} {\bibfnamefont {J.}~\bibnamefont
  {Martin}}, \bibinfo {author} {\bibfnamefont {A.}~\bibnamefont {Riazuelo}}, \
  and\ \bibinfo {author} {\bibfnamefont {M.}~\bibnamefont {Sakellariadou}},\
  }\href@noop {} {\bibfield  {journal} {\bibinfo  {journal} {Phys. Rev. D}\
  }\textbf {\bibinfo {volume} {61}},\ \bibinfo {pages} {083518} (\bibinfo
  {year} {2000})},\ \Eprint {http://arxiv.org/abs/astro-ph/9904167}
  {astro-ph/9904167} \BibitemShut {NoStop}%
\bibitem [{\citenamefont {Chung}\ \emph {et~al.}(2000)\citenamefont {Chung},
  \citenamefont {Kolb}, \citenamefont {Riotto},\ and\ \citenamefont
  {Tkachev}}]{chung2000probing}%
  \BibitemOpen
  \bibfield  {author} {\bibinfo {author} {\bibfnamefont {D.~J.~H.}\
  \bibnamefont {Chung}}, \bibinfo {author} {\bibfnamefont {E.~W.}\ \bibnamefont
  {Kolb}}, \bibinfo {author} {\bibfnamefont {A.}~\bibnamefont {Riotto}}, \ and\
  \bibinfo {author} {\bibfnamefont {I.~I.}\ \bibnamefont {Tkachev}},\
  }\href@noop {} {\bibfield  {journal} {\bibinfo  {journal} {Phys. Rev. D}\
  }\textbf {\bibinfo {volume} {62}},\ \bibinfo {pages} {043508} (\bibinfo
  {year} {2000})},\ \Eprint {http://arxiv.org/abs/hep-ph/9910437}
  {hep-ph/9910437} \BibitemShut {NoStop}%
\bibitem [{\citenamefont {Barnaby}\ and\ \citenamefont
  {Huang}(2009)}]{barnaby2009particle}%
  \BibitemOpen
  \bibfield  {author} {\bibinfo {author} {\bibfnamefont {N.}~\bibnamefont
  {Barnaby}}\ and\ \bibinfo {author} {\bibfnamefont {Z.}~\bibnamefont
  {Huang}},\ }\href@noop {} {\bibfield  {journal} {\bibinfo  {journal} {Phys.
  Rev. D}\ }\textbf {\bibinfo {volume} {80}},\ \bibinfo {pages} {126018}
  (\bibinfo {year} {2009})},\ \Eprint {http://arxiv.org/abs/0909.0751}
  {0909.0751} \BibitemShut {NoStop}%
\bibitem [{\citenamefont {Barnaby}(2010)}]{barnaby2010features}%
  \BibitemOpen
  \bibfield  {author} {\bibinfo {author} {\bibfnamefont {N.}~\bibnamefont
  {Barnaby}},\ }\href@noop {} {\bibfield  {journal} {\bibinfo  {journal} {Phys.
  Rev. D}\ }\textbf {\bibinfo {volume} {82}},\ \bibinfo {pages} {106009}
  (\bibinfo {year} {2010})},\ \Eprint {http://arxiv.org/abs/1006.4615}
  {1006.4615} \BibitemShut {NoStop}%
\bibitem [{\citenamefont {Bugaev}\ and\ \citenamefont
  {Klimai}(2011)}]{bugaev2011curvature}%
  \BibitemOpen
  \bibfield  {author} {\bibinfo {author} {\bibfnamefont {E.}~\bibnamefont
  {Bugaev}}\ and\ \bibinfo {author} {\bibfnamefont {P.}~\bibnamefont
  {Klimai}},\ }\href@noop {} {\bibfield  {journal} {\bibinfo  {journal} {J.
  Cosmol. Astropart. Phys.}\ }\textbf {\bibinfo {volume} {11}},\ \bibinfo
  {pages} {028} (\bibinfo {year} {2011})},\ \Eprint
  {http://arxiv.org/abs/1107.3754} {1107.3754} \BibitemShut {NoStop}%
\bibitem [{\citenamefont {Jungman}\ \emph {et~al.}(1996)\citenamefont
  {Jungman}, \citenamefont {Kamionkowski},\ and\ \citenamefont
  {Griest}}]{jungman1996supersymmetric}%
  \BibitemOpen
  \bibfield  {author} {\bibinfo {author} {\bibfnamefont {G.}~\bibnamefont
  {Jungman}}, \bibinfo {author} {\bibfnamefont {M.}~\bibnamefont
  {Kamionkowski}}, \ and\ \bibinfo {author} {\bibfnamefont {K.}~\bibnamefont
  {Griest}},\ }\href@noop {} {\bibfield  {journal} {\bibinfo  {journal} {Phys.
  Rep.}\ }\textbf {\bibinfo {volume} {267}},\ \bibinfo {pages} {195} (\bibinfo
  {year} {1996})},\ \Eprint {http://arxiv.org/abs/hep-ph/9506380}
  {hep-ph/9506380} \BibitemShut {NoStop}%
\bibitem [{\citenamefont {Bode}\ \emph {et~al.}(2001)\citenamefont {Bode},
  \citenamefont {Ostriker},\ and\ \citenamefont {Turok}}]{bode2001halo}%
  \BibitemOpen
  \bibfield  {author} {\bibinfo {author} {\bibfnamefont {P.}~\bibnamefont
  {Bode}}, \bibinfo {author} {\bibfnamefont {J.~P.}\ \bibnamefont {Ostriker}},
  \ and\ \bibinfo {author} {\bibfnamefont {N.}~\bibnamefont {Turok}},\
  }\href@noop {} {\bibfield  {journal} {\bibinfo  {journal} {Astrophys. J.}\
  }\textbf {\bibinfo {volume} {556}},\ \bibinfo {pages} {93} (\bibinfo {year}
  {2001})},\ \Eprint {http://arxiv.org/abs/astro-ph/0010389} {astro-ph/0010389}
  \BibitemShut {NoStop}%
\bibitem [{\citenamefont {Viel}\ \emph {et~al.}(2013)\citenamefont {Viel},
  \citenamefont {Becker}, \citenamefont {Bolton},\ and\ \citenamefont
  {Haehnelt}}]{viel2013warm}%
  \BibitemOpen
  \bibfield  {author} {\bibinfo {author} {\bibfnamefont {M.}~\bibnamefont
  {Viel}}, \bibinfo {author} {\bibfnamefont {G.~D.}\ \bibnamefont {Becker}},
  \bibinfo {author} {\bibfnamefont {J.~S.}\ \bibnamefont {Bolton}}, \ and\
  \bibinfo {author} {\bibfnamefont {M.~G.}\ \bibnamefont {Haehnelt}},\
  }\href@noop {} {\bibfield  {journal} {\bibinfo  {journal} {Phys. Rev. D}\
  }\textbf {\bibinfo {volume} {88}},\ \bibinfo {pages} {043502} (\bibinfo
  {year} {2013})},\ \Eprint {http://arxiv.org/abs/1306.2314} {1306.2314}
  \BibitemShut {NoStop}%
\bibitem [{\citenamefont {Ade}\ \emph {et~al.}(2016)\citenamefont {Ade} \emph
  {et~al.}}]{2016planck}%
  \BibitemOpen
  \bibfield  {author} {\bibinfo {author} {\bibfnamefont {P.~A.~R.}\
  \bibnamefont {Ade}} \emph {et~al.} (\bibinfo {collaboration} {Planck
  Collaboration}),\ }\href@noop {} {\bibfield  {journal} {\bibinfo  {journal}
  {Astron. Astrophys.}\ }\textbf {\bibinfo {volume} {594}},\ \bibinfo {pages}
  {A13} (\bibinfo {year} {2016})},\ \Eprint {http://arxiv.org/abs/1502.01589}
  {1502.01589} \BibitemShut {NoStop}%
\bibitem [{\citenamefont {Challinor}\ and\ \citenamefont
  {Lewis}(2011)}]{challinor2011linear}%
  \BibitemOpen
  \bibfield  {author} {\bibinfo {author} {\bibfnamefont {A.}~\bibnamefont
  {Challinor}}\ and\ \bibinfo {author} {\bibfnamefont {A.}~\bibnamefont
  {Lewis}},\ }\href@noop {} {\bibfield  {journal} {\bibinfo  {journal} {Phys.
  Rev. D}\ }\textbf {\bibinfo {volume} {84}},\ \bibinfo {pages} {043516}
  (\bibinfo {year} {2011})},\ \Eprint {http://arxiv.org/abs/1105.5292}
  {1105.5292} \BibitemShut {NoStop}%
\bibitem [{\citenamefont {Lewis}\ and\ \citenamefont
  {Challinor}(2007)}]{lewis200721}%
  \BibitemOpen
  \bibfield  {author} {\bibinfo {author} {\bibfnamefont {A.}~\bibnamefont
  {Lewis}}\ and\ \bibinfo {author} {\bibfnamefont {A.}~\bibnamefont
  {Challinor}},\ }\href@noop {} {\bibfield  {journal} {\bibinfo  {journal}
  {Phys. Rev. D}\ }\textbf {\bibinfo {volume} {76}},\ \bibinfo {pages} {083005}
  (\bibinfo {year} {2007})},\ \Eprint {http://arxiv.org/abs/astro-ph/0702600}
  {astro-ph/0702600} \BibitemShut {NoStop}%
\bibitem [{\citenamefont {Green}\ \emph {et~al.}(2004)\citenamefont {Green},
  \citenamefont {Hofmann},\ and\ \citenamefont {Schwarz}}]{green2004power}%
  \BibitemOpen
  \bibfield  {author} {\bibinfo {author} {\bibfnamefont {A.~M.}\ \bibnamefont
  {Green}}, \bibinfo {author} {\bibfnamefont {S.}~\bibnamefont {Hofmann}}, \
  and\ \bibinfo {author} {\bibfnamefont {D.~J.}\ \bibnamefont {Schwarz}},\
  }\href@noop {} {\bibfield  {journal} {\bibinfo  {journal} {Mon. Not. R.
  Astron. Soc.}\ }\textbf {\bibinfo {volume} {353}},\ \bibinfo {pages} {L23}
  (\bibinfo {year} {2004})},\ \Eprint {http://arxiv.org/abs/astro-ph/0309621}
  {astro-ph/0309621} \BibitemShut {NoStop}%
\bibitem [{\citenamefont {Hu}\ and\ \citenamefont
  {Sugiyama}(1996)}]{hu1996small}%
  \BibitemOpen
  \bibfield  {author} {\bibinfo {author} {\bibfnamefont {W.}~\bibnamefont
  {Hu}}\ and\ \bibinfo {author} {\bibfnamefont {N.}~\bibnamefont {Sugiyama}},\
  }\href@noop {} {\bibfield  {journal} {\bibinfo  {journal} {Astrophys. J.}\
  }\textbf {\bibinfo {volume} {471}},\ \bibinfo {pages} {542} (\bibinfo {year}
  {1996})},\ \Eprint {http://arxiv.org/abs/astro-ph/9510117} {astro-ph/9510117}
  \BibitemShut {NoStop}%
\bibitem [{\citenamefont {Springel}(2005)}]{springel2005cosmological}%
  \BibitemOpen
  \bibfield  {author} {\bibinfo {author} {\bibfnamefont {V.}~\bibnamefont
  {Springel}},\ }\href@noop {} {\bibfield  {journal} {\bibinfo  {journal} {Mon.
  Not. R. Astron. Soc.}\ }\textbf {\bibinfo {volume} {364}},\ \bibinfo {pages}
  {1105} (\bibinfo {year} {2005})},\ \Eprint
  {http://arxiv.org/abs/astro-ph/0505010} {astro-ph/0505010} \BibitemShut
  {NoStop}%
\bibitem [{\citenamefont {Springel}\ \emph {et~al.}(2001)\citenamefont
  {Springel}, \citenamefont {Yoshida},\ and\ \citenamefont
  {White}}]{springel2001gadget}%
  \BibitemOpen
  \bibfield  {author} {\bibinfo {author} {\bibfnamefont {V.}~\bibnamefont
  {Springel}}, \bibinfo {author} {\bibfnamefont {N.}~\bibnamefont {Yoshida}}, \
  and\ \bibinfo {author} {\bibfnamefont {S.~D.~M.}\ \bibnamefont {White}},\
  }\href@noop {} {\bibfield  {journal} {\bibinfo  {journal} {New Astron.}\
  }\textbf {\bibinfo {volume} {6}},\ \bibinfo {pages} {79} (\bibinfo {year}
  {2001})},\ \Eprint {http://arxiv.org/abs/astro-ph/0003162} {astro-ph/0003162}
  \BibitemShut {NoStop}%
\bibitem [{\citenamefont {Behroozi}\ \emph
  {et~al.}(2013{\natexlab{a}})\citenamefont {Behroozi}, \citenamefont
  {Wechsler},\ and\ \citenamefont {Wu}}]{behroozi2012rockstar}%
  \BibitemOpen
  \bibfield  {author} {\bibinfo {author} {\bibfnamefont {P.~S.}\ \bibnamefont
  {Behroozi}}, \bibinfo {author} {\bibfnamefont {R.~H.}\ \bibnamefont
  {Wechsler}}, \ and\ \bibinfo {author} {\bibfnamefont {H.-Y.}\ \bibnamefont
  {Wu}},\ }\href@noop {} {\bibfield  {journal} {\bibinfo  {journal} {Astrophys.
  J.}\ }\textbf {\bibinfo {volume} {762}},\ \bibinfo {pages} {109} (\bibinfo
  {year} {2013}{\natexlab{a}})},\ \Eprint {http://arxiv.org/abs/1110.4372}
  {1110.4372} \BibitemShut {NoStop}%
\bibitem [{\citenamefont {Angulo}\ \emph {et~al.}(2013)\citenamefont {Angulo},
  \citenamefont {Hahn},\ and\ \citenamefont {Abel}}]{angulo2013warm}%
  \BibitemOpen
  \bibfield  {author} {\bibinfo {author} {\bibfnamefont {R.~E.}\ \bibnamefont
  {Angulo}}, \bibinfo {author} {\bibfnamefont {O.}~\bibnamefont {Hahn}}, \ and\
  \bibinfo {author} {\bibfnamefont {T.}~\bibnamefont {Abel}},\ }\href@noop {}
  {\bibfield  {journal} {\bibinfo  {journal} {Mon. Not. R. Astron. Soc.}\
  }\textbf {\bibinfo {volume} {434}},\ \bibinfo {pages} {3337} (\bibinfo {year}
  {2013})},\ \Eprint {http://arxiv.org/abs/1304.2406} {1304.2406} \BibitemShut
  {NoStop}%
\bibitem [{\citenamefont {Lovell}\ \emph {et~al.}(2014)\citenamefont {Lovell},
  \citenamefont {Frenk}, \citenamefont {Eke}, \citenamefont {Jenkins},
  \citenamefont {Gao},\ and\ \citenamefont {Theuns}}]{lovell2014properties}%
  \BibitemOpen
  \bibfield  {author} {\bibinfo {author} {\bibfnamefont {M.~R.}\ \bibnamefont
  {Lovell}}, \bibinfo {author} {\bibfnamefont {C.~S.}\ \bibnamefont {Frenk}},
  \bibinfo {author} {\bibfnamefont {V.~R.}\ \bibnamefont {Eke}}, \bibinfo
  {author} {\bibfnamefont {A.}~\bibnamefont {Jenkins}}, \bibinfo {author}
  {\bibfnamefont {L.}~\bibnamefont {Gao}}, \ and\ \bibinfo {author}
  {\bibfnamefont {T.}~\bibnamefont {Theuns}},\ }\href@noop {} {\bibfield
  {journal} {\bibinfo  {journal} {Mon. Not. R. Astron. Soc.}\ }\textbf
  {\bibinfo {volume} {439}},\ \bibinfo {pages} {300} (\bibinfo {year}
  {2014})},\ \Eprint {http://arxiv.org/abs/1308.1399} {1308.1399} \BibitemShut
  {NoStop}%
\bibitem [{\citenamefont {Behroozi}\ \emph
  {et~al.}(2013{\natexlab{b}})\citenamefont {Behroozi}, \citenamefont
  {Wechsler}, \citenamefont {Wu}, \citenamefont {Busha}, \citenamefont
  {Klypin},\ and\ \citenamefont {Primack}}]{behroozi2012gravitationally}%
  \BibitemOpen
  \bibfield  {author} {\bibinfo {author} {\bibfnamefont {P.~S.}\ \bibnamefont
  {Behroozi}}, \bibinfo {author} {\bibfnamefont {R.~H.}\ \bibnamefont
  {Wechsler}}, \bibinfo {author} {\bibfnamefont {H.-Y.}\ \bibnamefont {Wu}},
  \bibinfo {author} {\bibfnamefont {M.~T.}\ \bibnamefont {Busha}}, \bibinfo
  {author} {\bibfnamefont {A.~A.}\ \bibnamefont {Klypin}}, \ and\ \bibinfo
  {author} {\bibfnamefont {J.~R.}\ \bibnamefont {Primack}},\ }\href@noop {}
  {\bibfield  {journal} {\bibinfo  {journal} {Astrophys. J.}\ }\textbf
  {\bibinfo {volume} {763}},\ \bibinfo {pages} {18} (\bibinfo {year}
  {2013}{\natexlab{b}})},\ \Eprint {http://arxiv.org/abs/1110.4370} {1110.4370}
  \BibitemShut {NoStop}%
\bibitem [{\citenamefont {Binney}\ and\ \citenamefont
  {Tremaine}(1987)}]{binney1987galactic}%
  \BibitemOpen
  \bibfield  {author} {\bibinfo {author} {\bibfnamefont {J.}~\bibnamefont
  {Binney}}\ and\ \bibinfo {author} {\bibfnamefont {S.}~\bibnamefont
  {Tremaine}},\ }\href@noop {} {\emph {\bibinfo {title} {Galactic Dynamics}}}\
  (\bibinfo  {publisher} {Princeton University Press},\ \bibinfo {address}
  {Princeton, NJ},\ \bibinfo {year} {1987})\BibitemShut {NoStop}%
\bibitem [{\citenamefont {Moore}\ \emph {et~al.}(1999)\citenamefont {Moore},
  \citenamefont {Quinn}, \citenamefont {Governato}, \citenamefont {Stadel},\
  and\ \citenamefont {Lake}}]{moore1999cold}%
  \BibitemOpen
  \bibfield  {author} {\bibinfo {author} {\bibfnamefont {B.}~\bibnamefont
  {Moore}}, \bibinfo {author} {\bibfnamefont {T.}~\bibnamefont {Quinn}},
  \bibinfo {author} {\bibfnamefont {F.}~\bibnamefont {Governato}}, \bibinfo
  {author} {\bibfnamefont {J.}~\bibnamefont {Stadel}}, \ and\ \bibinfo {author}
  {\bibfnamefont {G.}~\bibnamefont {Lake}},\ }\href@noop {} {\bibfield
  {journal} {\bibinfo  {journal} {Mon. Not. R. Astron. Soc.}\ }\textbf
  {\bibinfo {volume} {310}},\ \bibinfo {pages} {1147} (\bibinfo {year}
  {1999})},\ \Eprint {http://arxiv.org/abs/astro-ph/9903164} {astro-ph/9903164}
  \BibitemShut {NoStop}%
\bibitem [{\citenamefont {Berezinsky}\ \emph {et~al.}(1992)\citenamefont
  {Berezinsky}, \citenamefont {Gurevich},\ and\ \citenamefont
  {Zybin}}]{berezinsky1992distribution}%
  \BibitemOpen
  \bibfield  {author} {\bibinfo {author} {\bibfnamefont {V.~S.}\ \bibnamefont
  {Berezinsky}}, \bibinfo {author} {\bibfnamefont {A.~V.}\ \bibnamefont
  {Gurevich}}, \ and\ \bibinfo {author} {\bibfnamefont {K.~P.}\ \bibnamefont
  {Zybin}},\ }\href@noop {} {\bibfield  {journal} {\bibinfo  {journal} {Phys.
  Lett. B}\ }\textbf {\bibinfo {volume} {294}},\ \bibinfo {pages} {221}
  (\bibinfo {year} {1992})}\BibitemShut {NoStop}%
\bibitem [{\citenamefont {Lacey}\ and\ \citenamefont
  {Cole}(1993)}]{lacey1993merger}%
  \BibitemOpen
  \bibfield  {author} {\bibinfo {author} {\bibfnamefont {C.}~\bibnamefont
  {Lacey}}\ and\ \bibinfo {author} {\bibfnamefont {S.}~\bibnamefont {Cole}},\
  }\href@noop {} {\bibfield  {journal} {\bibinfo  {journal} {Mon. Not. R.
  Astron. Soc.}\ }\textbf {\bibinfo {volume} {262}},\ \bibinfo {pages} {627}
  (\bibinfo {year} {1993})}\BibitemShut {NoStop}%
\bibitem [{\citenamefont {Bond}\ and\ \citenamefont
  {Myers}(1996{\natexlab{b}})}]{bond1996peak2}%
  \BibitemOpen
  \bibfield  {author} {\bibinfo {author} {\bibfnamefont {J.}~\bibnamefont
  {Bond}}\ and\ \bibinfo {author} {\bibfnamefont {S.}~\bibnamefont {Myers}},\
  }\href@noop {} {\bibfield  {journal} {\bibinfo  {journal} {Astrophys. J.
  Suppl. Ser.}\ }\textbf {\bibinfo {volume} {103}},\ \bibinfo {pages} {41}
  (\bibinfo {year} {1996}{\natexlab{b}})}\BibitemShut {NoStop}%
\bibitem [{\citenamefont {Stein}\ \emph {et~al.}(2019)\citenamefont {Stein},
  \citenamefont {Alvarez},\ and\ \citenamefont {Bond}}]{stein2018mass}%
  \BibitemOpen
  \bibfield  {author} {\bibinfo {author} {\bibfnamefont {G.}~\bibnamefont
  {Stein}}, \bibinfo {author} {\bibfnamefont {M.~A.}\ \bibnamefont {Alvarez}},
  \ and\ \bibinfo {author} {\bibfnamefont {J.~R.}\ \bibnamefont {Bond}},\
  }\href@noop {} {\bibfield  {journal} {\bibinfo  {journal} {Mon. Not. R.
  Astron. Soc.}\ }\textbf {\bibinfo {volume} {483}},\ \bibinfo {pages} {2236}
  (\bibinfo {year} {2019})},\ \Eprint {http://arxiv.org/abs/1810.07727}
  {1810.07727} \BibitemShut {NoStop}%
\bibitem [{\citenamefont {Green}(2011)}]{green2011colour}%
  \BibitemOpen
  \bibfield  {author} {\bibinfo {author} {\bibfnamefont {D.~A.}\ \bibnamefont
  {Green}},\ }\href@noop {} {\bibfield  {journal} {\bibinfo  {journal} {Bull.
  Astron. Soc. India}\ }\textbf {\bibinfo {volume} {39}},\ \bibinfo {pages}
  {289} (\bibinfo {year} {2011})},\ \Eprint {http://arxiv.org/abs/1108.5083}
  {1108.5083} \BibitemShut {NoStop}%
\end{thebibliography}%

\end{document}